\documentclass[aps,pre,twocolumn,groupedaddress,superscriptaddress,showpacs]{revtex4-1}

\usepackage{graphicx}
\usepackage{color}
\usepackage{amsmath}
\usepackage{amsfonts}
\usepackage{amssymb}
\usepackage{dcolumn}
\usepackage{hyperref}
\usepackage{enumerate}
\usepackage{cancel}
\usepackage{mathtools}
\usepackage{multirow} 
\usepackage{siunitx}
\usepackage{changes}
\usepackage{booktabs} 
\usepackage{cleveref}
\Crefname{equation}{Eq.}{Eqs.}

\newcommand{\mb}{\mathbf}

\begin{document}

\title{Learning the Ising Model with Generative Neural Networks}
\author{Francesco D'Angelo}
\email{fdangelo@ethz.ch}
\affiliation{Institute of Neuroinformatics, University of Zurich and ETH Zurich, 8057, Zurich, Switzerland }
\author{Lucas B\"ottcher}
\email{lucasb@ethz.ch}
\affiliation{Computational Medicine, UCLA, 90095-1766, Los Angeles, United States}
\affiliation{Institute for Theoretical Physics, ETH Zurich, 8093, Zurich, Switzerland}
\affiliation{Center of Economic Research, ETH Zurich, 8092, Zurich, Switzerland} 
\date{\today}
\begin{abstract}
Recent advances in deep learning and neural networks have led to an increased interest in the application of generative models in statistical and condensed matter physics. In particular, restricted Boltzmann machines (RBMs) and variational autoencoders (VAEs) as specific classes of neural networks have been successfully applied in the context of physical feature extraction and representation learning. Despite these successes, however, there is only limited understanding of their representational properties and limitations. To better understand the representational characteristics of RBMs and VAEs, we study their ability to capture physical features of the Ising model at different temperatures. This approach allows us to quantitatively assess learned representations by comparing sample features with corresponding theoretical predictions. Our results suggest that the considered RBMs and convolutional VAEs are able to capture the temperature dependence of magnetization, energy, and spin-spin correlations. The samples generated by RBMs are more evenly distributed across temperature than those generated by VAEs. We also find that convolutional layers in VAEs are important to model spin correlations whereas RBMs achieve similar or even better performances without convolutional filters.
\end{abstract}
\maketitle
\section{Introduction}
After the successful application of deep learning and neural networks in speech and pattern recognition~\cite{lecun2015deep}, there is an increased interest in applying generative models in condensed matter physics and other fields~\cite{nguyen2017inverse,davidsen2019deep}. For example, restricted Boltzmann machines (RBMs)~\cite{ackley1985learning,hinton1989deterministic,hinton2002training}, as one class of stochastic neural networks, were used to study phase transitions~\cite{kim2018smallest,efthymiou2019super}, represent wave functions~\cite{carleo2017solving,torlai2018neural}, and extract features from physical systems~\cite{mehta2019high,wu2019toward}. Furthermore, variational autoencoders (VAEs)~\cite{kingma2013auto} have been applied to different physical representation learning problems~\cite{wetzel2017unsupervised}. In addition to the application of generative neural networks to physical systems, connections between statistical physics and the theoretical description of certain neural network models helped to gain insights into their learning dynamics~\cite{iso2018scale,li2018neural,koch2018mutual,mehta2019high}.

Despite these developments, the representational characteristics and limitations of neural networks have been only partially explored. To better understand the representational properties of RBMs and VAEs, we study their ability to capture the temperature dependence of physical features of the Ising model. This allows us to quantitatively compare learned distributions with corresponding theoretical predictions of a well-characterized physical system. We train both neural networks with realizations of Ising configurations and use the trained neural networks to generate new samples. Previous studies utilized single RBMs for each temperature to learn the distribution of Ising configurations that consist of a few dozen spins~\cite{torlai2016learning,morningstar2017deep,iso2018scale}. Our work complements Refs.~\cite{torlai2016learning,morningstar2017deep,iso2018scale} in three ways. First, we use system sizes that are about one order of magnitude larger than in previous studies. This has profound implications for monitoring the learning progress of RBMs. For the system sizes we consider, it is computationally not feasible anymore to directly evaluate the partition function as outlined in Ref.~\cite{torlai2016learning} and we therefore provide an overview of loss-function approximations that allow us to monitor RBMs during training. Second, instead of using single machines for each temperature, we focus on the training of one generative neural network for all temperatures and use classification networks to determine the temperatures of generated Ising configurations. Our results indicate that this type of training can improve the quality of VAE samples substantially. Third, in addition to RBMs, we also consider non-convolutional and convolutional VAEs and quantitatively compare different neural network architectures in terms of their ability to capture magnetization, energy, and spin-spin correlations. 

Our paper proceeds as follows. In Sec.~\ref{sec:models}, we give a brief introduction to generative models and summarize the concepts that are necessary to train RBMs and VAEs. To keep track of the learning progress of both models, we provide an overview of different monitoring techniques in Sec.~\ref{sec:monitoring}. In Sec.~\ref{sec:learning}, we apply RBMs and VAEs to learn the distribution of Ising spin configurations for different temperatures. We conclude our paper and discuss our results in Sec.~\ref{sec:discussion}.
\section{Generative models}
\label{sec:models}
Generative models are used to approximate the distribution of a dataset 
$\mathcal{D} = \{ x_1, \dots, x_m \}$ whose entities are samples $x_i \sim \hat{p}$ drawn from a distribution $\hat{p}$ ($1\leq i \leq m$). Usually, the distribution $\hat{p}$ is unknown and thus approximated by a generative model distribution $p(\theta)$, where $\theta$ is a corresponding parameter set. We can think of generative models as neural networks whose underlying weights and activation function parameters are described by $\theta$. Once trained, generative models are used to generate new samples similar to the ones of the considered dataset $\mathcal{D}$~\cite{mehta2019high}. To quantitatively study the representational power of generative models, we consider two specific architectures: (i) RBMs and (ii) VAEs. We introduce the basic concepts behind RBMs and VAEs in Secs.~\ref{sec:rbm} and \ref{sec:vae}.
\subsection{Restricted Boltzmann machine}
\label{sec:rbm}
\begin{figure}
\centering
\includegraphics[width=0.35\textwidth]{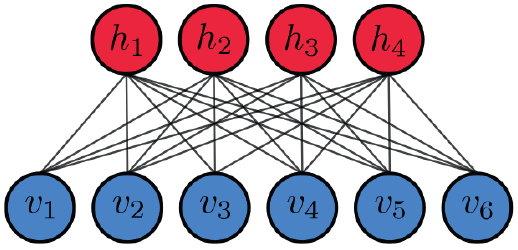}
\caption{\textbf{Restricted Boltzmann machine.} An RBM is composed of one visible layer (blue) and one hidden layer (red). In this example, the respective layers consist of six visible units $\left\{v_i\right\}_{i\in\{1,\dots,6\}}$ and four hidden units $\left\{h_i\right\}_{i\in \{1,\dots,4\}}$. The network structure underlying an RBM is bipartite.}
\label{fig:rbm}
\end{figure}
RBMs are a particular type of stochastic neural networks and were first introduced by Smolensky in 1986 under the name \emph{Harmonium}~\cite{smolensky1chapter}. In the early 2000s, Hinton developed some efficient training algorithms that made it possible to apply RBMs in different learning tasks~\cite{hinton2002training,hinton2012practical}. The network structure of an RBM is composed of one visible and one hidden layer. In contrast to Boltzmann machines~\cite{ackley1985learning}, no intra-layer connections are present in RBMs (i.e., the network structure of RBMs is bipartite). We show an illustration of an RBM network in Fig.~\ref{fig:rbm}. 

Mathematically, we describe the visible layer by a vector $\mathbf{v}=\left(v_1,\dots,v_m\right)^T$, which consists of $m$ visible units. Each visible unit represents one element of the dataset $\mathcal{D}$. Similarly, we represent the hidden layer by a vector $\mathbf{h}=\left(h_1,\dots,h_n\right)^T$, which is composed of $n$ hidden units. To model the distribution of a certain dataset, hidden units serve as additional degrees of freedom to capture the complex interaction between the original variables. Visible and hidden units are binary (i.e., $v_i\in\{0,1\}$ ($1\leq i \leq m$) and $h_j\in \{0,1\}$ ($1\leq j \leq n$)). We note that there also exist other formulations with continuous degrees of freedom~\cite{schmah2009generative}. In an RBM, connections between units $v_i$ and $h_j$ are undirected and we describe their weights by a weight matrix $W\in \mathbb{R}^{m\times n}$ with elements $w_{ij} \in \mathbb{R}$. Furthermore, we use $\mathbf{b} \in \mathbb{R}^m$ and $\mathbf{c} \in \mathbb{R}^n$ to account for biases in the visible and hidden layers. According to these definitions, there is an energy
\begin{equation}
E(\mb{v}, \mb{h}) = -\mb{b}^T \mb{v} - \mb{c}^T \mb{h} - \mb{v}^T W \mb{h}
\end{equation}
associated with every configuration $({\bf v}, {\bf h})$. The probability of the system to be found in a certain configuration $({\bf v}, {\bf h})$ is described by the Boltzmann distribution~\cite{ackley1985learning,mehta2019high}
\begin{equation}
p({\bf v},{\bf h}) = \frac{1}{Z} e^{-E(\mb{v}, \mb{h})},
\label{eq:boltzmann_prob}
\end{equation}
where $Z=\sum_{\{{\bf v},{\bf h}\}} e^{-E(\mb{v}, \mb{h})}$ is the canonical partition function. The sum $\sum_{\{{\bf v},{\bf h}\}}$ is taken over all possible configurations $\{{\bf v},{\bf h}\}$.

Due to the absence of intra-layer connections in RBMs, hidden (visible) variables are mutually independent given the visible (hidden) ones. Therefore, the corresponding conditional probabilities are~\cite{li2018exponential}
\begin{equation}
p(\mb{v}|\mb{h}) = \prod_{i=1}^m p(v_i|\mb{h}) \quad \text{and} \quad p(\mb{h}|\mb{v}) = \prod_{j=1}^n p(h_j|\mb{v})\,,
\label{eq:product_cond}
\end{equation}
where
\begin{align}
p(h_j=1| \mb{v}) &= \sigma \left( c_j+\sum_{i=1}^m w_{ij} v_i \right)
\,,\label{eqn:hidde_vis_prob1} 
\\  
p(v_i=1| \mb{h}) &=  \sigma \left(  b_i+\sum_{j=1}^n w_{ij} h_j \right)\, ,   
\label{eqn:hidde_vis_prob2} 
\end{align}
and $\sigma(x)=1/\left[1+\exp(-x)\right]$ denotes the sigmoid function.
\begin{figure*}
\centering
\includegraphics[width=\textwidth]{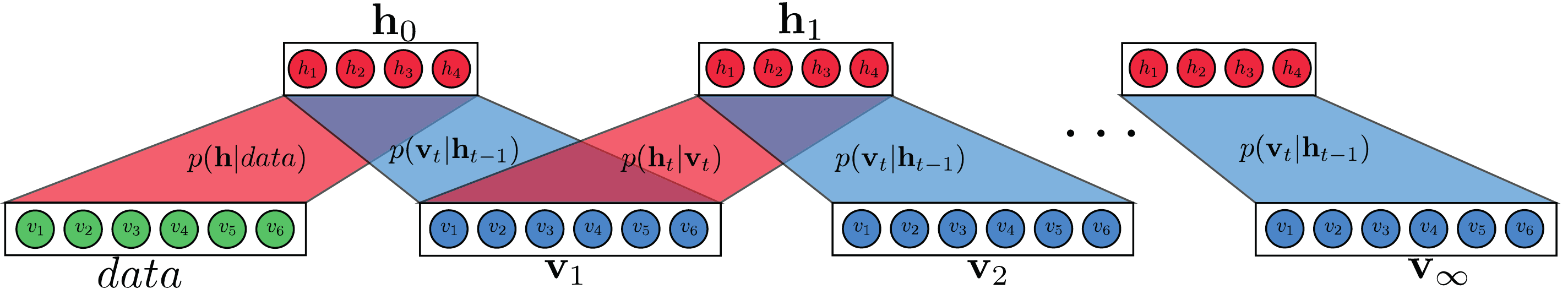}
\caption{\textbf{Block Gibbs sampling in an RBM.} We show an example of an RBM with a visible layer that consists of six visible units and a hidden layer that consists of four hidden units. Because of the bipartite network structure of RBMs, units within one layer can be grouped together and updated in parallel (\emph{block Gibbs sampling}). Initially, visible units (green) are determined by the dataset. Hidden (red) and visible (blue) units are then updated in an alternating manner.}
\label{fig:CD}
\end{figure*}

We now consider a dataset $\mathcal{D}$ that consists of $N$ i.i.d.~realizations $\left\{{\bf x}_1,\dots,{\bf x}_N\right\}$. Each realization ${\bf x}_k=\left(x_{k 1},\dots,x_{k m}\right)^T$ ($1\leq k \leq N$) in turn consists of $m$ binary elements. We associate these elements with the visible units of an RBM. The corresponding log-likelihood function is~\cite{li2018exponential}
\begin{equation}
\begin{split}
\log \mathcal{L}\left(\mb{x},\theta\right) &= \log p(\mb{x}_1, \dots, \mb{x}_N ; \theta)  = \log \prod_{k=1}^N p(\mb{x}_k ; \theta) \\ &= \sum_{k=1}^N \log p(\mb{x}_k ; \theta) = \sum_{k=1}^N \log \sum_{\{\mathbf{h}\}} \frac{e^{-E(\mathbf{x}_k, \mathbf{h})}}{Z} \\
&= \sum_{k=1}^N \log \left[  \frac{1}{Z} e^{\mb{b}^T \mb{x}_k} \prod_{j=1}^n \left(1 +  e^{(c_j + \sum_{i=1}^m w_{ij} x_{k i})}\right) \right] \\
	&= - N \log (Z) + \sum_{k=1}^N \mb{b}^T \mb{x}_k \\
	& + \sum_{k=1}^N\sum_{j=1}^n \log \left(1 +  e^{(c_j + \sum_{i=1}^m w_{ij} x_{k i})}\right)\,,
\end{split} 
\label{eq:log_l_RBM}
\end{equation}
where the parameter set $\theta$ describes weights and biases and the partition function is
\begin{equation}
Z  = \sum_{\{\mb{h}\}} e^{\mb{c}^T \mb{h}} \prod_{i=1}^m \left(1 + e^{b_i + \sum_{j=1}^n w_{ij} h_j } \right)\,.
\label{eqn:partition}
\end{equation}
Based on Eq.~\eqref{eq:log_l_RBM}, we perform a maximum-likelihood estimation of RBM parameters. We may express the log-likelihood function in terms of the free energy $\mathcal{F}=-\log Z$~\cite{huang2009introduction}. This yields $\log \mathcal{L} = N \mathcal{F}-\sum_{k=1}^N \mathcal{F}_c(\mb{x}_k)$, where $\mathcal{F}_c(\mb{x}_k) = - \log \left( \sum_{\{\mathbf{h}\}} e^{-E(\mathbf{x}_k, \mathbf{h})} \right)$ is the (clamped) free energy of an RBM whose visible states are clamped to the elements of ${\bf x}_k$~\cite{NIPS2015_5788}.

To train an RBM, we have to find a set of parameters $\theta$ that minimizes the difference between the distribution $p_\theta$ of the machine and the distribution $\hat{p}$ of the data. One possibility is to quantify the dissimilarity of these two distributions in terms of the Kullback-Leibler (KL) divergence (i.e., relative entropy)~\cite{ackley1985learning}
\begin{equation}
D_{\text{KL}}\big(\hat{p}(\mb{x}) || p_\theta(\mb{v})\big) = \sum_{\mb{x}} \hat{p}(\mb{x}) \log \frac{\hat{p}(\mb{x})}{p_\theta(\mb{v})} \geq 0 \, . 
\label{eqn:kl}
\end{equation}
For continuous distributions, the sum in Eq.~\eqref{eqn:kl} has to be replaced by a corresponding integral.
As an alternative to KL-divergence minimization, it is possible to minimize the negative log-likelihood.
This optimization problem can be solved using a stochastic gradient descent algorithm~\cite{bottou2018optimization}, so that the model parameters are iteratively updated: 
\begin{equation}
\theta^{t+1}_i = \theta^t_i - \eta \frac{\partial \mathcal{R}(\mathcal{D}, \theta^t)}{\partial \theta_i^t} \,,
\label{eq:grad}
\end{equation}
where $\mathcal{R}(\mathcal{D}, \theta)$ denotes the loss function (i.e., negative log-likelihood or KL divergence) and $\eta$ is the corresponding learning rate. 
In the case of an RBM, the derivatives of the loss function are
\begin{align}
\frac{\partial \mathcal{R}(\mathcal{D}, \theta)}{\partial \theta} &= \mathbb{E}_{\mb{x} \sim \hat{p}(\mb{x})} \bigg[ \frac{\partial E}{\partial \theta}  \bigg]- \mathbb{E}_{\mb{v}\sim p(\mb{v})} \bigg[  \frac{\partial{E}}{\partial \theta} \bigg] \,, \label{eq:gradients_gen}\\
	\frac{\partial \mathcal{R}(\mathcal{D}, \theta)}{\partial w_{ij}} &= \bigg\langle \hat{p}(h_j = 1| \mb{x}) x_i  \bigg\rangle_{\mb{x} \sim \hat{p}(\mb{x})} \nonumber \\
	&-\bigg\langle  p(h_j = 1| \mb{v}) v_i \bigg\rangle_{\mb{v} \sim p(\mb{v})} \,, \label{eq:gradients_w}\\
	\frac{\partial \mathcal{R}(\mathcal{D}, \theta)}{\partial b_j} &= \big\langle  x_i  \big\rangle_{\mb{x} \sim \hat{p}(\mb{x})} - \big\langle  v_i \big\rangle_{\mb{v} \sim p(\mb{v})} \,, \label{eq:gradients_b}\\
	\frac{\partial \mathcal{R}(\mathcal{D}, \theta)}{\partial c_{i}} &= \big\langle \hat{p}(h_j = 1| \mb{x}) \big\rangle_{\mb{x} \sim \hat{p}(\mb{x})} - \ \big\langle  p(h_j = 1| \mb{v})  \big\rangle_{\mb{v} \sim p(\mb{v})} \,, \label{eq:gradients_c}
\end{align} 
where $\langle \cdot \rangle$ indicates the ensemble average over different realizations of the quantities of interest.
According to \Cref{eq:gradients_w,eq:gradients_b,eq:gradients_c}, weights and biases are updated to minimize the differences between data and model distributions. 

The bipartite network structure of RBMs allows to update units within one layer in parallel (see \Cref{eqn:hidde_vis_prob1,eqn:hidde_vis_prob2}). We illustrate this so-called \emph{block Gibbs sampling} procedure in Fig.~\ref{fig:CD}. After a certain number of updates of the visible and hidden layers, the model distribution $p_\theta$ equilibrates. It has been shown empirically~\cite{hinton2002training} that a small number of updates may be sufficient to train an RBM according to \Cref{eq:gradients_w,eq:gradients_b,eq:gradients_c}. This technique is also known as \emph{$k$-contrastive divergence} (CD-$k$), where $k$ denotes the number of Gibbs sampling steps. An alternative to CD-$k$ methods is \emph{persistent contrastive divergence} (PCD)~\cite{tieleman2008training}. Instead of using the environment data as initial condition for the sampling process (see Fig.~\ref{fig:CD}), PCD uses the final state of the model from the previous sampling process to initialize the current one. The idea behind PCD is that the model distribution only changes slightly after updating the model parameters $\theta$ according to Eq.~\eqref{eq:grad} when using a small learning rate. For further details on the RBM training process, see Sec.~\ref{sec:sep_temps}.
\subsection{Variational autoencoder}
\label{sec:vae}
A variational autoencoder (VAE) is a graphical model for variational inference and consists of an encoder and a decoder network. It belongs to the class of latent variable models, which are based on the assumption that the unknown data distribution $\hat{p}$ can be described by random vectors $\mb{z} \in \mathbb{R}^{d}$ and an appropriate prior $p(\mb{z})$ in a low-dimensional and unobserved (latent) space. For a given prior $p(\mb{z})$, we need a corresponding probabilistic model $p_\theta(\mb{x}, \mb{z})$ with parameter set $\theta$ to describe the data-generation process. To infer $p_\theta(\mb{x}, \mb{z})$, we need to maximize the marginal log-likelihood $\log p_\theta(\mb{x})$ that we can rewrite as follows~\cite{blei2017variational}: 
\begin{equation}
\begin{split}
	\log p_\theta(\mb{x}) &= \log \int p_\theta(\mb{x},\mb{z}) \, \mathrm{d}\mathbf{z} = \log \int p_\theta(\mb{x}| \mb{z}) p(\mb{z}) \, \mathrm{d}\mathbf{z}\\ 
	&= \log \int p_\theta(\mb{x}| \mb{z}) p(\mb{z}) \frac{q_\phi(\mb{z}|\mb{x})}{q_\phi(\mb{z}|\mb{x})}\, \mathrm{d}\mathbf{z} \\ 
	&\geq \mathbb{E}_{\mb{z} \sim q_\phi(\mb{z}|\mb{x})} \left[ \log p_\theta(\mb{x}|\mb{z}) \right] -  D_{\text{KL}} \left(q_\phi(\mb{z}|\mb{x}) || p (\mb{z}) \right) \\
	&\coloneqq \mathcal{E}\left(\mb{x},\theta,\phi\right)\,,
\end{split}
\label{eq:ELBO}
\end{equation}
where we applied Jensen's inequality in the fourth step. The integration over the latent variable $\mb{z}$ in the second step is usually intractable and we therefore introduced the approximate posterior distribution $q_\phi(\mb{z}|\mb{x})$ with parameter set $\phi$ and use the variational-inference principle to obtain a tractable bound of the marginal log-likelihood, the so-called \emph{evidence lower bound} (ELBO)~\cite{kingma2013auto}. We use $p(\mb{z}) = \mathcal{N}(\mb{0},\mathbb{I})$ as the latent variable prior and $\mathcal{E}\left(\mb{x},\theta,\phi\right)$. The ELBO is a tractable lower bound of the log-likelihood and can thus be maximized to infer $p_\theta(\mb{x}| \mb{z})$. The term $\mathbb{E} \left[ \log p_\theta(\mb{x}|\mb{z}) \right]$ may be interpreted as a reconstruction error, because maximizing it makes the output of the decoder more similar to the input of the encoder and the term $D_{\text{KL}} \left(q_\phi(\mb{z}|\mb{x}) || p (\mb{z}) \right)$ ensures that the latent representation is Gaussian such that data points with similar features have a similar Gaussian representation.

\begin{figure}
\centering
\includegraphics[width=0.5\textwidth]{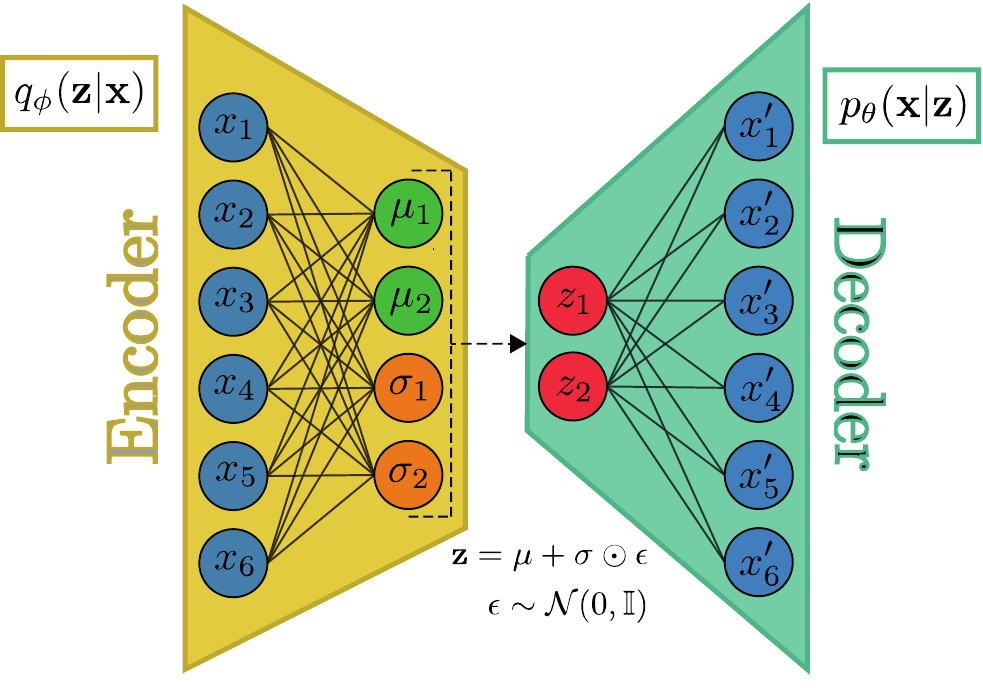}
\caption{\textbf{Variational autoencoder.} A VAE is composed of two neural networks: the encoder (yellow) and decoder (green). In this example, each network consists of an input and an output layer (multiple layers are also possible). The dotted black arrow between the two networks represents the reparameterization trick (see Eq.~\eqref{eq:repam_trick}). Each unit in the input layer of the decoder requires a corresponding mean $\mu$ and variance $\sigma$ as input.}
\label{fig:vae}
\end{figure}

So far, we outlined the general idea behind latent variable models but we still need to specify the approximate (variational) posterior $q_\phi(\mb{z}|\mb{x})$ and model likelihood $\log p_\theta(\mb{x}| \mb{z})$. A common choice for the approximate posterior is a factorized Gaussian distribution
\begin{equation}
q_\phi(\mb{z} | \mb{x}) = \mathcal{N} (\mb{z}|\boldsymbol{\mu}, \text{diag}(\boldsymbol{\sigma}^2)) = \prod_{j=1}^{d} \mathcal{N} (z_j| \mu_j, \sigma_j^2)\,,
\end{equation}
where $\boldsymbol{\mu}=\boldsymbol{\mu}(\mb{x})$ and $\boldsymbol{\sigma}=\boldsymbol{\sigma}(\mb{x})$ denote mean and standard deviation vectors of the model. Given the binary nature of our data, we describe the model likelihood $p_\theta(\mb{x}|\mb{z})$ by a Bernoulli distribution with $\mathrm{Pr}(x_i=1) =p_i$: 
\begin{equation}
\begin{split}
	\log p_\theta(\mb{x}|\mb{z}) &= \log \prod_{i=1}^m p_i^{x_i}(1-p_i)^{1-x_i} \\
	&= \sum_{i=1}^m x_i \log p_i + (1-x_i) \log (1-p_i) \,.
\end{split}	
\end{equation}
Standard inference frameworks would determine the variational parameters per sample $\{\boldsymbol{\mu}_k,\boldsymbol{\sigma}_k,\mb{p}_k\}$ ($1\leq k \leq N$). Instead, variational autoencoders utilize encoder and decoder networks to infer $q_\phi(\mb{z} | \mb{x})$ and $p_\theta(\mb{x}|\mb{z})$ (see Fig.~\ref{fig:vae}). The encoder maps the input data to the variational parameters of the approximate posterior (i.e., $\mb{x}_k \longrightarrow \{\boldsymbol{\mu}_k(\mb{x}),\boldsymbol{\sigma}_k(\mb{x})\}$) and the decoder maps a certain point in the latent parameter space to the corresponding likelihood (i.e., $\mb{z} \longrightarrow \mb{p}(\mb{z})$). This procedure allows us to amortize the cost of inference by only learning the parameters of the two neural networks.

To summarize, a VAE is based on an encoder network that outputs the parameters $(\boldsymbol{\mu}(\mb{x}), \boldsymbol{\sigma} (\mb{x}))$ of the latent Gaussian representation $q_\phi (\mb{z}|\mb{x})$ of $p (\mb{z}|\mb{x})$. The second neural network in a VAE is a decoder that uses samples of the Gaussian $q_\phi (\mb{z}|\mb{x})$ as input to generate new samples according to the distribution $p_\theta(\mb{x}|\mb{z})$ (see Fig.~\ref{fig:vae}). We estimate all parameters by maximizing the ELBO using backpropagation~\cite{kingma2013auto} and therefore need a deterministic and differentiable map between the output of the encoder and the input of the decoder w.r.t.~$\boldsymbol{\mu}$ and $\boldsymbol{\sigma}$. To calculate the corresponding derivatives in the ELBO, we need to express the random variable $\mb{z}$ as some differentiable and invertible transformation $g$ of an other auxiliary random variable $\boldsymbol{\epsilon}$ (i.e., $\mb{z} = g(\boldsymbol{\mu},\boldsymbol{\sigma},\boldsymbol{\epsilon})$). We employ the so-called \emph{reparameterization trick}~\cite{kingma2013auto} that uses $g(\boldsymbol{\mu},\boldsymbol{\sigma},\boldsymbol{\epsilon})=\boldsymbol{\mu} + \boldsymbol{\sigma} \odot \boldsymbol{\epsilon}$ such that 
\begin{equation}
\mb{z} = \boldsymbol{\mu} + \boldsymbol{\sigma} \odot \boldsymbol{\epsilon}\,,
\label{eq:repam_trick}
\end{equation}
where $\boldsymbol{\epsilon} \sim \mathcal{N}(\mb{0},    \mathbb{I})$ and $\odot$ is the element-wise product. 

To learn the distributions of physical features of the Ising model for different temperatures (i.e., classes), we use a conditional VAE (cVAE)~\cite{NIPS2015_5775} (see Sec.~\ref{sec:VAE} for further details). The advantage of cVAEs over standard VAEs is that they allow to condition the encoder and decoder on the label of the data. We use $l$ to denote the total number of classes and $c \in \{1,\dots,l \}$ is the label of a certain class. The data in a certain class is then distributed according to $p(\mb{z}|c)$ and the model distributions of the cVAE are given by $q_\phi(\mb{z}|\mb{x})$ and $p_\theta(\mb{x} | \mb{z}, c)$. These modified distributions are used in the ELBO of a cVAE. From a practical point of view, we have to provide extra dimensions in the input layers of the decoder of cVAEs to account for the label of a certain class.
\subsection{Comparison of the two models} 
In Secs.~\ref{sec:rbm} and \ref{sec:vae}, we outlined the basic ideas behind RBMs and VAEs and discussed how they can be used as generative neural networks that are able to extract and learn features of a dataset. Both models have in common that they create a low-dimensional representation of a given dataset and approximate the underlying distribution minimizing the error in the reconstruction (see Eqs.~\eqref{eqn:kl} and \eqref{eq:ELBO}).

One major difference between the two models is that the compression and decompression is performed by the same network in the case of an RBM. For a VAE, however, two networks (encoder and decoder) are used to perform these tasks. A second important difference is that the latent distribution is a product of Bernoulli distributions for RBMs and a product of Gaussians for VAEs. Therefore, the representational power of RBMs is restricted to $2^m$, where $m$ is the number of hidden units. In the case of VAEs, it is $\mathbb{R}^{d}$ where $d$ is the dimension of the latent space. The greater representational power of a VAE may be an advantage, but it could also result in overfitting the data.
\section{Monitoring learning}
\label{sec:monitoring}
To monitor the learning progress of RBMs and VAEs, we can keep track of the \emph{reconstruction error} (i.e., the squared Euclidean distance between the original data and its reconstruction). Another monitoring approach uses the \emph{binary cross entropy} (BCE)~\cite{bengio2007greedy}
\begin{equation}
R(\mb{x})=-\sum_{i=1}^m x_i \log p_i(\mb{x})+(1-x_i) \log(1-p_i(\mb{x})) \,,
\end{equation}
where $x_i\in\{0,1\}$ and $p_i(\mb{x})$ is the reconstruction probability of bit $i$. In the following subsections, we describe additional methods that allow us to monitor approximations of log-likelihood (see Eq.~\eqref{eq:log_l_RBM}) and KL divergence (see Eq.~\eqref{eqn:kl}). Approximations are important since log-likelihood and KL divergence can only be calculated exactly for small system sizes. In Ref.~\cite{torlai2016learning}, the partition function of RBMs has been computed directly since the considered system sizes contained only a few dozen spins.
\subsection{Pseudo log-likelihood}
One possibility to monitor the learning progress of an RBM is the so-called pseudo log-likelihood~\cite{besag1975statistical}. We consider a general approximation of an $m$-dimensional parametric distribution: 
\begin{equation}
	\begin{split}
		p(\mathbf{x}; \theta) & = \prod_{i=1}^m p\left(x_i \Bigg| \bigcap_{j=1}^{i-1} x_j; \theta\right) \\
		& \approx \prod_{i=1}^m p(x_i| x_1, \dots, x_{i-1}, x_{i+1}, \dots, x_m; \theta ) \\ 
		& = \prod_{i=1}^m p(x_i| \mathbf{x}_{-i}; \theta) \coloneqq \mathcal{P}(\mathbf{x}; \theta) \, ,
	\end{split} 
\end{equation}
where $\mathbf{x}_{-i}$ is the set of all variables apart form $x_i$. We now take the logarithm of $\mathcal{P}(\mathbf{x}; \theta)$ and obtain the \emph{pseudo log-likelihood}
\begin{equation}
\log \mathcal{P}(\mathbf{x} ; \theta) = \sum_{i=1}^m \log p(x_i| \mathbf{x}_{-i}) \, .
\label{eq:log_P}
\end{equation}
Based on \Cref{eq:log_P}, we conclude that the pseudo log-likelihood is the sum of the log-probabilities of each $x_i$ conditioned on all other states apart from $x_i$. The summation over all states $x_i$ can be computationally expensive, so we approximate $\log \mathcal{P}(\mathbf{x} ; \theta)$ by the log-likelihood estimate~\cite{deeplearning}
\begin{equation}
\log \tilde{\mathcal{P}} (\mathbf{x} ; \theta) = m \log p(x_i| \mathbf{x}_{-i})\,,\quad i \sim \mathcal{U}(0,m)\,,
 \end{equation}
where we sample uniformly to efficiently approximate the pseudo log-likelihood. For RBMs, the approximated pseudo log-likelihood is
\begin{equation}
	 \log \tilde{\mathcal{P}} (\mathbf{x} ; \theta)  = m \log \left( \frac{e^{-\mathcal{F}_c(\mathbf{x})}}{e^{-\mathcal{F}_c(\mathbf{\tilde{x}})} + e^{-\mathcal{F}_c(\mathbf{x})}} \right) \, ,
\label{eq:approx_log_likeli}
\end{equation}
where $\mathbf{\tilde{x}}$ corresponds to the vector $\mathbf{x}$ but with a flipped $i$-th component. Equation \eqref{eq:approx_log_likeli} follows from the identity
\begin{equation}
\begin{split}
p(x_i|\mathbf{x}_{-i}) &= \frac{p(x_i,\mathbf{x}_{-i})}{p(\mathbf{x}_{-i})} = \frac{p(x_i,\mathbf{x}_{-i})}{\sum_{x_i\in\{0,1\}} p(x_i,\mathbf{x}_{-i})} \\
&= \frac{p(\mathbf{x})}{p(\mathbf{\tilde{x}})+ p(\mathbf{x})} = \frac{e^{-\mathcal{F}_c(\mathbf{x})}}{e^{-\mathcal{F}_c(\mathbf{\tilde{x}})} + e^{-\mathcal{F}_c(\mathbf{x})}}\,.
\end{split}
\end{equation} 
\subsection{Estimating KL divergence}
\label{sec:kl}
In most unsupervised learning problems, except some synthetic problems such as Gaussian mixture models~\cite{rasmussen2000infinite}, we do not know the distribution underlying a certain dataset. We are only given samples and can therefore not directly determine the KL divergence (see \Cref{eqn:kl}). Instead, we have to use some alternative methods. According to Ref.~\cite{wang2009divergence}, we use the nearest-neighbor (NN) estimation of the KL divergence. Given two continuous distributions $p$ and $q$ and $N$ and $\tilde{N}$ corresponding i.i.d.~$m$-dimensional samples $\{\mb{x}_1,\dots, \mb{x}_N\}$ and $\{\mb{y}_1,\dots, \mb{y}_{\tilde{N}} \}$, we want to estimate $D_{\text{KL}}(p||q)$. For two consistent estimators~\cite{loftsgaarden1965nonparametric} $p'$ and $q'$, the KL divergence can be approximated by 
\begin{equation}
\begin{split}
D_{\text{KL}}(p||q) &= \sum_{\mb{x} \sim p(\mb{x})} p(\mb{x}) \log \frac{p (\mb{x})}{q(\mb{x})} \\ 
 &\approx \frac{1}{N} \sum_{i=1}^N \log \frac{p'(\mb{x}_i)}{q'(\mb{x}_i)}\,.
\end{split} 
\end{equation}
\begin{figure*}
\includegraphics[width=0.32\textwidth]{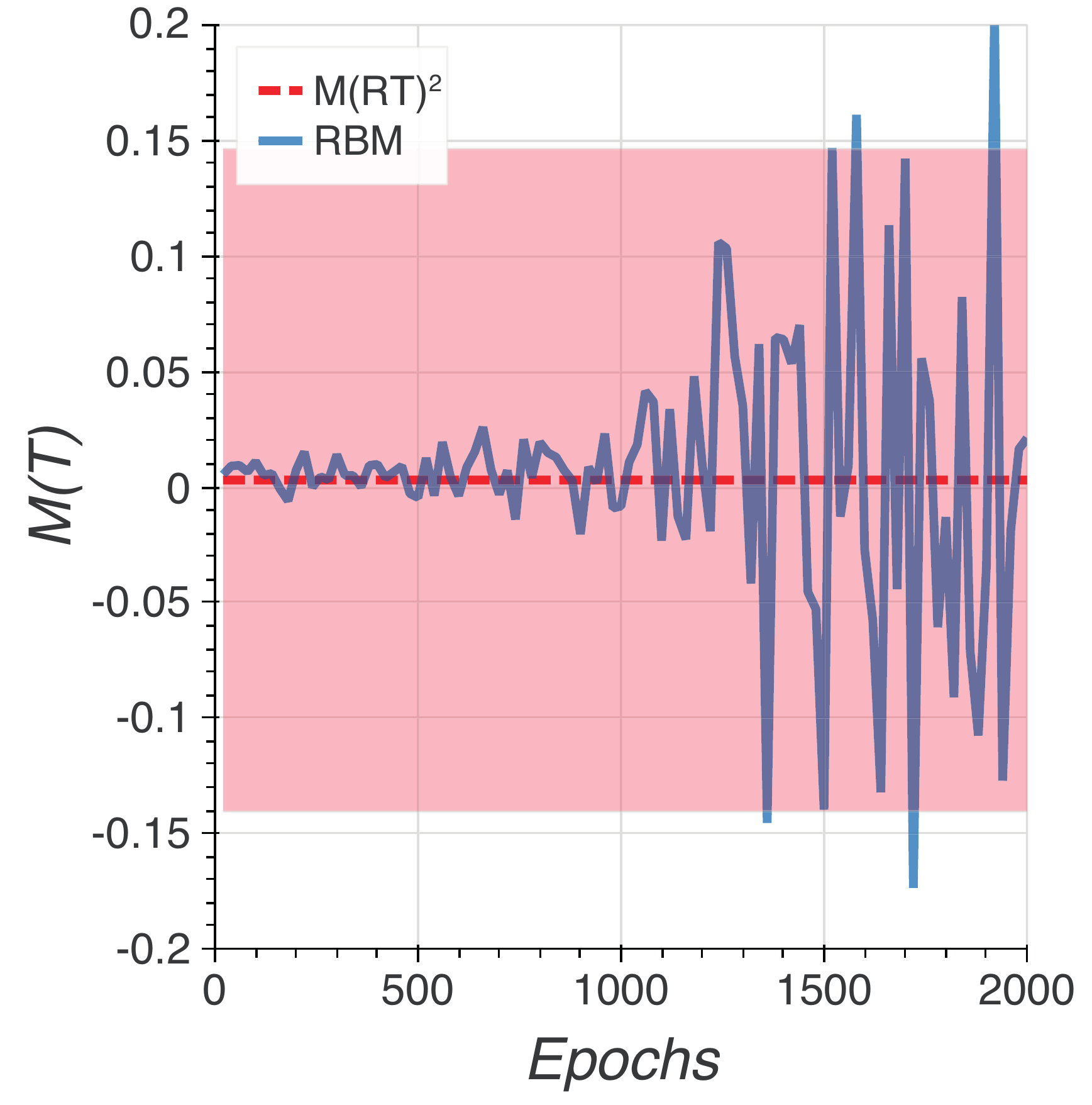}
\includegraphics[width=0.32\textwidth]{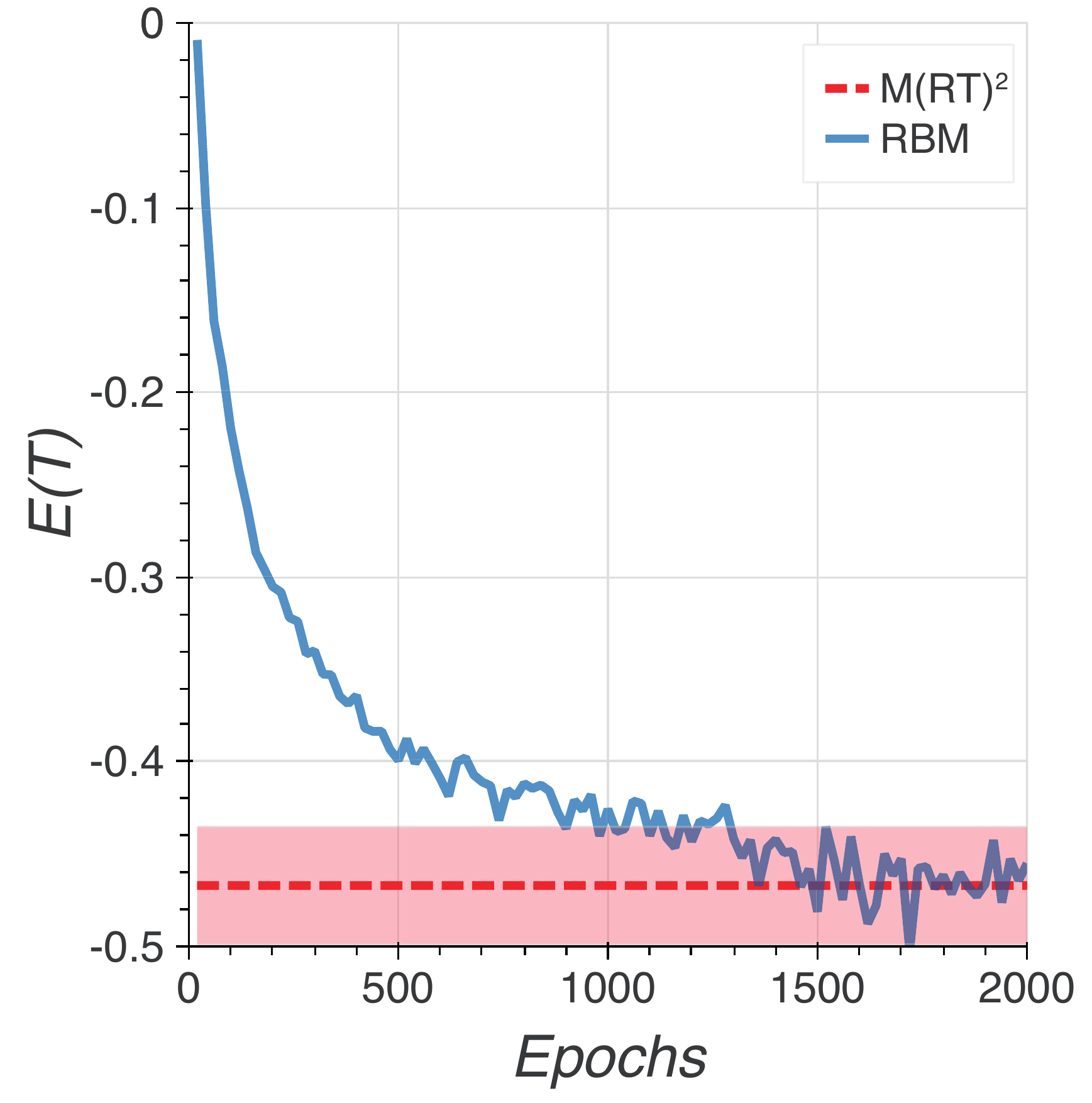}
\includegraphics[width=0.32\textwidth]{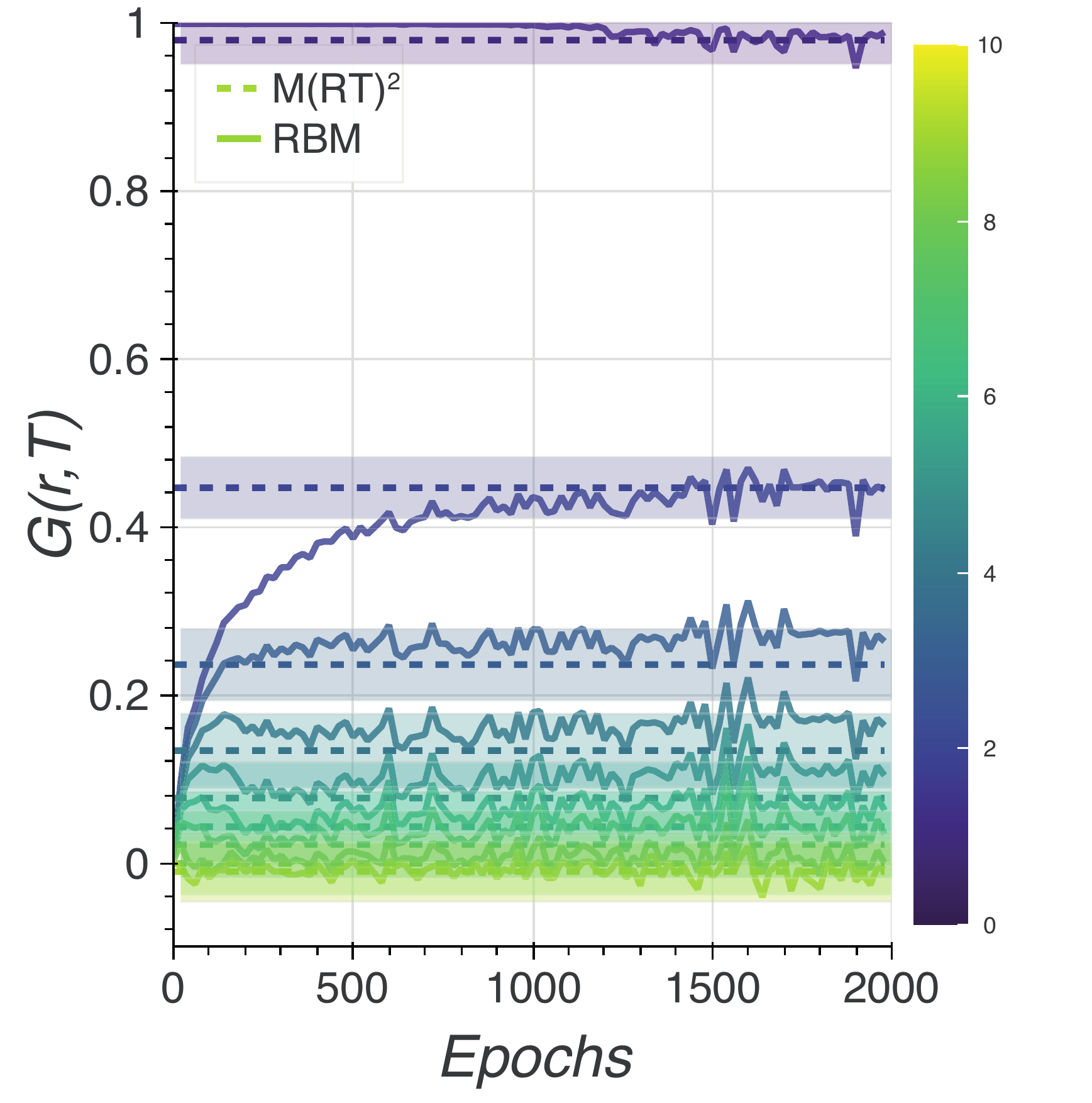}
\caption{\textbf{Evolution of physical features during training.} We show the evolution of the magnetization $M(T)$, energy $E(T)$, and correlation function $G(r,T)$ for the training of a single RBM at $T=2.75$. Dashed lines and shaded regions indicate the mean and standard deviation of the corresponding M(RT)$^2$ samples and different colors in the plot of $G(r,T)$ represent different radii $r$.}
\label{fig:learning}
\end{figure*}
We construct the consistent estimators as follows. Let $\rho_k(\mb{x}_i)$ be the Jaccard distance~\footnote{The choice of the distance metric depends on the nature of the samples. We use the Jaccard distance for the binary data in our models.} between $\mb{x}_i$ and its k-NN in the subset $\{\mb{x}_j\}$ for $j \neq i$ and $\nu_k(\mb{x}_i)$ the Jaccard distance between $\mb{x}_i$ and its k-NN in the set $\{\mb{y}_l\}$. We obtain the KL divergence estimate
\begin{equation}
\hat{D}(p||q) = \frac{m}{N} \sum_{i=1}^N \log{\frac{\nu_k(\mb{x}_i)}{\rho_k{(\mb{x}_i)}} + \log{\frac{\tilde{N}}{N-1}}}\,.
\end{equation}
In the context of generative models this means that we can calculate the KL divergence in terms of the Jaccard distances $\rho_k(\mb{x}_i)$ and $\nu_k(\mb{x}_i)$.
\section{Learning the Ising model}
\label{sec:learning}
We now use the generative models of Sec.~\ref{sec:models} to learn the distribution of Ising configurations at different temperatures. This enables us to examine the ability of RBMs and cVAEs to capture the temperature-dependence of physical features such as energy, magnetization, and spin-spin correlations. We first focus on the training of RBMs in Sec.~\ref{sec:sep_temps} and in Sec.~\ref{sec:VAE} we describe the training of cVAEs. However, before focusing on the training, we give a brief overview of some relevant properties of the Ising model.

The Ising model is a mathematical model of magnetism and describes the interactions of spins $\sigma_i \in \{-1,1\}$ ($1\leq i \leq m$) according to the Hamiltonian~\cite{landau-binder2009}
\begin{equation}
\mathcal{H}\left(\{\sigma\}\right)=-J \sum_{\langle i,j \rangle} \sigma_i \sigma_j-H \sum_i \sigma_i\,,
\label{eq:ising_hamiltonian}
\end{equation}
where $\langle i,j \rangle$ denotes the summation over nearest neighbors and $H$ is an external magnetic field. We consider the ferromagnetic case with $J=1$ on a two-dimensional lattice with $32\times 32$ sites. In an infinite two-dimensional lattice, the Ising model exhibits a second-order phase transition at $T=T_c=2/\log(1+\sqrt{2})\approx 2.269$~\cite{onsager1944crystal,landau-binder2009}. To train RBMs and cVAEs, we generate $20\times 10^4$ realizations of spin configurations at temperatures $T\in\{1.5,2,2.5,2.75,3,4\}$ using the M(RT)$^2$ algorithm~\cite{landau-binder2009,kalos2009monte}. In previous studies, RBMs were applied to smaller systems with only about 100 spins to learn the distribution of spin configurations with one RBM per temperature (see Refs.~\cite{torlai2016learning,iso2018scale} for further details).

We also consider the training of one RBM per temperature and compare the results with those obtained by a single RBM and cVAE that were trained for all temperatures.
If RBMs and cVAEs are able to learn the distribution of Ising configurations $\{\sigma\}$ at a certain temperature, the model-generated samples should have the same (spontaneous) magnetization, energy, and spin-spin correlations as the M(RT)$^2$ samples. We compute the spontaneous magnetization according to
\begin{equation}
M_S(T)=\lim_{H\rightarrow 0^+} \left\langle \frac{1}{m} \sum_{i=1}^m \sigma_i \right\rangle\,.
\end{equation}
Furthermore, we determine the spin-spin correlations between two spins $\sigma_i$ and $\sigma_j$ at positions $r_i$ and $r_j$: 
\begin{equation}
G(r_i,r_j) = \langle (\sigma_i - \langle \sigma_i \rangle) (\sigma_j - \langle  \sigma_j\rangle)\rangle  = \langle \sigma_i \sigma_j \rangle - \langle \sigma_i \rangle \langle \sigma_j \rangle\,.
\label{eq:correl_1}
\end{equation}
We use translational-invariance and rewrite \Cref{eq:correl_1} as
\begin{equation}
G(r_i, r_j,T) = G(r,T) = \langle s_i s_{i+r} \rangle - M_S^2(T) \,,
\end{equation}
where $r=|r_i-r_j|$ is the Euclidean distance between spins $\sigma_i$ and $\sigma_j$. The correlation function $G(r,T)$ quantifies the influence of one spin on another at distance $r$.
\subsection{Training RBMs}
\label{sec:sep_temps}
\begin{figure*}
\includegraphics[width=0.32\textwidth]{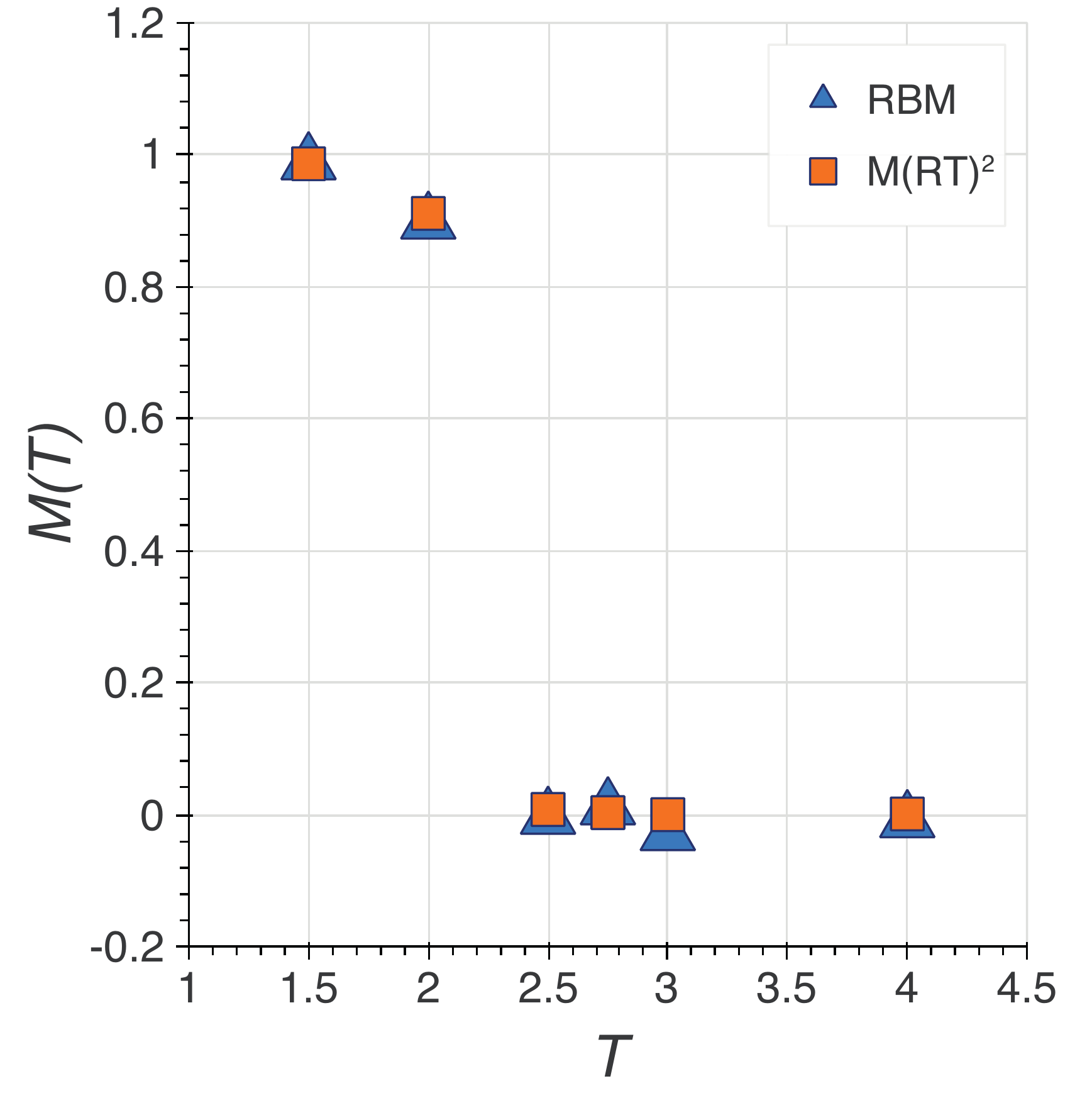}
\includegraphics[width=0.32\textwidth]{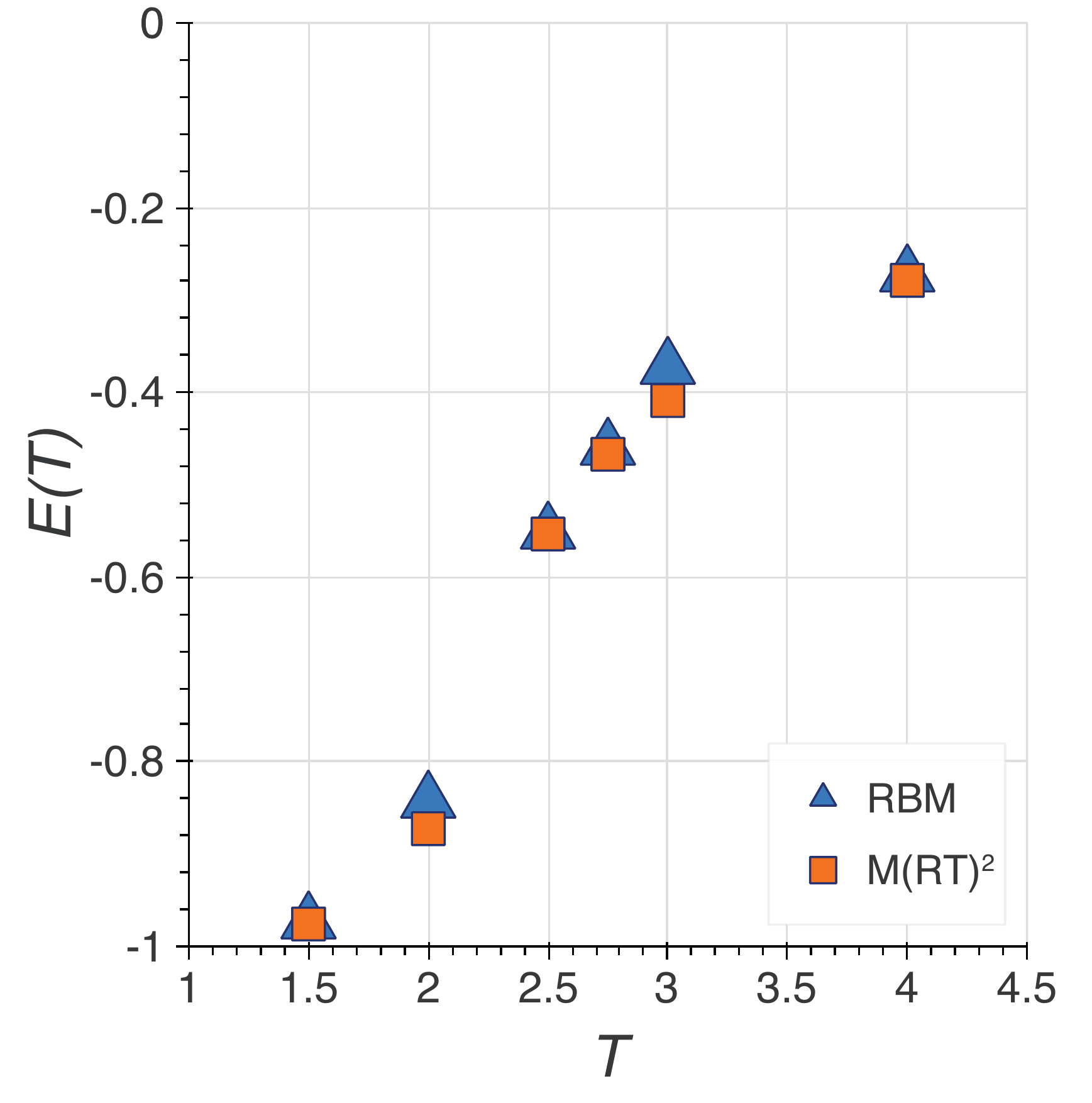}
\includegraphics[width=0.32\textwidth]{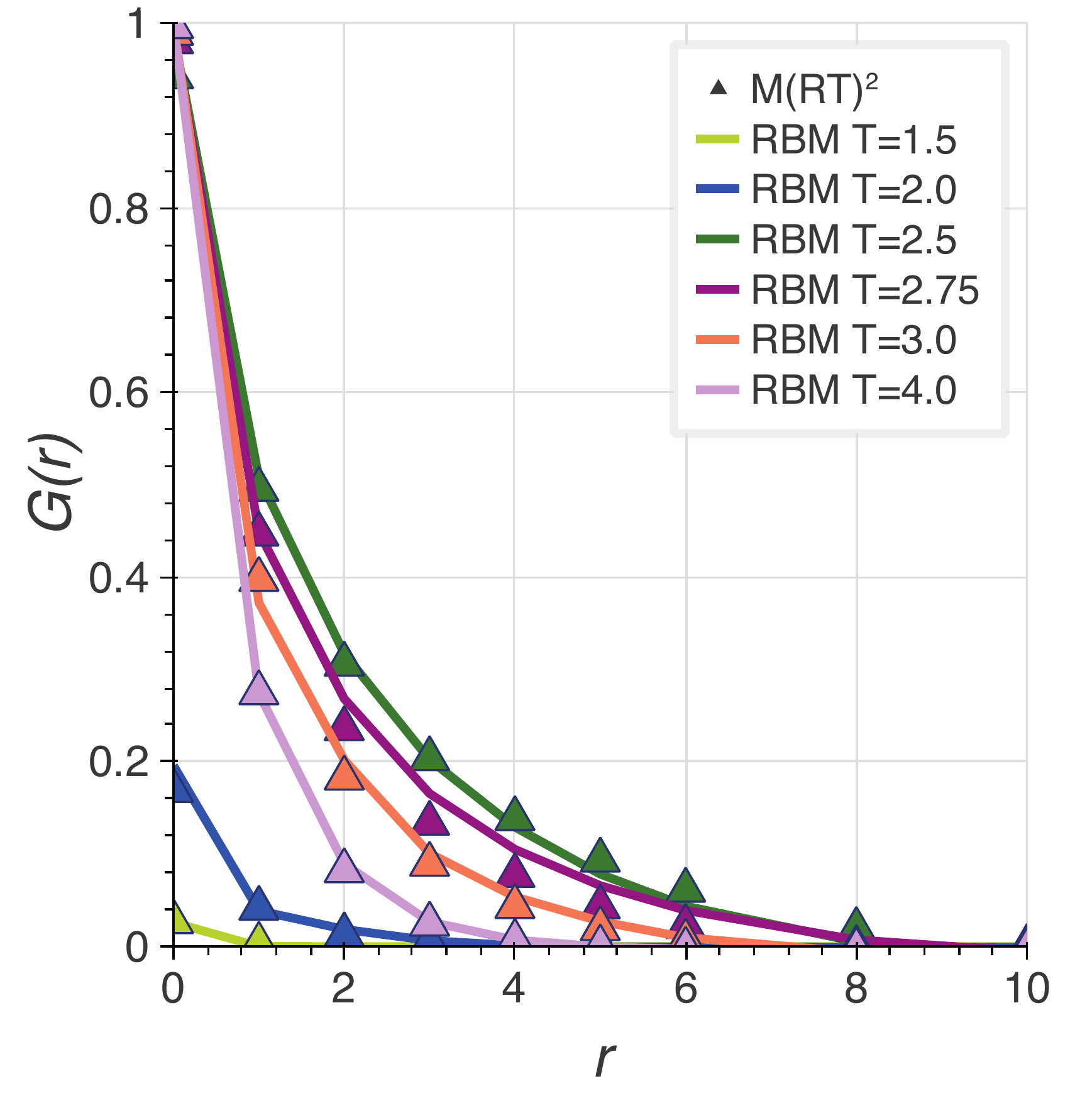}
\includegraphics[width=0.32\textwidth]{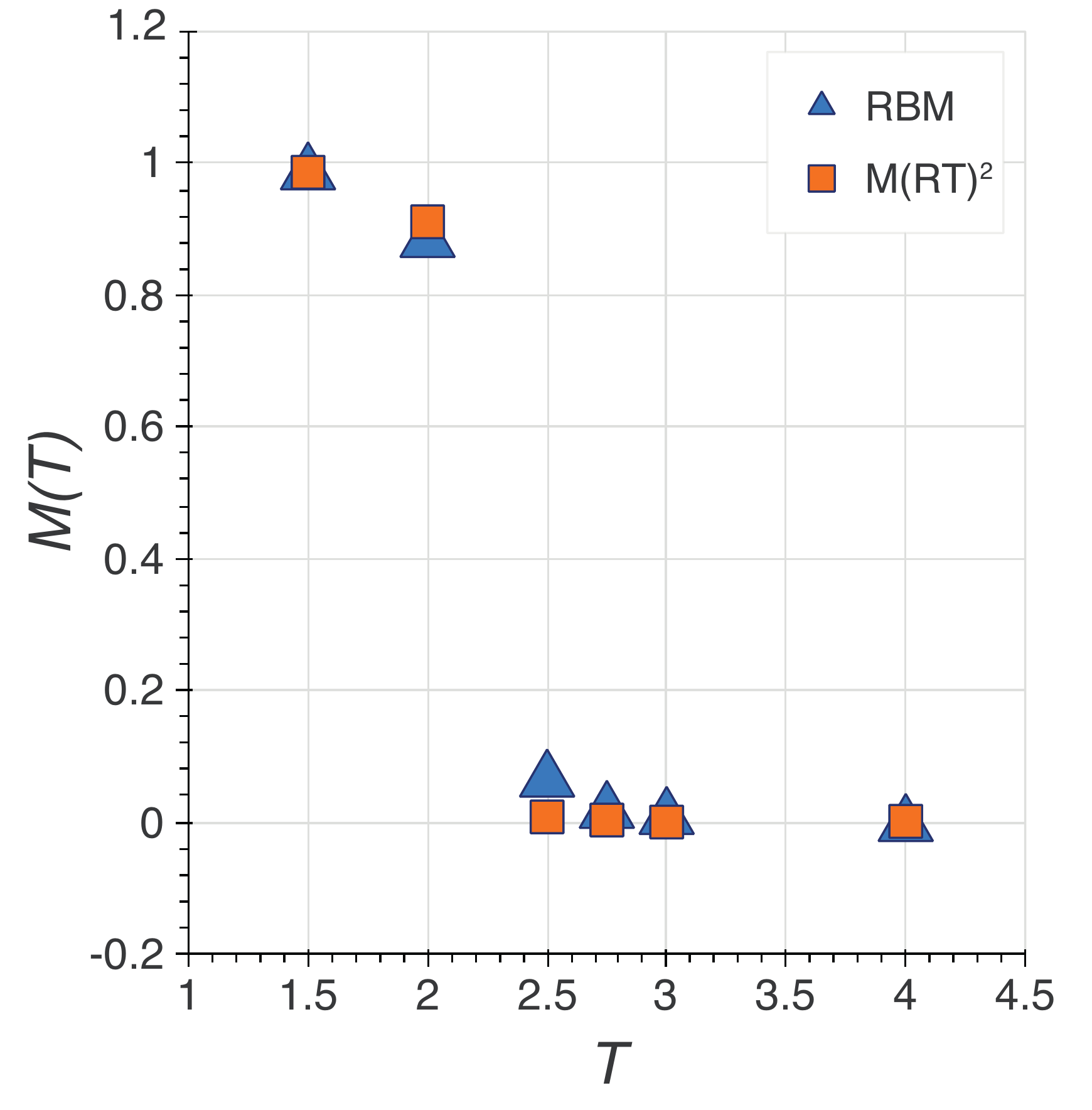}
\includegraphics[width=0.32\textwidth]{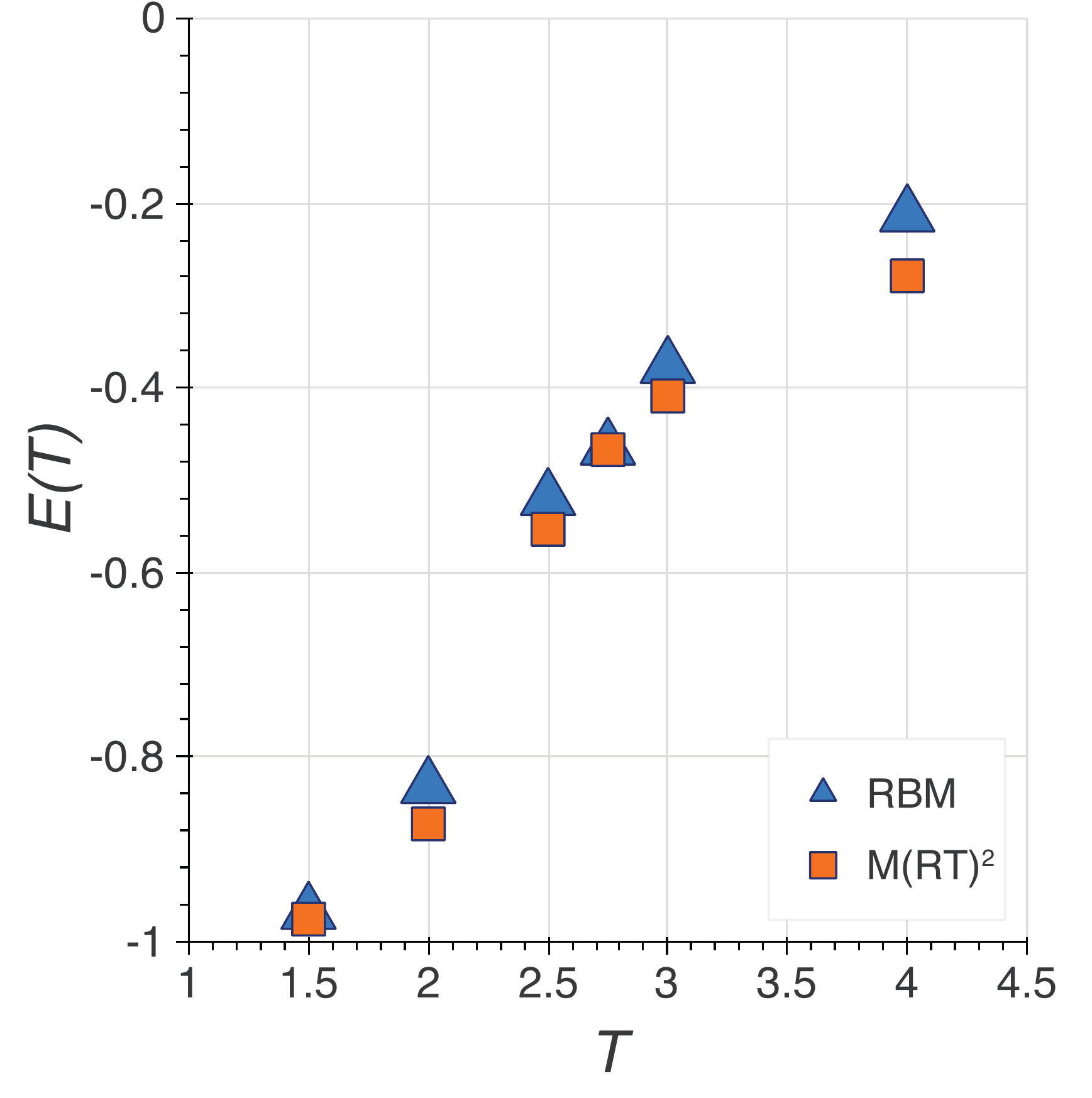}
\includegraphics[width=0.32\textwidth]{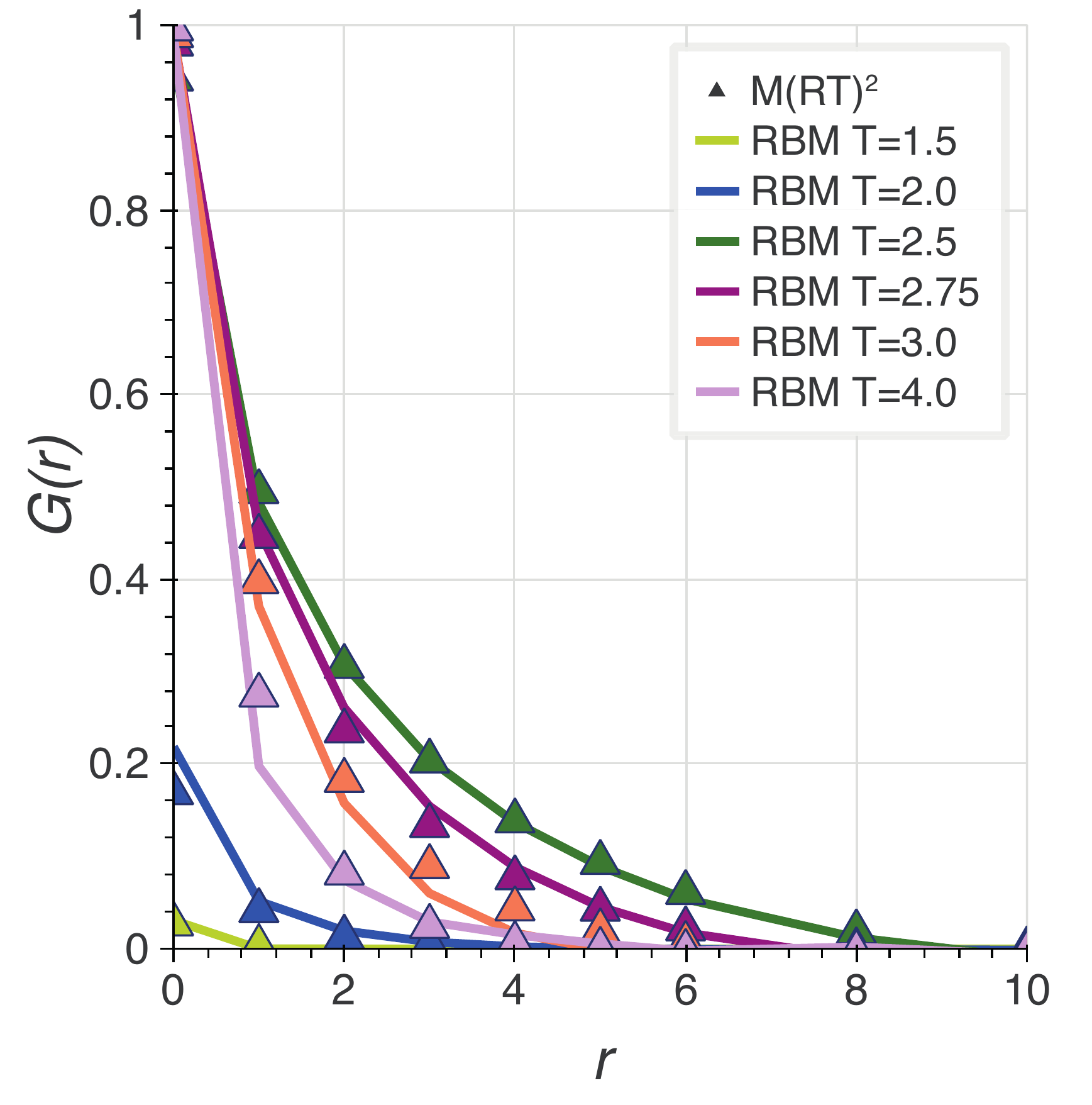}
\includegraphics[width=0.32\textwidth]{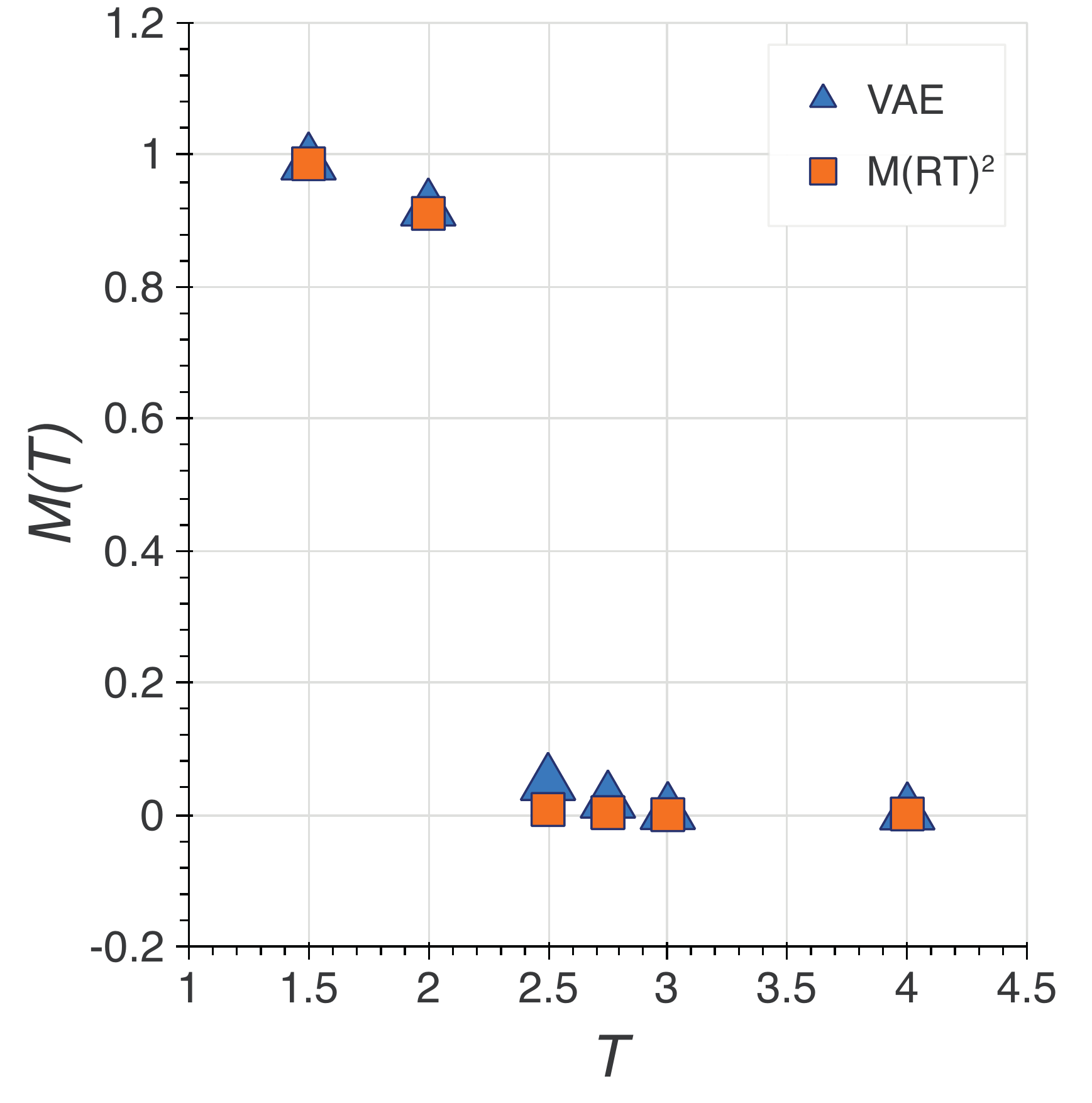}
\includegraphics[width=0.32\textwidth]{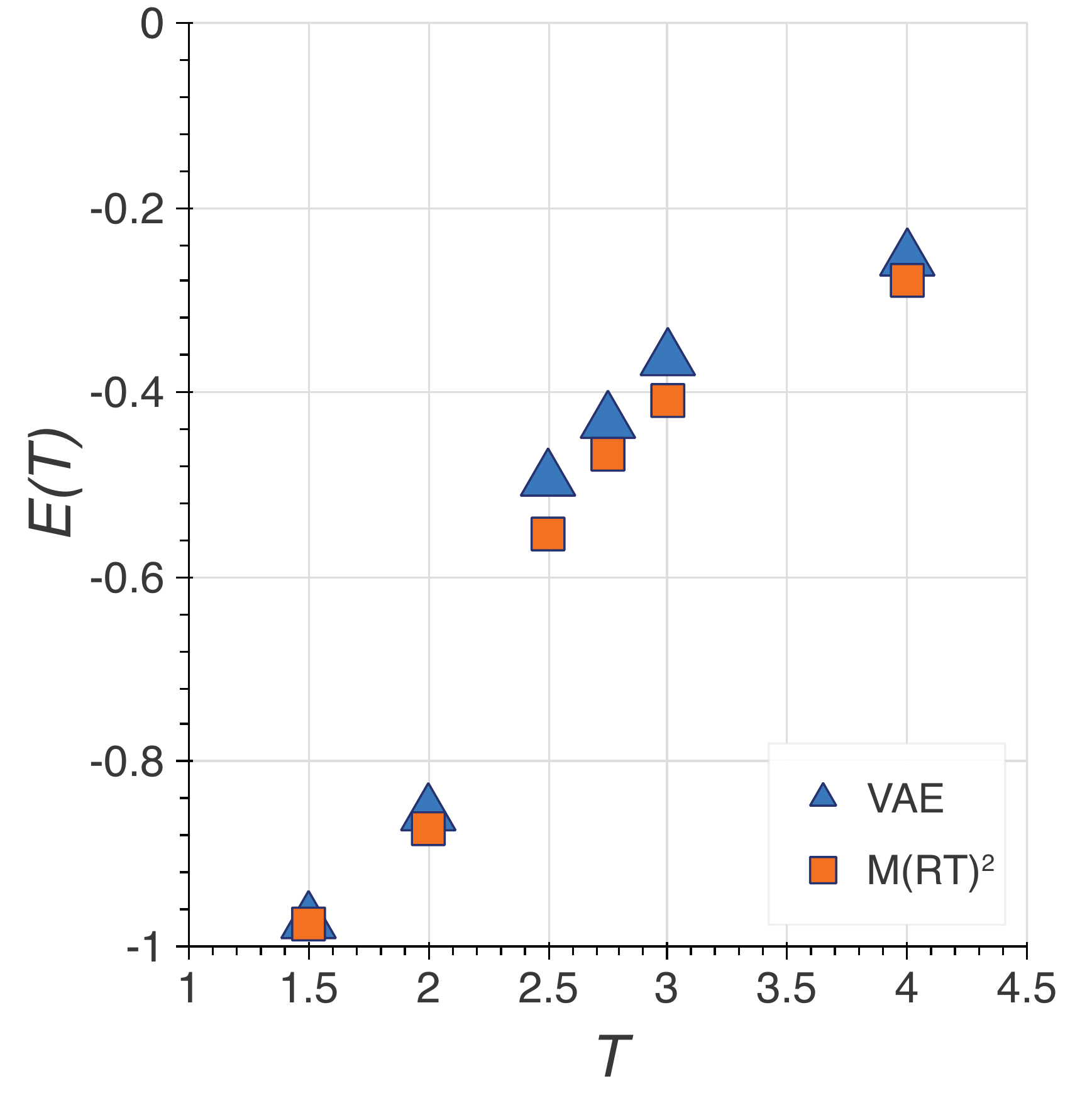}
\includegraphics[width=0.32\textwidth]{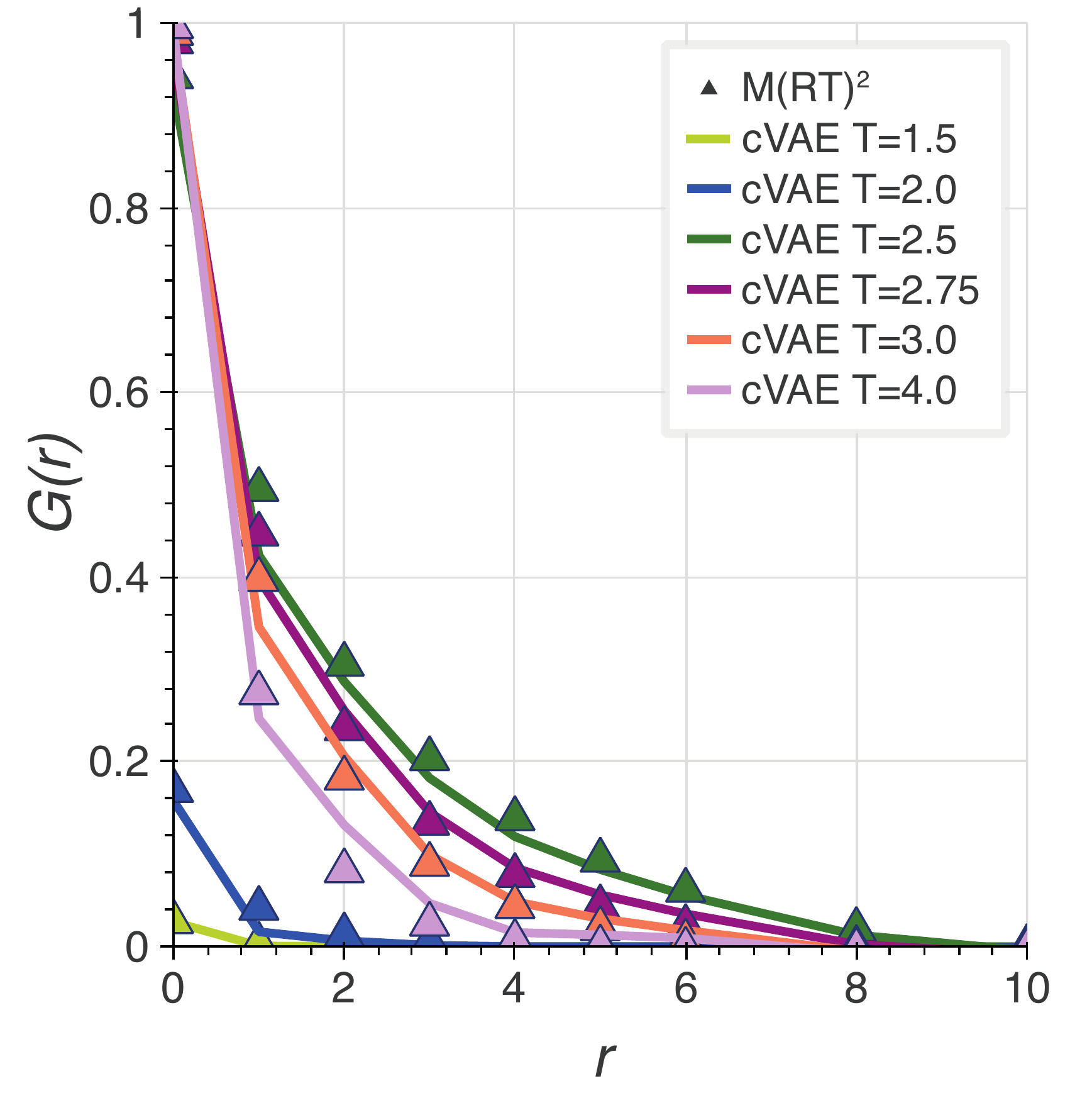}
\caption{\textbf{Physical features of RBM and convolutional VAE Ising samples}. We use RBMs and convolutional cVAEs to generate $20\times 10^4$ samples of Ising configurations with $32\times 32$ spins for temperatures $T\in\{1.5,2,2.5,2.75,3,4\}$. We show the magnetization $M(T)$, energy $E(T)$, and correlation function $G(r,T)$ for neural network and corresponding M(RT)$^2$ samples. In the top panel, we separately trained one RBM per temperature, whereas we used a single RBM for all temperatures in the middle panel. In the bottom panel, we show the behavior of $M(T)$, $E(T)$, and $G(r,T)$ for samples that are generated with a convolutional cVAE that was trained for all temperatures. Error bars are smaller than the markers.}
\label{fig:physical_features}
\end{figure*}
We consider all $20\times 10^4$ samples for $T \in \{ 1.50,2.0, 2.5,2.75,3.0, 4.0\}$ and first train one RBM per temperature using the \emph{Adam} optimizer~\cite{kingma2014adam} with a 5-step PCD approach (see Sec.~\ref{sec:rbm}). For all machines, we use the same network structure and training parameters and set the number of hidden units to $n=900$. The number of visible units is equal to the number of spins (i.e., $m=32^2=1024$). We initialize each Markov chain with uniformly at random distributed binary states $\{0,1\}$, train over 200 epochs, and repeat this procedure 10 times. We set the learning rate to $\eta = 10^{-4}$ and consider minibatches of size $128$. In addition, we encourage sparsity of the weight matrix $W$ by applying L1 regularization with a regularization coefficient of $\lambda=10^{-4}$. That is, we consider an additional contribution $-\lambda ||W||_1$ in the log-likelihood and use
\begin{equation}
\log \mathcal{L}'\left(\mb{x},\theta,\lambda\right)=\log \mathcal{L}\left(\mb{x},\theta\right)-\lambda ||W||_1
\label{eq:regul_L}
\end{equation}
instead of Eq.~\eqref{eq:log_l_RBM}. Thus, minimizing $-\log \mathcal{L}'\left(\mb{x},\theta,\lambda\right)$ means that we are also minimizing the 1-norm $||W||_1$ (i.e., the maximum absolute column sum) of the weight matrix $W$. We initialize all weights in $\textbf{W}$ according to a Gaussian distribution with the \emph{glorot} initializer~\cite{glorot2010understanding} and biases $\mb{b}$ and $\mb{c}$ with uniformly distributed numbers in the interval $[0,0.1]$. In Fig.~\ref{fig:learning}, we show the evolution of magnetization, energy, and correlations during training of an RBM for Ising samples at $T=2.75$. We observe that it takes about 1500 epochs of training to properly capture magnetization fluctuations, energy, and spin correlations. 

In addition to using one RBM per temperature, we also train one RBM for all temperatures. We use the same samples and training parameters as before. However, we add additional visible units to encode the temperature information of each training sample. For the training of this RBM, we start from a random initial configuration, train over 400 epochs, and repeat this procedure 10 times.

To monitor the training progress, we tested the considered models every 20 epochs on $10\%$ of the data by monitoring reconstruction error, binary cross entropy, pseudo log-likelihood, and KL divergence and its inverse (see Sec.~\ref{sec:monitoring}). We compute reconstruction error and binary cross entropy based on a single Gibbs sampling step. For the KL divergence, we generated $12 \times 10^3$ samples by performing $10^3$ sampling steps and keeping the last $10^2$ samples. We repeated this procedure $20$ times for each temperature and averaged over minibatches of $256$ samples to have better estimators.
\begin{figure}
\includegraphics[width=0.49\textwidth]{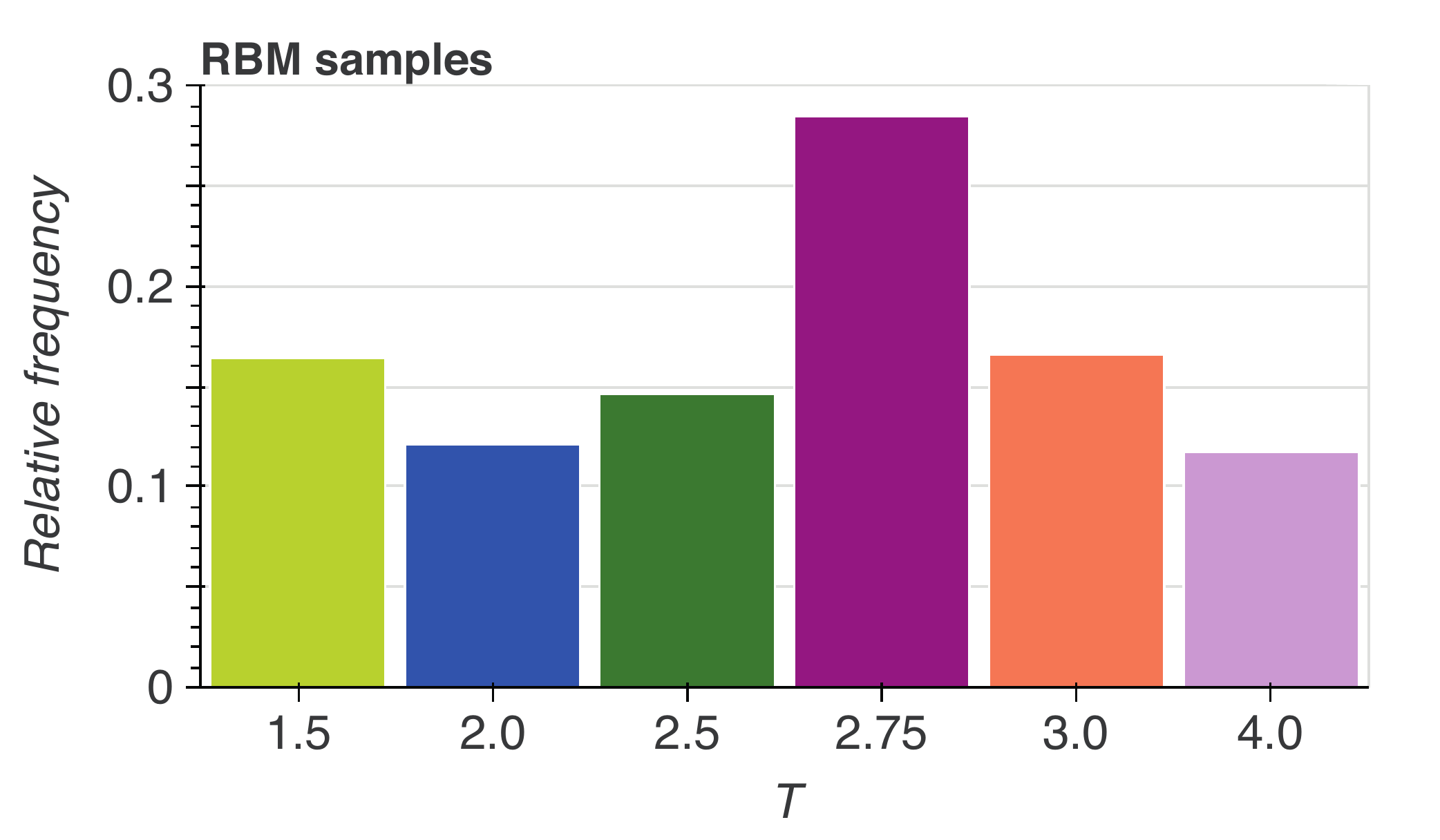}
\includegraphics[width=0.49\textwidth]{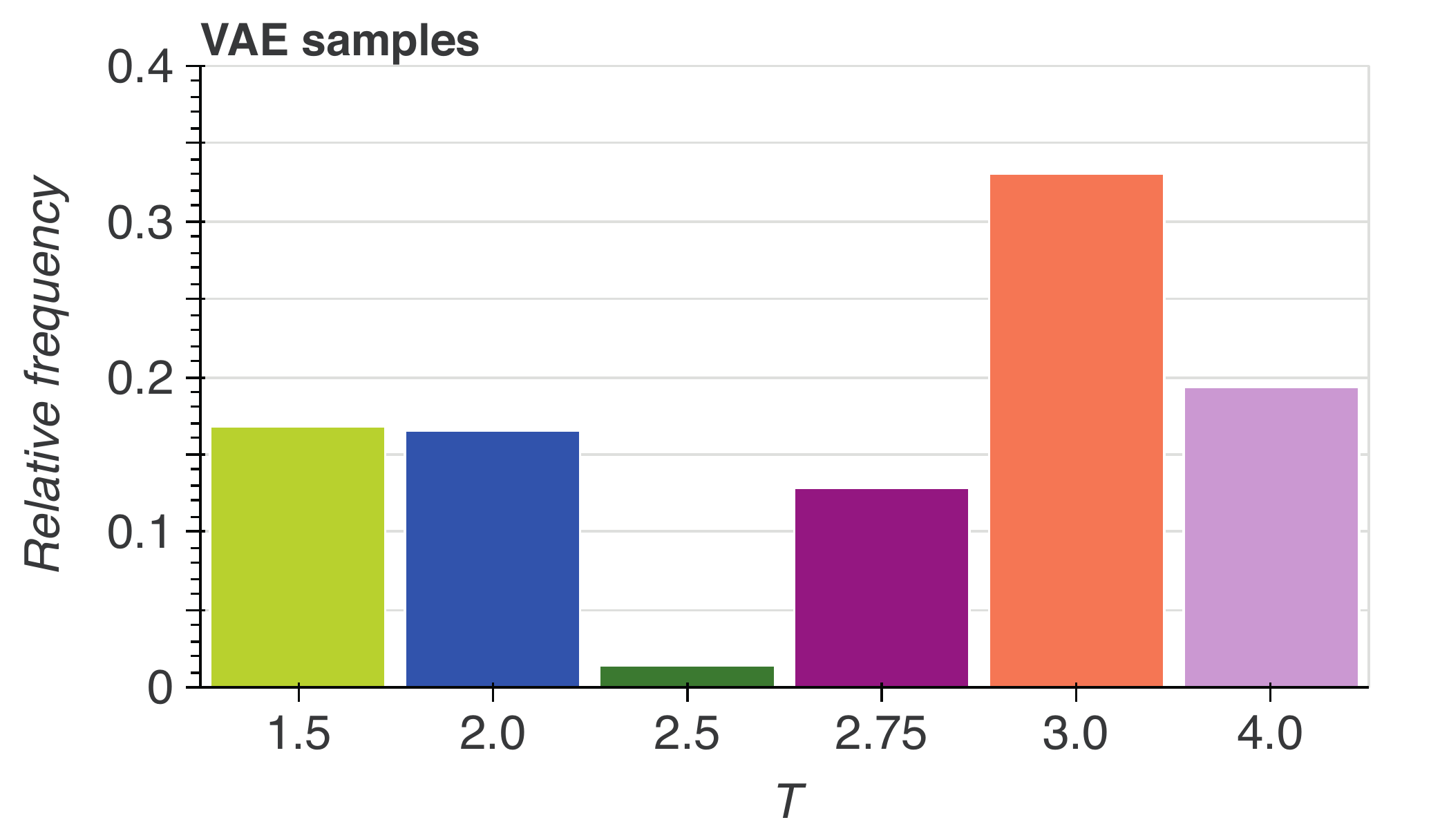}
\caption{\textbf{Relative frequencies of samples at different temperatures.} We show the relative frequencies of samples that are obtained by starting trained RBMs (top) and convoluational cVAEs (bottom) from random initial configurations. The data are based on $20\times 10^4$ samples for each temperature.}
\label{fig:counts}
\end{figure}
\subsection{Training cVAEs}
\label{sec:VAE}
We train one cVAE for all temperatures and consider an encoder that consists of three convolutional layers with $32$, $64$, and $64$ kernels (or filters) of size 3~\footnote{For filter sizes that approach the size of the underlying system, we would recover a non-convolutional network architecture. If the filters are too small, we can only account for local connectivity in very small neighborhoods. In accordance with Ref.~\cite{simonyan2014very}, we use filters of size 3. We also performed simulations for filter sizes 2 and 4 and found similar performances of the considered convolutional VAEs.} and a \emph{rectified linear unit} (ReLU) as activation function. Each convolutional layer is followed by a \emph{maxpool} layer to downsample the ``pictures''. In a maxpool layer, only the maximum values of subregions of an initial representation are considered. To regularize the cVAE, we use \emph{batchNormalization}~\cite{ioffe2015batch} and a dropout rate of $0.5$~\cite{srivastava2014dropout}. The last layer is a fully connected flat layer with 400 units (200 mean and 200 variance) to represent the 200 dimensional Gaussian latent variable $\mathbf{z}$.

For the decoder, we use an input layer that consists of 200 units to represent the latent variable $\mathbf{z}$ and concatenate it with the additional temperature labels. We upsample their dimension with a fully connected layer of 2048 units that are reshaped to be fed into three deconvolutional layers. The number of filters in these deconvolutional layers are $64$, $64$, and $32$ with filter size 3 and each deconvolutional layer is followed by an upsampling layer to upsample the dimension of the ``pictures''. The last layer of the encoder is a deconvolutional layer with one single filter to have a single sample as output. We train the complete architecture over $10^3$ epochs using the \textit{Adam} optimizer~\cite{kingma2014adam} and a learning rate of $\eta = 10^{-4}$. During training, we keep track of the ELBO (see Eq.~\eqref{eq:ELBO}) and the KL divergence and its inverse (see Sec.~\ref{sec:monitoring}) for $10 \%$ of the data.

We also consider non-convolutional VAE architectures in App.~\ref{sec:non_conVAE}. However, in the absence of convolutional layers, we found that some physical features are not well-captured anymore (see App.~\ref{sec:non_conVAE} for further details).
\subsection{Training classifier}
\label{sec:classifier}
In the case of single RBMs and cVAEs that we train for all temperatures, we have to use a classifier to determine the temperature of the generated samples. We consider a classifier network that consists of two convolutional layers with respectively 32 and 64 filters and four fully connected layers of $[256,128,64,32]$ units with a \emph{ReLU} activation function and stride 2 convolution. Similar to Eq.~\eqref{eq:regul_L}, we regularize all layers using a L2 regularizer with a regularization coefficient of $0.01$. The last layer is a softmax fully connected layer that outputs the probability of a sample to belong to each temperature class and we determine the sample temperature by identifying the highest probability class. We trained the model for 100 epochs with the \emph{Adam} optimizer~\cite{kingma2014adam} and a learning rate of $10^{-4}$. We monitor both the test loss (i.e., categorical cross entropy) and accuracy. With this model, we achieve a test accuracy of $89 \%$ for the considered six temperature classes.
\subsection{Generating Ising samples}
After training both generative neural networks with $20\times 10^4$ realizations of Ising configurations at different temperatures~\footnote{Training the considered RBMs for $10^3$ epochs required 7 days of computing time on an Intel Xeon E7-8867v3 (2.5 GHz) CPU. For cVAEs, the training took 20 hours on a GeForce GTX 1080 Ti GPU.}, we can now study the ability of RBMs and cVAEs to capture physical features of the Ising model. To generate samples from an RBM, we start 200 Markov chains from random initial configurations whose mean corresponds to the mean of spins in the data at the desired temperature. This specific initialization together with the encoding of corresponding temperature labels helps sampling from single RBMs that were trained for all temperatures. We perform $10^3$ sampling steps, keep the last $10^2$ samples, and repeat this procedure $2\times 10^2$ times for each temperature. The total number of samples is therefore $20\times 10^4$. For the trained cVAE, we sample from the decoder using $20\times 10^3$ 200-dimensional vectors with normally-distributed components for each temperature. We concatenate these vectors with the respective temperature labels and use them as input of the decoder. For the single RBM and cVAE that we trained for all temperatures, we use the classifier (see Sec.~\ref{sec:classifier}) to identify the temperature of the newly generated patterns. 

We show the distribution of RBM and cVAE samples over different temperature classes in Fig.~\ref{fig:counts}. For both neural networks, the relative frequency of samples with temperature $T\geq 2.5$ is larger than for lower temperatures. In addition, the results of Fig.~\ref{fig:counts} suggest that the considered convolutional cVAE has difficulties to generate samples for $T\approx T_c\approx2.269$. Non-convolutional VAEs suffer from the same problems and also have difficulties to properly capture physical features (see App.~\ref{sec:non_conVAE}). Interestingly, we do not observe this issue for the RBM samples in Fig.~\ref{fig:counts}.

In general, patterns at criticality are difficult to learn and generate since they exhibit long-range correlation effects. One possibility to enforce the generation of VAE samples at $T_c$ is to determine the latent representation of samples at the critical temperature and use slightly perturbed versions of these samples as input in the decoder.

Next, we determine magnetization, energy, and correlation functions of the generated samples at different temperatures (see Fig.~\ref{fig:physical_features}). In the top panel of Fig.~\ref{fig:physical_features}, we observe that RBMs that were trained for single temperatures are able to capture the temperature dependence of magnetization, energy, and spin correlations. In the case of single RBMs and convolutional cVAEs that were trained for all temperatures, we also find that the aforementioned physical features are captured well. However, our results indicate that non-convolutional VAEs have difficulties in capturing magnetization, energy, and spin correlations (see App.~\ref{sec:non_conVAE} for further details). 

Convolutional layers have been proven effective in image classification~\cite{krizhevsky2012imagenet,zeiler2014visualizing,szegedy2015going} and generation~\cite{radford2015unsupervised}. The advantage of convolutional architectures is their ability to model coarse-grained (``global'') and fine-grained (``local'') features of a given dataset. The application of convolutional layers allows us to account for the two-dimensional structure of the input data and corresponding (local) interactions of spatially-neighboring spins by dividing the original $32\times 32$ lattice into smaller overlapping regions. These regions provide the basis for a partitioned feature extraction/reconstruction that preserves local spatial properties of the data~\cite{mitchell2012deep}. The max-pooling operations that we use in our convolutional VAE compress the input data and also preserve their two-dimensional structure. For the outlined reasons, convolutional VAEs have a clear advantage over non-convolutional VAEs in terms of local feature extraction. This is particularly evident when comparing their ability to capture spin-spin correlations and energy (see Fig.~\ref{fig:physical_features} (bottom) and App.~\ref{sec:non_conVAE}).

Interestingly, the RBMs we consider are able to capture correlation effects without additional convolutional layers. In general, such performance differences originate from several factors including: (i) the intrinsic properties of the chosen model (e.g., network structure, data representation, etc.), and (ii) the underlying optimization algorithm. Thus, even if convolutional layers provide an efficient way to capture local information, this does not imply that non-convolutional models cannot achieve the same results. In our case, both models rely on a maximum-likelihood parameter estimation. However, in the case of VAEs, we can only maximize the ELBO (see Eq.~\eqref{eq:ELBO}) without any guarantee that this function shares the same local maxima with the actual likelihood. For RBMs, our results suggest that contrastive-divergence learning can find solutions that preserve spatial information without using convolutional layers. 

\begin{figure}
\centering
\includegraphics[width=0.32\textwidth]{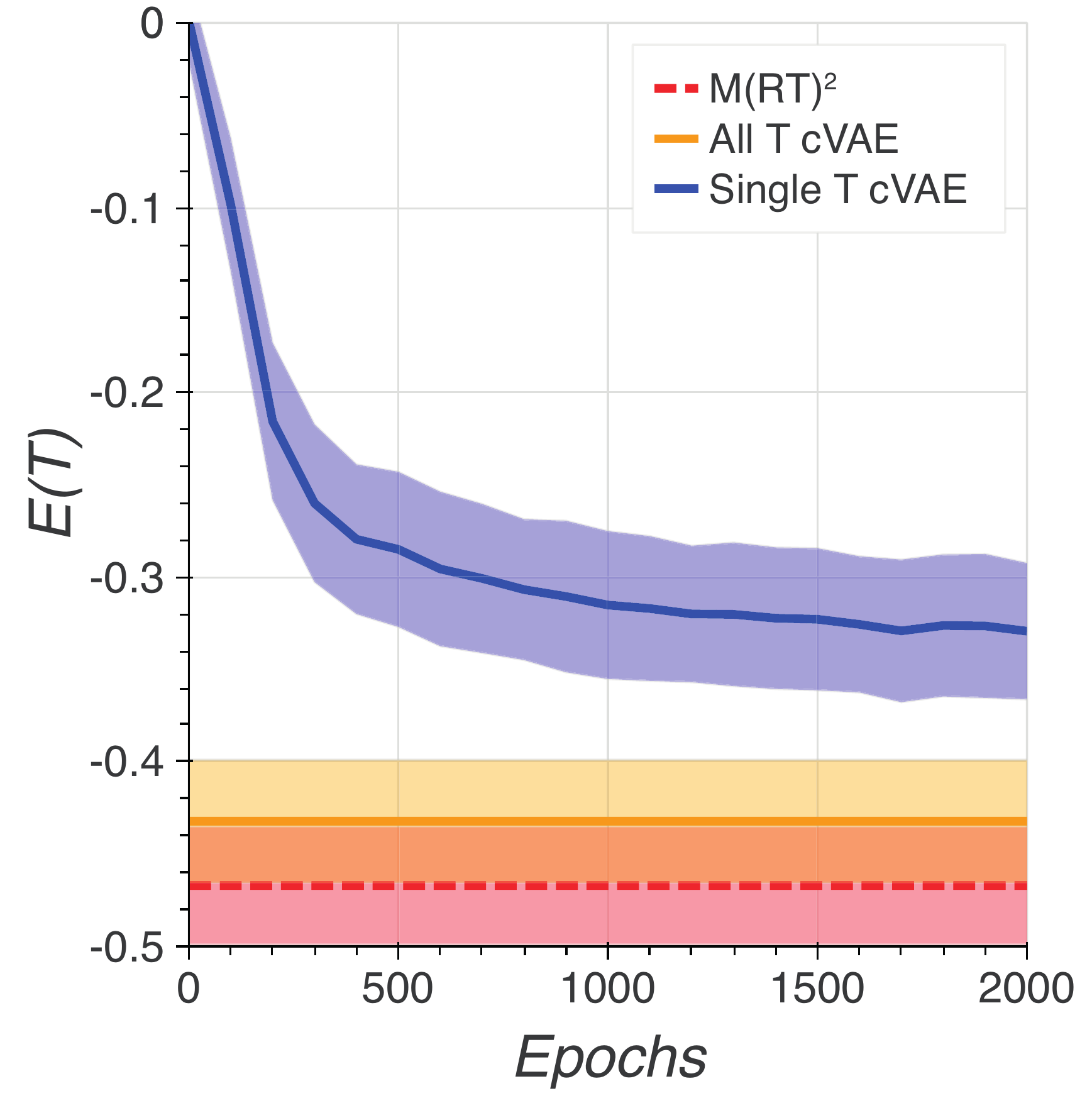}
\caption{\textbf{Evolution of energy during training of a convolutional VAE.} We show the evolution of the energy $E(T)$ for the training of a single convolutional VAE at $T=2.75$ (blue line). The dashed red line indicates the mean energy of the corresponding $\text{M(RT)}^2$ samples. The yellow line shows the mean energy of the convolutional cVAE trained on all temperatures for $1000$ epochs.}
\label{fig:one_vs_allcvae}
\end{figure}

In our analysis, we also noticed another important difference between the learning dynamics of RBMs and VAEs. Restricted Boltzmann machines produce similar results independent of whether they were trained on single or multiple temperatures (see Fig.~\ref{fig:physical_features}) whereas cVAEs exhibit a better performance if trained on all temperatures (see Fig.~\ref{fig:one_vs_allcvae}). Patterns at low and high temperatures are easier to learn and seem to serve as reference points for the encoder during the training of VAEs on all temperatures. This type of training may be useful in other VAE-learning applications.

\begin{figure}
\centering
\includegraphics[width=0.5\textwidth]{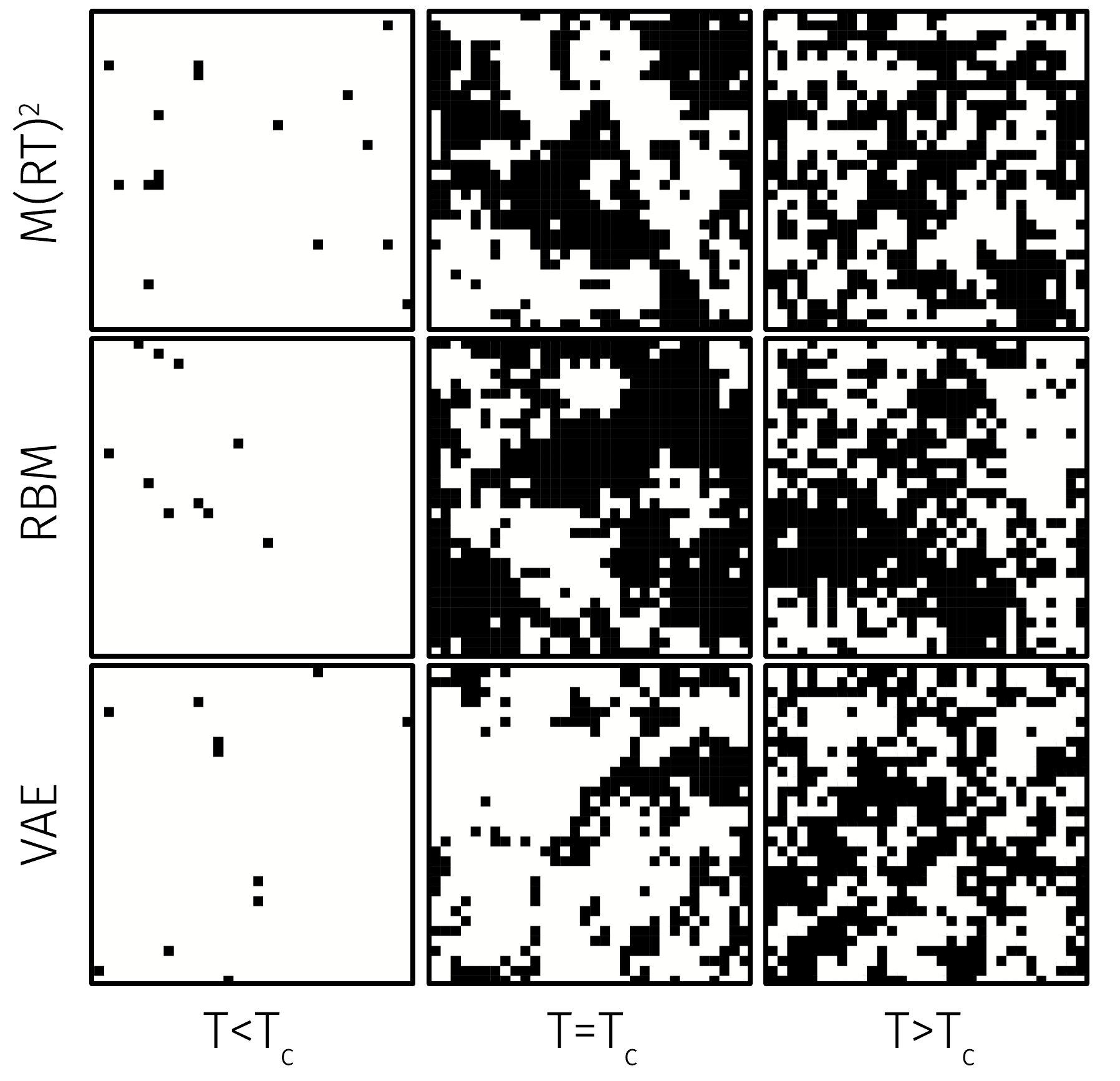}
\caption{\textbf{Snapshots of Ising configurations.} We show snapshots of Ising configurations for $T\in \{1.5,2.5,4\}$. The configurations in the top, middle, and bottom panels are based on M(RT)$^2$, RBM, and convolutional cVAE samples, respectively.}
\label{fig:snapshots}
\end{figure}

For a more visual comparison, we show snapshots of Ising configurations in the original data and compare them to RBM and VAE-generated spin configurations in Fig.~\ref{fig:snapshots}. We also visualize the differences between original data and model-generated Ising samples in terms of two-dimensional energy-magnetization scatter plots and magnetization distributions in App.~\ref{app:scatter_plots}. We observe that the magnetization distributions of the training data are qualitatively well-captured by the corresponding neural-network-generated magnetization distributions. To make a quantitative comparison between model and training data distributions, we determine their corresponding Wasserstein distances $\mathcal{W}$~\cite{wasserstein}. Smaller values of $\mathcal{W}$ indicate a higher similarity of two distributions. Based on the results that we show in Tab.~\ref{tab:tableW}, we conclude that, in terms of a smaller Wasserstein distance, the considered cVAEs are able to capture the magnetization (and energy) distributions of the training data better than the considered RBMs. We, however, note that the mean values may be better captured by RBMs than cVAEs for certain temperatures (see Fig.~\ref{fig:physical_features}).

As a further extension of our framework, we also study the ability of RBMs and cVAEs to capture magnetization, energy, and spin correlations of Ising systems with heterogeneous coupling distributions (see App.~\ref{app:heterogeneous}). Using the example of an Ising model with a Gaussian coupling distribution, we find that both neural network are also able to qualitatively reproduce the temperature dependence of the mentioned thermodynamic quantities. Our results indicate that RBMs perform slightly better than cVAEs in this task for the considered network architecture and training parameters. We note that alternative approaches such as teacher-student frameworks can be also used to generate spin configurations after inferring spin couplings from a given training dataset (inverse Ising problem)~\cite{nguyen2017inverse,abbara2019learning}. Teacher-student frameworks are useful tools if the model family is known and, in the case of the Ising model, system sizes are small enough if spin couplings are heterogeneous. In contrast, the RBMs and cVAEs that we consider in this paper are applicable to homogeneous and heterogeneous Ising systems.
\begin{table*}
  \begin{tabular}{l|c|c|c|c}
    temperature & $\mathbf{\mathcal{W}(M_{RBM}|M_{M(RT)^2})}$ & $\mathbf{\mathcal{W}(M_{cVAE}|M_{M(RT)^2})}$ & $\mathbf{\mathcal{W}(E_{RBM}|E_{M(RT)^2})}$ & $\mathbf{\mathcal{W}(E_{cVAE}|E_{M(RT)^2})}$\\
    \hline
      $T=1.50$ & 0.002 & 0.0003 & 0.006& 0.002 \\
      $T=2.00$ & 0.014 & 0.0054 & 0.06 & 0.02\\
      $T=2.50$ & 0.099 & 0.0471 & 0.02 & 0.06\\
      $T=2.75$ & 0.042 & 0.0402 & 0.01 & 0.04\\
      $T=3.00$ & 0.024 & 0.0225 & 0.04 & 0.04\\
      $T=4.00$ & 0.021 & 0.0243 & 0.13 & 0.02\\
    \hline
      \textbf{mean} &  \textbf{0.034}& \textbf{0.023} & \textbf{0.045} &  \textbf{0.029} \\
  \end{tabular}
  \caption{\textbf{Wasserstein distance between neural-network and training data magnetization and energy distributions.} Wasserstein distance $\mathcal{W}$~\cite{wasserstein} between the magnetization and energy distributions of M(RT)$^2$ samples and corresponding RBM and cVAE samples. Smaller values of $\mathcal{W}$ indicate a higher similarity of distributions.}
  \label{tab:tableW}
\end{table*}
\section{Conclusions and Discussion}
\label{sec:discussion}

In comparison with other generative neural networks, the training of RBMs is challenging and computationally expensive~\cite{schulz2010investigating}. Moreover, conventional RBMs can only handle binary data and generalized architectures are necessary to describe continuous input variables~\cite{schmah2009generative}. For these reasons and the good performance that convolutional models have demonstrated in image classification~\cite{krizhevsky2012imagenet, zeiler2014visualizing, szegedy2015going} and generation~\cite{radford2015unsupervised}, RBM-based representation learning is being more and more replaced by models like VAEs and generative adversarial networks (GANs)~\cite{fisher2018boltzmann,goodfellow2014generative,liu2017simulating}. However, performance assessments of RBMs and VAEs are mainly based on a few common test data sets (e.g., MNIST) that are frequently used in the deep learning community~\cite{mehta2019high}. This approach can only provide limited insights into the representational properties of neural network models. 

To better understand the representational characteristics of RBMs and VAEs, we studied their ability to learn the distribution of physical features of Ising configurations at different temperatures. This approach allowed us to quantitatively compare the distributions learned by RBMs and VAEs in terms of features of an exactly solvable physical system. We found that physical features are well-captured by the considered RBM and VAE models. However, samples generated by the employed VAE are less evenly distributed across temperatures than is the case for samples that we generated with an RBM. In particular, close to the critical point of the Ising model, the considered VAEs has difficulties to generate corresponding samples. In addition, our results also showed that convolutional layers improve the performance of VAEs in terms of their ability to capture magnetization, energy, and spin correlations (see App.~\ref{sec:non_conVAE} for more information on non-convolutional VAEs). Such layers are useful to model and preserve local information~\cite{long2015fully}. Interestingly, RBMs achieve a similar or even better performance without additional convolutional layers. One possible reason for the observed behavior is that the training of VAEs relies on maximizing the ELBO (see Eq.~\eqref{eq:ELBO}), which may not share the same local maxima with the actual log-likelihood. Another distinctive feature between the considered models is that RBMs show a similar performance independent of whether they were trained on one or all temperatures, whereas the performance of VAEs improves if trained on all temperatures. Due to long-range correlation effects in the vicinity of the critical temperature, features of Ising configurations at criticality are more difficult to capture than at low and high temperatures. In the case of VAEs that were trained on all temperatures, low- and high-temperature samples seem to provide reference points that facilitate feature extraction at criticality. To summarize, in the context of physical feature extraction, our results suggest that RBMs are still able to compete with the more recently developed VAEs. 

Possible directions for future research include the study of representational characteristics of RBMs, VAEs, and other generative neural networks by applying them to alternative physical systems (e.g., disordered magnetic systems~\cite{hugli2012artificial})~\cite{westerhout2019neural}. Furthermore, it might be useful to explore connections between the learning capabilities of generative neural networks and corresponding phase-space approximation methods for Ising-like systems such as cluster variational approximation methods (CVAMs)~\cite{cirillo1999correlation,tanaka2002methods,pelizzola2005cluster}.

Future studies may also focus on a comparison between the sampling performance of neural networks and conventional methods. It could be of interest to compare the training and sampling performance between maximum-entropy models and generative neural networks~\cite{nguyen2017inverse}. In the case of one RBM (or cVAE) that was trained for all temperatures, the sampling always involves a classification and it might be the case that one has to discard many samples until one has generated a sample with the desired temperature. Even if the physical features are well-captured, this kind of sampling is not as efficient as M(RT)$^2$, cluster algorithms, or histogram methods. If one uses single machines for each temperature, the sampling process is more efficient. For cVAEs, one just has to generate different vectors with normally-distributed components and for RBMs the sampling is based on parallel updates of the visible units.
\section*{Code availability}
All codes and further detailed descriptions are available on GitHub~\cite{github}.
\acknowledgements{We thank Gonzalo Ruz and Lenka Zdeborova for helpful discussions and comments. Numerical calculations were performed on the ETH Euler cluster.  LB acknowledges financial support from the SNF Early Postdoc.Mobility fellowship on ``Multispecies interacting stochastic systems in biology'' and the Army Research Office (W911NF-18-1-0345).}
\bibliography{refs}

\begin{thebibliography}{63}%
\makeatletter
\providecommand \@ifxundefined [1]{%
 \@ifx{#1\undefined}
}%
\providecommand \@ifnum [1]{%
 \ifnum #1\expandafter \@firstoftwo
 \else \expandafter \@secondoftwo
 \fi
}%
\providecommand \@ifx [1]{%
 \ifx #1\expandafter \@firstoftwo
 \else \expandafter \@secondoftwo
 \fi
}%
\providecommand \natexlab [1]{#1}%
\providecommand \enquote  [1]{``#1''}%
\providecommand \bibnamefont  [1]{#1}%
\providecommand \bibfnamefont [1]{#1}%
\providecommand \citenamefont [1]{#1}%
\providecommand \href@noop [0]{\@secondoftwo}%
\providecommand \href [0]{\begingroup \@sanitize@url \@href}%
\providecommand \@href[1]{\@@startlink{#1}\@@href}%
\providecommand \@@href[1]{\endgroup#1\@@endlink}%
\providecommand \@sanitize@url [0]{\catcode `\\12\catcode `\$12\catcode
  `\&12\catcode `\#12\catcode `\^12\catcode `\_12\catcode `\%12\relax}%
\providecommand \@@startlink[1]{}%
\providecommand \@@endlink[0]{}%
\providecommand \url  [0]{\begingroup\@sanitize@url \@url }%
\providecommand \@url [1]{\endgroup\@href {#1}{\urlprefix }}%
\providecommand \urlprefix  [0]{URL }%
\providecommand \Eprint [0]{\href }%
\providecommand \doibase [0]{http://dx.doi.org/}%
\providecommand \selectlanguage [0]{\@gobble}%
\providecommand \bibinfo  [0]{\@secondoftwo}%
\providecommand \bibfield  [0]{\@secondoftwo}%
\providecommand \translation [1]{[#1]}%
\providecommand \BibitemOpen [0]{}%
\providecommand \bibitemStop [0]{}%
\providecommand \bibitemNoStop [0]{.\EOS\space}%
\providecommand \EOS [0]{\spacefactor3000\relax}%
\providecommand \BibitemShut  [1]{\csname bibitem#1\endcsname}%
\let\auto@bib@innerbib\@empty
\bibitem [{\citenamefont {LeCun}\ \emph {et~al.}(2015)\citenamefont {LeCun},
  \citenamefont {Bengio},\ and\ \citenamefont {Hinton}}]{lecun2015deep}%
  \BibitemOpen
  \bibfield  {author} {\bibinfo {author} {\bibfnamefont {Y.}~\bibnamefont
  {LeCun}}, \bibinfo {author} {\bibfnamefont {Y.}~\bibnamefont {Bengio}}, \
  and\ \bibinfo {author} {\bibfnamefont {G.}~\bibnamefont {Hinton}},\
  }\href@noop {} {\bibfield  {journal} {\bibinfo  {journal} {Nature}\ }\textbf
  {\bibinfo {volume} {521}},\ \bibinfo {pages} {436} (\bibinfo {year}
  {2015})}\BibitemShut {NoStop}%
\bibitem [{\citenamefont {Nguyen}\ \emph {et~al.}(2017)\citenamefont {Nguyen},
  \citenamefont {Zecchina},\ and\ \citenamefont {Berg}}]{nguyen2017inverse}%
  \BibitemOpen
  \bibfield  {author} {\bibinfo {author} {\bibfnamefont {H.~C.}\ \bibnamefont
  {Nguyen}}, \bibinfo {author} {\bibfnamefont {R.}~\bibnamefont {Zecchina}}, \
  and\ \bibinfo {author} {\bibfnamefont {J.}~\bibnamefont {Berg}},\ }\href@noop
  {} {\bibfield  {journal} {\bibinfo  {journal} {Adv. Phys.}\ }\textbf
  {\bibinfo {volume} {66}},\ \bibinfo {pages} {197} (\bibinfo {year}
  {2017})}\BibitemShut {NoStop}%
\bibitem [{\citenamefont {Davidsen}\ \emph {et~al.}(2019)\citenamefont
  {Davidsen}, \citenamefont {Olson}, \citenamefont {DeWitt~III}, \citenamefont
  {Feng}, \citenamefont {Harkins}, \citenamefont {Bradley},\ and\ \citenamefont
  {Matsen~IV}}]{davidsen2019deep}%
  \BibitemOpen
  \bibfield  {author} {\bibinfo {author} {\bibfnamefont {K.}~\bibnamefont
  {Davidsen}}, \bibinfo {author} {\bibfnamefont {B.~J.}\ \bibnamefont {Olson}},
  \bibinfo {author} {\bibfnamefont {W.~S.}\ \bibnamefont {DeWitt~III}},
  \bibinfo {author} {\bibfnamefont {J.}~\bibnamefont {Feng}}, \bibinfo {author}
  {\bibfnamefont {E.}~\bibnamefont {Harkins}}, \bibinfo {author} {\bibfnamefont
  {P.}~\bibnamefont {Bradley}}, \ and\ \bibinfo {author} {\bibfnamefont
  {F.~A.}\ \bibnamefont {Matsen~IV}},\ }\href@noop {} {\bibfield  {journal}
  {\bibinfo  {journal} {Elife}\ }\textbf {\bibinfo {volume} {8}} (\bibinfo
  {year} {2019})}\BibitemShut {NoStop}%
\bibitem [{\citenamefont {Ackley}\ \emph {et~al.}(1985)\citenamefont {Ackley},
  \citenamefont {Hinton},\ and\ \citenamefont
  {Sejnowski}}]{ackley1985learning}%
  \BibitemOpen
  \bibfield  {author} {\bibinfo {author} {\bibfnamefont {D.~H.}\ \bibnamefont
  {Ackley}}, \bibinfo {author} {\bibfnamefont {G.~E.}\ \bibnamefont {Hinton}},
  \ and\ \bibinfo {author} {\bibfnamefont {T.~J.}\ \bibnamefont {Sejnowski}},\
  }\href@noop {} {\bibfield  {journal} {\bibinfo  {journal} {Cogn. Sci.}\
  }\textbf {\bibinfo {volume} {9}},\ \bibinfo {pages} {147} (\bibinfo {year}
  {1985})}\BibitemShut {NoStop}%
\bibitem [{\citenamefont {Hinton}(1989)}]{hinton1989deterministic}%
  \BibitemOpen
  \bibfield  {author} {\bibinfo {author} {\bibfnamefont {G.~E.}\ \bibnamefont
  {Hinton}},\ }\href@noop {} {\bibfield  {journal} {\bibinfo  {journal} {Neural
  Comp.}\ }\textbf {\bibinfo {volume} {1}},\ \bibinfo {pages} {143} (\bibinfo
  {year} {1989})}\BibitemShut {NoStop}%
\bibitem [{\citenamefont {Hinton}(2002)}]{hinton2002training}%
  \BibitemOpen
  \bibfield  {author} {\bibinfo {author} {\bibfnamefont {G.~E.}\ \bibnamefont
  {Hinton}},\ }\href@noop {} {\bibfield  {journal} {\bibinfo  {journal} {Neural
  Comp.}\ }\textbf {\bibinfo {volume} {14}},\ \bibinfo {pages} {1771} (\bibinfo
  {year} {2002})}\BibitemShut {NoStop}%
\bibitem [{\citenamefont {Kim}\ and\ \citenamefont
  {Kim}(2018)}]{kim2018smallest}%
  \BibitemOpen
  \bibfield  {author} {\bibinfo {author} {\bibfnamefont {D.}~\bibnamefont
  {Kim}}\ and\ \bibinfo {author} {\bibfnamefont {D.-H.}\ \bibnamefont {Kim}},\
  }\href@noop {} {\bibfield  {journal} {\bibinfo  {journal} {Phys. Rev. E}\
  }\textbf {\bibinfo {volume} {98}},\ \bibinfo {pages} {022138} (\bibinfo
  {year} {2018})}\BibitemShut {NoStop}%
\bibitem [{\citenamefont {Efthymiou}\ \emph {et~al.}(2019)\citenamefont
  {Efthymiou}, \citenamefont {Beach},\ and\ \citenamefont
  {Melko}}]{efthymiou2019super}%
  \BibitemOpen
  \bibfield  {author} {\bibinfo {author} {\bibfnamefont {S.}~\bibnamefont
  {Efthymiou}}, \bibinfo {author} {\bibfnamefont {M.~J.}\ \bibnamefont
  {Beach}}, \ and\ \bibinfo {author} {\bibfnamefont {R.~G.}\ \bibnamefont
  {Melko}},\ }\href@noop {} {\bibfield  {journal} {\bibinfo  {journal} {Phys.
  Rev. B}\ }\textbf {\bibinfo {volume} {99}},\ \bibinfo {pages} {075113}
  (\bibinfo {year} {2019})}\BibitemShut {NoStop}%
\bibitem [{\citenamefont {Carleo}\ and\ \citenamefont
  {Troyer}(2017)}]{carleo2017solving}%
  \BibitemOpen
  \bibfield  {author} {\bibinfo {author} {\bibfnamefont {G.}~\bibnamefont
  {Carleo}}\ and\ \bibinfo {author} {\bibfnamefont {M.}~\bibnamefont
  {Troyer}},\ }\href@noop {} {\bibfield  {journal} {\bibinfo  {journal}
  {Science}\ }\textbf {\bibinfo {volume} {355}},\ \bibinfo {pages} {602}
  (\bibinfo {year} {2017})}\BibitemShut {NoStop}%
\bibitem [{\citenamefont {Torlai}\ \emph {et~al.}(2018)\citenamefont {Torlai},
  \citenamefont {Mazzola}, \citenamefont {Carrasquilla}, \citenamefont
  {Troyer}, \citenamefont {Melko},\ and\ \citenamefont
  {Carleo}}]{torlai2018neural}%
  \BibitemOpen
  \bibfield  {author} {\bibinfo {author} {\bibfnamefont {G.}~\bibnamefont
  {Torlai}}, \bibinfo {author} {\bibfnamefont {G.}~\bibnamefont {Mazzola}},
  \bibinfo {author} {\bibfnamefont {J.}~\bibnamefont {Carrasquilla}}, \bibinfo
  {author} {\bibfnamefont {M.}~\bibnamefont {Troyer}}, \bibinfo {author}
  {\bibfnamefont {R.}~\bibnamefont {Melko}}, \ and\ \bibinfo {author}
  {\bibfnamefont {G.}~\bibnamefont {Carleo}},\ }\href@noop {} {\bibfield
  {journal} {\bibinfo  {journal} {Nature Phys.}\ }\textbf {\bibinfo {volume}
  {14}},\ \bibinfo {pages} {447} (\bibinfo {year} {2018})}\BibitemShut
  {NoStop}%
\bibitem [{\citenamefont {Mehta}\ \emph {et~al.}(2019)\citenamefont {Mehta},
  \citenamefont {Bukov}, \citenamefont {Wang}, \citenamefont {Day},
  \citenamefont {Richardson}, \citenamefont {Fisher},\ and\ \citenamefont
  {Schwab}}]{mehta2019high}%
  \BibitemOpen
  \bibfield  {author} {\bibinfo {author} {\bibfnamefont {P.}~\bibnamefont
  {Mehta}}, \bibinfo {author} {\bibfnamefont {M.}~\bibnamefont {Bukov}},
  \bibinfo {author} {\bibfnamefont {C.-H.}\ \bibnamefont {Wang}}, \bibinfo
  {author} {\bibfnamefont {A.~G.}\ \bibnamefont {Day}}, \bibinfo {author}
  {\bibfnamefont {C.}~\bibnamefont {Richardson}}, \bibinfo {author}
  {\bibfnamefont {C.~K.}\ \bibnamefont {Fisher}}, \ and\ \bibinfo {author}
  {\bibfnamefont {D.~J.}\ \bibnamefont {Schwab}},\ }\href@noop {} {\bibfield
  {journal} {\bibinfo  {journal} {Phys. Rep.}\ } (\bibinfo {year}
  {2019})}\BibitemShut {NoStop}%
\bibitem [{\citenamefont {Wu}\ and\ \citenamefont
  {Tegmark}(2019)}]{wu2019toward}%
  \BibitemOpen
  \bibfield  {author} {\bibinfo {author} {\bibfnamefont {T.}~\bibnamefont
  {Wu}}\ and\ \bibinfo {author} {\bibfnamefont {M.}~\bibnamefont {Tegmark}},\
  }\href@noop {} {\bibfield  {journal} {\bibinfo  {journal} {Phys. Rev. E}\
  }\textbf {\bibinfo {volume} {100}},\ \bibinfo {pages} {033311} (\bibinfo
  {year} {2019})}\BibitemShut {NoStop}%
\bibitem [{\citenamefont {Kingma}\ and\ \citenamefont
  {Welling}(2013)}]{kingma2013auto}%
  \BibitemOpen
  \bibfield  {author} {\bibinfo {author} {\bibfnamefont {D.~P.}\ \bibnamefont
  {Kingma}}\ and\ \bibinfo {author} {\bibfnamefont {M.}~\bibnamefont
  {Welling}},\ }\href@noop {} {\bibfield  {journal} {\bibinfo  {journal} {arXiv
  preprint arXiv:1312.6114}\ } (\bibinfo {year} {2013})}\BibitemShut {NoStop}%
\bibitem [{\citenamefont {Wetzel}(2017)}]{wetzel2017unsupervised}%
  \BibitemOpen
  \bibfield  {author} {\bibinfo {author} {\bibfnamefont {S.~J.}\ \bibnamefont
  {Wetzel}},\ }\href@noop {} {\bibfield  {journal} {\bibinfo  {journal} {{Phys.
  Rev. E}}\ }\textbf {\bibinfo {volume} {96}},\ \bibinfo {pages} {022140}
  (\bibinfo {year} {2017})}\BibitemShut {NoStop}%
\bibitem [{\citenamefont {Iso}\ \emph {et~al.}(2018)\citenamefont {Iso},
  \citenamefont {Shiba},\ and\ \citenamefont {Yokoo}}]{iso2018scale}%
  \BibitemOpen
  \bibfield  {author} {\bibinfo {author} {\bibfnamefont {S.}~\bibnamefont
  {Iso}}, \bibinfo {author} {\bibfnamefont {S.}~\bibnamefont {Shiba}}, \ and\
  \bibinfo {author} {\bibfnamefont {S.}~\bibnamefont {Yokoo}},\ }\href@noop {}
  {\bibfield  {journal} {\bibinfo  {journal} {Phys. Rev. E}\ }\textbf {\bibinfo
  {volume} {97}},\ \bibinfo {pages} {053304} (\bibinfo {year}
  {2018})}\BibitemShut {NoStop}%
\bibitem [{\citenamefont {Li}\ and\ \citenamefont {Wang}(2018)}]{li2018neural}%
  \BibitemOpen
  \bibfield  {author} {\bibinfo {author} {\bibfnamefont {S.-H.}\ \bibnamefont
  {Li}}\ and\ \bibinfo {author} {\bibfnamefont {L.}~\bibnamefont {Wang}},\
  }\href@noop {} {\bibfield  {journal} {\bibinfo  {journal} {Phys. Rev. Lett.}\
  }\textbf {\bibinfo {volume} {121}},\ \bibinfo {pages} {260601} (\bibinfo
  {year} {2018})}\BibitemShut {NoStop}%
\bibitem [{\citenamefont {Koch-Janusz}\ and\ \citenamefont
  {Ringel}(2018)}]{koch2018mutual}%
  \BibitemOpen
  \bibfield  {author} {\bibinfo {author} {\bibfnamefont {M.}~\bibnamefont
  {Koch-Janusz}}\ and\ \bibinfo {author} {\bibfnamefont {Z.}~\bibnamefont
  {Ringel}},\ }\href@noop {} {\bibfield  {journal} {\bibinfo  {journal} {Nat.
  Phys.}\ }\textbf {\bibinfo {volume} {14}},\ \bibinfo {pages} {578} (\bibinfo
  {year} {2018})}\BibitemShut {NoStop}%
\bibitem [{\citenamefont {Torlai}\ and\ \citenamefont
  {Melko}(2016)}]{torlai2016learning}%
  \BibitemOpen
  \bibfield  {author} {\bibinfo {author} {\bibfnamefont {G.}~\bibnamefont
  {Torlai}}\ and\ \bibinfo {author} {\bibfnamefont {R.~G.}\ \bibnamefont
  {Melko}},\ }\href@noop {} {\bibfield  {journal} {\bibinfo  {journal} {Phys.
  Rev. B}\ }\textbf {\bibinfo {volume} {94}},\ \bibinfo {pages} {165134}
  (\bibinfo {year} {2016})}\BibitemShut {NoStop}%
\bibitem [{\citenamefont {Morningstar}\ and\ \citenamefont
  {Melko}(2017)}]{morningstar2017deep}%
  \BibitemOpen
  \bibfield  {author} {\bibinfo {author} {\bibfnamefont {A.}~\bibnamefont
  {Morningstar}}\ and\ \bibinfo {author} {\bibfnamefont {R.~G.}\ \bibnamefont
  {Melko}},\ }\href@noop {} {\bibfield  {journal} {\bibinfo  {journal} {J.
  Mach. Learn. Res.}\ }\textbf {\bibinfo {volume} {18}},\ \bibinfo {pages}
  {5975} (\bibinfo {year} {2017})}\BibitemShut {NoStop}%
\bibitem [{\citenamefont {Smolensky}()}]{smolensky1chapter}%
  \BibitemOpen
  \bibfield  {author} {\bibinfo {author} {\bibfnamefont {P.}~\bibnamefont
  {Smolensky}},\ }\href@noop {} {\bibfield  {journal} {\bibinfo  {journal}
  {Parallel Distributed Processing: Explorations in the Microstructure of
  Cognition}\ }\textbf {\bibinfo {volume} {1}}}\BibitemShut {NoStop}%
\bibitem [{\citenamefont {Hinton}(2012)}]{hinton2012practical}%
  \BibitemOpen
  \bibfield  {author} {\bibinfo {author} {\bibfnamefont {G.~E.}\ \bibnamefont
  {Hinton}},\ }in\ \href@noop {} {\emph {\bibinfo {booktitle} {{Neural
  networks: Tricks of the trade}}}}\ (\bibinfo  {publisher} {Springer},\
  \bibinfo {year} {2012})\ pp.\ \bibinfo {pages} {599--619}\BibitemShut
  {NoStop}%
\bibitem [{\citenamefont {Schmah}\ \emph {et~al.}(2009)\citenamefont {Schmah},
  \citenamefont {Hinton}, \citenamefont {Small}, \citenamefont {Strother},\
  and\ \citenamefont {Zemel}}]{schmah2009generative}%
  \BibitemOpen
  \bibfield  {author} {\bibinfo {author} {\bibfnamefont {T.}~\bibnamefont
  {Schmah}}, \bibinfo {author} {\bibfnamefont {G.~E.}\ \bibnamefont {Hinton}},
  \bibinfo {author} {\bibfnamefont {S.~L.}\ \bibnamefont {Small}}, \bibinfo
  {author} {\bibfnamefont {S.}~\bibnamefont {Strother}}, \ and\ \bibinfo
  {author} {\bibfnamefont {R.~S.}\ \bibnamefont {Zemel}},\ }in\ \href@noop {}
  {\emph {\bibinfo {booktitle} {Adv. Neural Inf. Process. Syst.}}}\ (\bibinfo
  {year} {2009})\ pp.\ \bibinfo {pages} {1409--1416}\BibitemShut {NoStop}%
\bibitem [{\citenamefont {Li}\ and\ \citenamefont
  {Zhu}(2018)}]{li2018exponential}%
  \BibitemOpen
  \bibfield  {author} {\bibinfo {author} {\bibfnamefont {Y.}~\bibnamefont
  {Li}}\ and\ \bibinfo {author} {\bibfnamefont {X.}~\bibnamefont {Zhu}},\ }in\
  \href@noop {} {\emph {\bibinfo {booktitle} {2018 International Joint
  Conference on Neural Networks (IJCNN)}}}\ (\bibinfo {organization} {IEEE},\
  \bibinfo {year} {2018})\ pp.\ \bibinfo {pages} {1--10}\BibitemShut {NoStop}%
\bibitem [{\citenamefont {Huang}(2009)}]{huang2009introduction}%
  \BibitemOpen
  \bibfield  {author} {\bibinfo {author} {\bibfnamefont {K.}~\bibnamefont
  {Huang}},\ }\href@noop {} {\emph {\bibinfo {title} {Introduction to
  statistical physics}}}\ (\bibinfo  {publisher} {{Chapman and Hall/CRC,
  London}},\ \bibinfo {year} {2009})\BibitemShut {NoStop}%
\bibitem [{\citenamefont {Gabrie}\ \emph {et~al.}(2015)\citenamefont {Gabrie},
  \citenamefont {Tramel},\ and\ \citenamefont {Krzakala}}]{NIPS2015_5788}%
  \BibitemOpen
  \bibfield  {author} {\bibinfo {author} {\bibfnamefont {M.}~\bibnamefont
  {Gabrie}}, \bibinfo {author} {\bibfnamefont {E.~W.}\ \bibnamefont {Tramel}},
  \ and\ \bibinfo {author} {\bibfnamefont {F.}~\bibnamefont {Krzakala}},\ }in\
  \href
  {http://papers.nips.cc/paper/5788-training-restricted-boltzmann-machine-via-the-thouless-anderson-palmer-free-energy.pdf}
  {\emph {\bibinfo {booktitle} {Advances in Neural Information Processing
  Systems 28}}},\ \bibinfo {editor} {edited by\ \bibinfo {editor}
  {\bibfnamefont {C.}~\bibnamefont {Cortes}}, \bibinfo {editor} {\bibfnamefont
  {N.~D.}\ \bibnamefont {Lawrence}}, \bibinfo {editor} {\bibfnamefont {D.~D.}\
  \bibnamefont {Lee}}, \bibinfo {editor} {\bibfnamefont {M.}~\bibnamefont
  {Sugiyama}}, \ and\ \bibinfo {editor} {\bibfnamefont {R.}~\bibnamefont
  {Garnett}}}\ (\bibinfo  {publisher} {Curran Associates, Inc.},\ \bibinfo
  {year} {2015})\ pp.\ \bibinfo {pages} {640--648}\BibitemShut {NoStop}%
\bibitem [{\citenamefont {Bottou}\ \emph {et~al.}(2018)\citenamefont {Bottou},
  \citenamefont {Curtis},\ and\ \citenamefont
  {Nocedal}}]{bottou2018optimization}%
  \BibitemOpen
  \bibfield  {author} {\bibinfo {author} {\bibfnamefont {L.}~\bibnamefont
  {Bottou}}, \bibinfo {author} {\bibfnamefont {F.~E.}\ \bibnamefont {Curtis}},
  \ and\ \bibinfo {author} {\bibfnamefont {J.}~\bibnamefont {Nocedal}},\
  }\href@noop {} {\bibfield  {journal} {\bibinfo  {journal} {{SIAM Rev.}}\
  }\textbf {\bibinfo {volume} {60}},\ \bibinfo {pages} {223} (\bibinfo {year}
  {2018})}\BibitemShut {NoStop}%
\bibitem [{\citenamefont {Tieleman}(2008)}]{tieleman2008training}%
  \BibitemOpen
  \bibfield  {author} {\bibinfo {author} {\bibfnamefont {T.}~\bibnamefont
  {Tieleman}},\ }in\ \href@noop {} {\emph {\bibinfo {booktitle} {Proceedings of
  the 25th international conference on Machine learning}}}\ (\bibinfo
  {organization} {ACM},\ \bibinfo {year} {2008})\ pp.\ \bibinfo {pages}
  {1064--1071}\BibitemShut {NoStop}%
\bibitem [{\citenamefont {Blei}\ \emph {et~al.}(2017)\citenamefont {Blei},
  \citenamefont {Kucukelbir},\ and\ \citenamefont
  {McAuliffe}}]{blei2017variational}%
  \BibitemOpen
  \bibfield  {author} {\bibinfo {author} {\bibfnamefont {D.~M.}\ \bibnamefont
  {Blei}}, \bibinfo {author} {\bibfnamefont {A.}~\bibnamefont {Kucukelbir}}, \
  and\ \bibinfo {author} {\bibfnamefont {J.~D.}\ \bibnamefont {McAuliffe}},\
  }\href@noop {} {\bibfield  {journal} {\bibinfo  {journal} {J. Am. Stat.
  Assoc.}\ }\textbf {\bibinfo {volume} {112}},\ \bibinfo {pages} {859}
  (\bibinfo {year} {2017})}\BibitemShut {NoStop}%
\bibitem [{\citenamefont {Sohn}\ \emph {et~al.}(2015)\citenamefont {Sohn},
  \citenamefont {Lee},\ and\ \citenamefont {Yan}}]{NIPS2015_5775}%
  \BibitemOpen
  \bibfield  {author} {\bibinfo {author} {\bibfnamefont {K.}~\bibnamefont
  {Sohn}}, \bibinfo {author} {\bibfnamefont {H.}~\bibnamefont {Lee}}, \ and\
  \bibinfo {author} {\bibfnamefont {X.}~\bibnamefont {Yan}},\ }in\ \href
  {http://papers.nips.cc/paper/5775-learning-structured-output-representation-using-deep-conditional-generative-models.pdf}
  {\emph {\bibinfo {booktitle} {Advances in Neural Information Processing
  Systems 28}}},\ \bibinfo {editor} {edited by\ \bibinfo {editor}
  {\bibfnamefont {C.}~\bibnamefont {Cortes}}, \bibinfo {editor} {\bibfnamefont
  {N.~D.}\ \bibnamefont {Lawrence}}, \bibinfo {editor} {\bibfnamefont {D.~D.}\
  \bibnamefont {Lee}}, \bibinfo {editor} {\bibfnamefont {M.}~\bibnamefont
  {Sugiyama}}, \ and\ \bibinfo {editor} {\bibfnamefont {R.}~\bibnamefont
  {Garnett}}}\ (\bibinfo  {publisher} {Curran Associates, Inc.},\ \bibinfo
  {year} {2015})\ pp.\ \bibinfo {pages} {3483--3491}\BibitemShut {NoStop}%
\bibitem [{\citenamefont {Bengio}\ \emph {et~al.}(2007)\citenamefont {Bengio},
  \citenamefont {Lamblin}, \citenamefont {Popovici},\ and\ \citenamefont
  {Larochelle}}]{bengio2007greedy}%
  \BibitemOpen
  \bibfield  {author} {\bibinfo {author} {\bibfnamefont {Y.}~\bibnamefont
  {Bengio}}, \bibinfo {author} {\bibfnamefont {P.}~\bibnamefont {Lamblin}},
  \bibinfo {author} {\bibfnamefont {D.}~\bibnamefont {Popovici}}, \ and\
  \bibinfo {author} {\bibfnamefont {H.}~\bibnamefont {Larochelle}},\ }in\
  \href@noop {} {\emph {\bibinfo {booktitle} {Adv. Neural Inf. Process.
  Syst.}}}\ (\bibinfo {year} {2007})\ pp.\ \bibinfo {pages}
  {153--160}\BibitemShut {NoStop}%
\bibitem [{\citenamefont {Besag}(1975)}]{besag1975statistical}%
  \BibitemOpen
  \bibfield  {author} {\bibinfo {author} {\bibfnamefont {J.}~\bibnamefont
  {Besag}},\ }\href@noop {} {\bibfield  {journal} {\bibinfo  {journal} {Journal
  of the Royal Statistical Society: Series D (The Statistician)}\ }\textbf
  {\bibinfo {volume} {24}},\ \bibinfo {pages} {179} (\bibinfo {year}
  {1975})}\BibitemShut {NoStop}%
\bibitem [{dee(2020)}]{deeplearning}%
  \BibitemOpen
  \href@noop {} {\enquote {\bibinfo {title} {{DeepLearning 0.1 Documentation}
  \url{http://deeplearning.net/tutorial/rbm.html}},}\ } (\bibinfo {year}
  {2020})\BibitemShut {NoStop}%
\bibitem [{\citenamefont {Rasmussen}(2000)}]{rasmussen2000infinite}%
  \BibitemOpen
  \bibfield  {author} {\bibinfo {author} {\bibfnamefont {C.~E.}\ \bibnamefont
  {Rasmussen}},\ }in\ \href@noop {} {\emph {\bibinfo {booktitle} {Advances in
  neural information processing systems}}}\ (\bibinfo {year} {2000})\ pp.\
  \bibinfo {pages} {554--560}\BibitemShut {NoStop}%
\bibitem [{\citenamefont {Wang}\ \emph {et~al.}(2009)\citenamefont {Wang},
  \citenamefont {Kulkarni},\ and\ \citenamefont
  {Verd{\'u}}}]{wang2009divergence}%
  \BibitemOpen
  \bibfield  {author} {\bibinfo {author} {\bibfnamefont {Q.}~\bibnamefont
  {Wang}}, \bibinfo {author} {\bibfnamefont {S.~R.}\ \bibnamefont {Kulkarni}},
  \ and\ \bibinfo {author} {\bibfnamefont {S.}~\bibnamefont {Verd{\'u}}},\
  }\href@noop {} {\bibfield  {journal} {\bibinfo  {journal} {IEEE Transactions
  on Information Theory}\ }\textbf {\bibinfo {volume} {55}},\ \bibinfo {pages}
  {2392} (\bibinfo {year} {2009})}\BibitemShut {NoStop}%
\bibitem [{\citenamefont {Loftsgaarden}\ \emph {et~al.}(1965)\citenamefont
  {Loftsgaarden}, \citenamefont {Quesenberry} \emph
  {et~al.}}]{loftsgaarden1965nonparametric}%
  \BibitemOpen
  \bibfield  {author} {\bibinfo {author} {\bibfnamefont {D.~O.}\ \bibnamefont
  {Loftsgaarden}}, \bibinfo {author} {\bibfnamefont {C.~P.}\ \bibnamefont
  {Quesenberry}},  \emph {et~al.},\ }\href@noop {} {\bibfield  {journal}
  {\bibinfo  {journal} {Ann. Math. Stat.}\ }\textbf {\bibinfo {volume} {36}},\
  \bibinfo {pages} {1049} (\bibinfo {year} {1965})}\BibitemShut {NoStop}%
\bibitem [{Note1()}]{Note1}%
  \BibitemOpen
  \bibinfo {note} {The choice of the distance metric depends on the nature of
  the samples. We use the Jaccard distance for the binary data in our
  models.}\BibitemShut {Stop}%
\bibitem [{\citenamefont {Landau}\ and\ \citenamefont
  {Binder}(2009)}]{landau-binder2009}%
  \BibitemOpen
  \bibfield  {author} {\bibinfo {author} {\bibfnamefont {D.~P.}\ \bibnamefont
  {Landau}}\ and\ \bibinfo {author} {\bibfnamefont {K.}~\bibnamefont
  {Binder}},\ }\href@noop {} {\emph {\bibinfo {title} {{A Guide to Monte Carlo
  Simulations in Statistical Physics}}}}\ (\bibinfo  {publisher} {Cambridge
  University Press, Cambridge},\ \bibinfo {year} {2009})\BibitemShut {NoStop}%
\bibitem [{\citenamefont {Onsager}(1944)}]{onsager1944crystal}%
  \BibitemOpen
  \bibfield  {author} {\bibinfo {author} {\bibfnamefont {L.}~\bibnamefont
  {Onsager}},\ }\href@noop {} {\bibfield  {journal} {\bibinfo  {journal} {Phys.
  Rev.}\ }\textbf {\bibinfo {volume} {65}},\ \bibinfo {pages} {117} (\bibinfo
  {year} {1944})}\BibitemShut {NoStop}%
\bibitem [{\citenamefont {Kalos}\ and\ \citenamefont
  {Whitlock}(2009)}]{kalos2009monte}%
  \BibitemOpen
  \bibfield  {author} {\bibinfo {author} {\bibfnamefont {M.~H.}\ \bibnamefont
  {Kalos}}\ and\ \bibinfo {author} {\bibfnamefont {P.~A.}\ \bibnamefont
  {Whitlock}},\ }\href@noop {} {\emph {\bibinfo {title} {Monte carlo
  methods}}}\ (\bibinfo  {publisher} {John Wiley \& Sons},\ \bibinfo {year}
  {2009})\BibitemShut {NoStop}%
\bibitem [{\citenamefont {Kingma}\ and\ \citenamefont
  {Ba}(2014)}]{kingma2014adam}%
  \BibitemOpen
  \bibfield  {author} {\bibinfo {author} {\bibfnamefont {D.~P.}\ \bibnamefont
  {Kingma}}\ and\ \bibinfo {author} {\bibfnamefont {J.}~\bibnamefont {Ba}},\
  }\href@noop {} {\bibfield  {journal} {\bibinfo  {journal} {arXiv preprint
  arXiv:1412.6980}\ } (\bibinfo {year} {2014})}\BibitemShut {NoStop}%
\bibitem [{\citenamefont {Glorot}\ and\ \citenamefont
  {Bengio}(2010)}]{glorot2010understanding}%
  \BibitemOpen
  \bibfield  {author} {\bibinfo {author} {\bibfnamefont {X.}~\bibnamefont
  {Glorot}}\ and\ \bibinfo {author} {\bibfnamefont {Y.}~\bibnamefont
  {Bengio}},\ }in\ \href@noop {} {\emph {\bibinfo {booktitle} {Proceedings of
  the thirteenth international conference on artificial intelligence and
  statistics}}}\ (\bibinfo {year} {2010})\ pp.\ \bibinfo {pages}
  {249--256}\BibitemShut {NoStop}%
\bibitem [{Note2()}]{Note2}%
  \BibitemOpen
  \bibinfo {note} {For filter sizes that approach the size of the underlying
  system, we would recover a non-convolutional network architecture. If the
  filters are too small, we can only account for local connectivity in very
  small neighborhoods. In accordance with Ref.~\cite {simonyan2014very}, we use
  filters of size 3. We also performed simulations for filter sizes 2 and 4 and
  found similar performances of the considered convolutional VAEs.}\BibitemShut
  {Stop}%
\bibitem [{\citenamefont {Ioffe}\ and\ \citenamefont
  {Szegedy}(2015)}]{ioffe2015batch}%
  \BibitemOpen
  \bibfield  {author} {\bibinfo {author} {\bibfnamefont {S.}~\bibnamefont
  {Ioffe}}\ and\ \bibinfo {author} {\bibfnamefont {C.}~\bibnamefont
  {Szegedy}},\ }\href@noop {} {\bibfield  {journal} {\bibinfo  {journal} {arXiv
  preprint arXiv:1502.03167}\ } (\bibinfo {year} {2015})}\BibitemShut {NoStop}%
\bibitem [{\citenamefont {Srivastava}\ \emph {et~al.}(2014)\citenamefont
  {Srivastava}, \citenamefont {Hinton}, \citenamefont {Krizhevsky},
  \citenamefont {Sutskever},\ and\ \citenamefont
  {Salakhutdinov}}]{srivastava2014dropout}%
  \BibitemOpen
  \bibfield  {author} {\bibinfo {author} {\bibfnamefont {N.}~\bibnamefont
  {Srivastava}}, \bibinfo {author} {\bibfnamefont {G.}~\bibnamefont {Hinton}},
  \bibinfo {author} {\bibfnamefont {A.}~\bibnamefont {Krizhevsky}}, \bibinfo
  {author} {\bibfnamefont {I.}~\bibnamefont {Sutskever}}, \ and\ \bibinfo
  {author} {\bibfnamefont {R.}~\bibnamefont {Salakhutdinov}},\ }\href@noop {}
  {\bibfield  {journal} {\bibinfo  {journal} {J. Mach. Learn. Res.}\ }\textbf
  {\bibinfo {volume} {15}},\ \bibinfo {pages} {1929} (\bibinfo {year}
  {2014})}\BibitemShut {NoStop}%
\bibitem [{Note3()}]{Note3}%
  \BibitemOpen
  \bibinfo {note} {Training the considered RBMs for $10^3$ epochs required 7
  days of computing time on an Intel Xeon E7-8867v3 (2.5 GHz) CPU. For cVAEs,
  the training took 20 hours on a GeForce GTX 1080 Ti GPU.}\BibitemShut {Stop}%
\bibitem [{\citenamefont {Krizhevsky}\ \emph {et~al.}(2012)\citenamefont
  {Krizhevsky}, \citenamefont {Sutskever},\ and\ \citenamefont
  {Hinton}}]{krizhevsky2012imagenet}%
  \BibitemOpen
  \bibfield  {author} {\bibinfo {author} {\bibfnamefont {A.}~\bibnamefont
  {Krizhevsky}}, \bibinfo {author} {\bibfnamefont {I.}~\bibnamefont
  {Sutskever}}, \ and\ \bibinfo {author} {\bibfnamefont {G.~E.}\ \bibnamefont
  {Hinton}},\ }\bibfield  {booktitle} {\emph {\bibinfo {booktitle} {Advances in
  neural information processing systems}},\ }\href@noop {} {\ ,\ \bibinfo
  {pages} {1097} (\bibinfo {year} {2012})}\BibitemShut {NoStop}%
\bibitem [{\citenamefont {Zeiler}\ and\ \citenamefont
  {Fergus}(2014)}]{zeiler2014visualizing}%
  \BibitemOpen
  \bibfield  {author} {\bibinfo {author} {\bibfnamefont {M.~D.}\ \bibnamefont
  {Zeiler}}\ and\ \bibinfo {author} {\bibfnamefont {R.}~\bibnamefont
  {Fergus}},\ }\bibfield  {booktitle} {\emph {\bibinfo {booktitle} {European
  conference on computer vision}},\ }\href@noop {} {\ ,\ \bibinfo {pages} {818}
  (\bibinfo {year} {2014})}\BibitemShut {NoStop}%
\bibitem [{\citenamefont {Szegedy}\ \emph {et~al.}(2015)\citenamefont
  {Szegedy}, \citenamefont {Liu}, \citenamefont {Jia}, \citenamefont
  {Sermanet}, \citenamefont {Reed}, \citenamefont {Anguelov}, \citenamefont
  {Erhan}, \citenamefont {Vanhoucke},\ and\ \citenamefont
  {Rabinovich}}]{szegedy2015going}%
  \BibitemOpen
  \bibfield  {author} {\bibinfo {author} {\bibfnamefont {C.}~\bibnamefont
  {Szegedy}}, \bibinfo {author} {\bibfnamefont {W.}~\bibnamefont {Liu}},
  \bibinfo {author} {\bibfnamefont {Y.}~\bibnamefont {Jia}}, \bibinfo {author}
  {\bibfnamefont {P.}~\bibnamefont {Sermanet}}, \bibinfo {author}
  {\bibfnamefont {S.}~\bibnamefont {Reed}}, \bibinfo {author} {\bibfnamefont
  {D.}~\bibnamefont {Anguelov}}, \bibinfo {author} {\bibfnamefont
  {D.}~\bibnamefont {Erhan}}, \bibinfo {author} {\bibfnamefont
  {V.}~\bibnamefont {Vanhoucke}}, \ and\ \bibinfo {author} {\bibfnamefont
  {A.}~\bibnamefont {Rabinovich}},\ }\bibfield  {booktitle} {\emph {\bibinfo
  {booktitle} {Proceedings of the IEEE conference on computer vision and
  pattern recognition}},\ }\href@noop {} {\ ,\ \bibinfo {pages} {1} (\bibinfo
  {year} {2015})}\BibitemShut {NoStop}%
\bibitem [{\citenamefont {Radford}\ \emph {et~al.}(2015)\citenamefont
  {Radford}, \citenamefont {Metz},\ and\ \citenamefont
  {Chintala}}]{radford2015unsupervised}%
  \BibitemOpen
  \bibfield  {author} {\bibinfo {author} {\bibfnamefont {A.}~\bibnamefont
  {Radford}}, \bibinfo {author} {\bibfnamefont {L.}~\bibnamefont {Metz}}, \
  and\ \bibinfo {author} {\bibfnamefont {S.}~\bibnamefont {Chintala}},\
  }\href@noop {} {\bibfield  {journal} {\bibinfo  {journal} {arXiv preprint
  arXiv:1511.06434}\ } (\bibinfo {year} {2015})}\BibitemShut {NoStop}%
\bibitem [{\citenamefont {Mitchell}\ and\ \citenamefont
  {Sheppard}(2012)}]{mitchell2012deep}%
  \BibitemOpen
  \bibfield  {author} {\bibinfo {author} {\bibfnamefont {B.}~\bibnamefont
  {Mitchell}}\ and\ \bibinfo {author} {\bibfnamefont {J.}~\bibnamefont
  {Sheppard}},\ }in\ \href@noop {} {\emph {\bibinfo {booktitle} {{2012 11th
  International Conference on Machine Learning and Applications}}}},\
  Vol.~\bibinfo {volume} {1}\ (\bibinfo {organization} {IEEE},\ \bibinfo {year}
  {2012})\ pp.\ \bibinfo {pages} {162--167}\BibitemShut {NoStop}%
\bibitem [{\citenamefont {Rueshendorff}()}]{wasserstein}%
  \BibitemOpen
  \bibfield  {author} {\bibinfo {author} {\bibfnamefont {L.}~\bibnamefont
  {Rueshendorff}},\ }\href
  {http://www.encyclopediaofmath.org/index.php?title=Wasserstein_metric&oldid=32292}
  {\enquote {\bibinfo {title} {Wasserstein metric},}\ }\BibitemShut {NoStop}%
\bibitem [{\citenamefont {Abbara}\ \emph {et~al.}(2019)\citenamefont {Abbara},
  \citenamefont {Kabashima}, \citenamefont {Obuchi},\ and\ \citenamefont
  {Xu}}]{abbara2019learning}%
  \BibitemOpen
  \bibfield  {author} {\bibinfo {author} {\bibfnamefont {A.}~\bibnamefont
  {Abbara}}, \bibinfo {author} {\bibfnamefont {Y.}~\bibnamefont {Kabashima}},
  \bibinfo {author} {\bibfnamefont {T.}~\bibnamefont {Obuchi}}, \ and\ \bibinfo
  {author} {\bibfnamefont {Y.}~\bibnamefont {Xu}},\ }\href@noop {} {\bibfield
  {journal} {\bibinfo  {journal} {arXiv preprint arXiv:1912.11591}\ } (\bibinfo
  {year} {2019})}\BibitemShut {NoStop}%
\bibitem [{\citenamefont {Schulz}\ \emph {et~al.}(2010)\citenamefont {Schulz},
  \citenamefont {M{\"u}ller},\ and\ \citenamefont
  {Behnke}}]{schulz2010investigating}%
  \BibitemOpen
  \bibfield  {author} {\bibinfo {author} {\bibfnamefont {H.}~\bibnamefont
  {Schulz}}, \bibinfo {author} {\bibfnamefont {A.}~\bibnamefont {M{\"u}ller}},
  \ and\ \bibinfo {author} {\bibfnamefont {S.}~\bibnamefont {Behnke}},\ }in\
  \href@noop {} {\emph {\bibinfo {booktitle} {NIPS 2010 Workshop on Deep
  Learning and Unsupervised Feature Learning}}}\ (\bibinfo {year}
  {2010})\BibitemShut {NoStop}%
\bibitem [{\citenamefont {Fisher}\ \emph {et~al.}(2018)\citenamefont {Fisher},
  \citenamefont {Smith},\ and\ \citenamefont {Walsh}}]{fisher2018boltzmann}%
  \BibitemOpen
  \bibfield  {author} {\bibinfo {author} {\bibfnamefont {C.~K.}\ \bibnamefont
  {Fisher}}, \bibinfo {author} {\bibfnamefont {A.~M.}\ \bibnamefont {Smith}}, \
  and\ \bibinfo {author} {\bibfnamefont {J.~R.}\ \bibnamefont {Walsh}},\
  }\href@noop {} {\bibfield  {journal} {\bibinfo  {journal} {arXiv preprint
  arXiv:1804.08682}\ } (\bibinfo {year} {2018})}\BibitemShut {NoStop}%
\bibitem [{\citenamefont {Goodfellow}\ \emph {et~al.}(2014)\citenamefont
  {Goodfellow}, \citenamefont {Pouget-Abadie}, \citenamefont {Mirza},
  \citenamefont {Xu}, \citenamefont {Warde-Farley}, \citenamefont {Ozair},
  \citenamefont {Courville},\ and\ \citenamefont
  {Bengio}}]{goodfellow2014generative}%
  \BibitemOpen
  \bibfield  {author} {\bibinfo {author} {\bibfnamefont {I.}~\bibnamefont
  {Goodfellow}}, \bibinfo {author} {\bibfnamefont {J.}~\bibnamefont
  {Pouget-Abadie}}, \bibinfo {author} {\bibfnamefont {M.}~\bibnamefont
  {Mirza}}, \bibinfo {author} {\bibfnamefont {B.}~\bibnamefont {Xu}}, \bibinfo
  {author} {\bibfnamefont {D.}~\bibnamefont {Warde-Farley}}, \bibinfo {author}
  {\bibfnamefont {S.}~\bibnamefont {Ozair}}, \bibinfo {author} {\bibfnamefont
  {A.}~\bibnamefont {Courville}}, \ and\ \bibinfo {author} {\bibfnamefont
  {Y.}~\bibnamefont {Bengio}},\ }in\ \href@noop {} {\emph {\bibinfo {booktitle}
  {Advances in neural information processing systems}}}\ (\bibinfo {year}
  {2014})\ pp.\ \bibinfo {pages} {2672--2680}\BibitemShut {NoStop}%
\bibitem [{\citenamefont {Liu}\ \emph {et~al.}(2017)\citenamefont {Liu},
  \citenamefont {Rodrigues},\ and\ \citenamefont {Cai}}]{liu2017simulating}%
  \BibitemOpen
  \bibfield  {author} {\bibinfo {author} {\bibfnamefont {Z.}~\bibnamefont
  {Liu}}, \bibinfo {author} {\bibfnamefont {S.~P.}\ \bibnamefont {Rodrigues}},
  \ and\ \bibinfo {author} {\bibfnamefont {W.}~\bibnamefont {Cai}},\
  }\href@noop {} {\bibfield  {journal} {\bibinfo  {journal} {arXiv preprint
  arXiv:1710.04987}\ } (\bibinfo {year} {2017})}\BibitemShut {NoStop}%
\bibitem [{\citenamefont {Long}\ \emph {et~al.}(2015)\citenamefont {Long},
  \citenamefont {Shelhamer},\ and\ \citenamefont {Darrell}}]{long2015fully}%
  \BibitemOpen
  \bibfield  {author} {\bibinfo {author} {\bibfnamefont {J.}~\bibnamefont
  {Long}}, \bibinfo {author} {\bibfnamefont {E.}~\bibnamefont {Shelhamer}}, \
  and\ \bibinfo {author} {\bibfnamefont {T.}~\bibnamefont {Darrell}},\ }in\
  \href@noop {} {\emph {\bibinfo {booktitle} {{Proceedings of the IEEE
  conference on computer vision and pattern recognition}}}}\ (\bibinfo {year}
  {2015})\ pp.\ \bibinfo {pages} {3431--3440}\BibitemShut {NoStop}%
\bibitem [{\citenamefont {H{\"u}gli}\ \emph {et~al.}(2012)\citenamefont
  {H{\"u}gli}, \citenamefont {Duff}, \citenamefont {O'Conchuir}, \citenamefont
  {Mengotti}, \citenamefont {Rodr{\'\i}guez}, \citenamefont {Nolting},
  \citenamefont {Heyderman},\ and\ \citenamefont
  {Braun}}]{hugli2012artificial}%
  \BibitemOpen
  \bibfield  {author} {\bibinfo {author} {\bibfnamefont {R.}~\bibnamefont
  {H{\"u}gli}}, \bibinfo {author} {\bibfnamefont {G.}~\bibnamefont {Duff}},
  \bibinfo {author} {\bibfnamefont {B.}~\bibnamefont {O'Conchuir}}, \bibinfo
  {author} {\bibfnamefont {E.}~\bibnamefont {Mengotti}}, \bibinfo {author}
  {\bibfnamefont {A.~F.}\ \bibnamefont {Rodr{\'\i}guez}}, \bibinfo {author}
  {\bibfnamefont {F.}~\bibnamefont {Nolting}}, \bibinfo {author} {\bibfnamefont
  {L.}~\bibnamefont {Heyderman}}, \ and\ \bibinfo {author} {\bibfnamefont
  {H.}~\bibnamefont {Braun}},\ }\href@noop {} {\bibfield  {journal} {\bibinfo
  {journal} {{Philosophical Transactions of the Royal Society A: Mathematical,
  Physical and Engineering Sciences}}\ }\textbf {\bibinfo {volume} {370}},\
  \bibinfo {pages} {5767} (\bibinfo {year} {2012})}\BibitemShut {NoStop}%
\bibitem [{\citenamefont {Westerhout}\ \emph {et~al.}(2019)\citenamefont
  {Westerhout}, \citenamefont {Astrakhantsev}, \citenamefont {Tikhonov},
  \citenamefont {Katsnelson},\ and\ \citenamefont
  {Bagrov}}]{westerhout2019neural}%
  \BibitemOpen
  \bibfield  {author} {\bibinfo {author} {\bibfnamefont {T.}~\bibnamefont
  {Westerhout}}, \bibinfo {author} {\bibfnamefont {N.}~\bibnamefont
  {Astrakhantsev}}, \bibinfo {author} {\bibfnamefont {K.~S.}\ \bibnamefont
  {Tikhonov}}, \bibinfo {author} {\bibfnamefont {M.}~\bibnamefont
  {Katsnelson}}, \ and\ \bibinfo {author} {\bibfnamefont {A.~A.}\ \bibnamefont
  {Bagrov}},\ }\href@noop {} {\bibfield  {journal} {\bibinfo  {journal} {arXiv
  preprint arXiv:1907.08186}\ } (\bibinfo {year} {2019})}\BibitemShut {NoStop}%
\bibitem [{\citenamefont {Cirillo}\ \emph {et~al.}(1999)\citenamefont
  {Cirillo}, \citenamefont {Gonnella}, \citenamefont {Troccoli},\ and\
  \citenamefont {Maritan}}]{cirillo1999correlation}%
  \BibitemOpen
  \bibfield  {author} {\bibinfo {author} {\bibfnamefont {E.~N.}\ \bibnamefont
  {Cirillo}}, \bibinfo {author} {\bibfnamefont {G.}~\bibnamefont {Gonnella}},
  \bibinfo {author} {\bibfnamefont {M.}~\bibnamefont {Troccoli}}, \ and\
  \bibinfo {author} {\bibfnamefont {A.}~\bibnamefont {Maritan}},\ }\href@noop
  {} {\bibfield  {journal} {\bibinfo  {journal} {{J. Stat. Phys.}}\ }\textbf
  {\bibinfo {volume} {94}},\ \bibinfo {pages} {67} (\bibinfo {year}
  {1999})}\BibitemShut {NoStop}%
\bibitem [{\citenamefont {Tanaka}(2002)}]{tanaka2002methods}%
  \BibitemOpen
  \bibfield  {author} {\bibinfo {author} {\bibfnamefont {T.}~\bibnamefont
  {Tanaka}},\ }\href@noop {} {\emph {\bibinfo {title} {Methods of statistical
  physics}}}\ (\bibinfo  {publisher} {{Cambridge University Press}},\ \bibinfo
  {year} {2002})\BibitemShut {NoStop}%
\bibitem [{\citenamefont {Pelizzola}(2005)}]{pelizzola2005cluster}%
  \BibitemOpen
  \bibfield  {author} {\bibinfo {author} {\bibfnamefont {A.}~\bibnamefont
  {Pelizzola}},\ }\href@noop {} {\bibfield  {journal} {\bibinfo  {journal} {J.
  Phys. A}\ }\textbf {\bibinfo {volume} {38}},\ \bibinfo {pages} {R309}
  (\bibinfo {year} {2005})}\BibitemShut {NoStop}%
\bibitem [{git()}]{github}%
  \BibitemOpen
  \href@noop {} {\enquote {\bibinfo {title}
  {\url{https://github.com/dngfra/Restricted-Boltzmann-Machines}},}\
  }\BibitemShut {NoStop}%
\end{thebibliography}%
\bibliographystyle{apsrev4-1}
\clearpage

\onecolumngrid

\appendix
\renewcommand\thefigure{\thesection.\arabic{figure}}
\setcounter{figure}{0}    
\section{Non-convolutional variational autoencoder}
\label{sec:non_conVAE}
\begin{figure*}[htp!]
\includegraphics[width=0.32\textwidth]{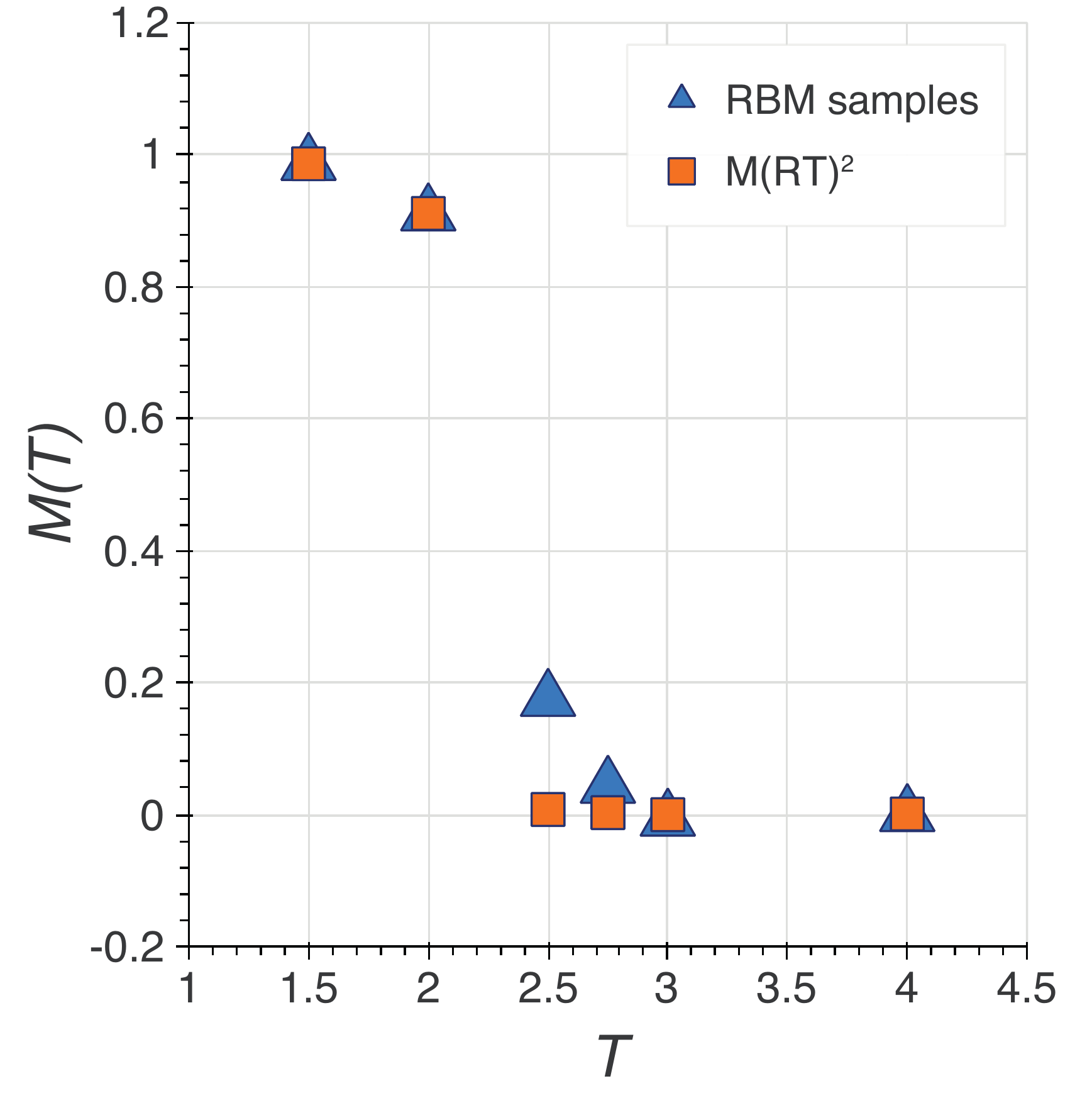}
\includegraphics[width=0.32\textwidth]{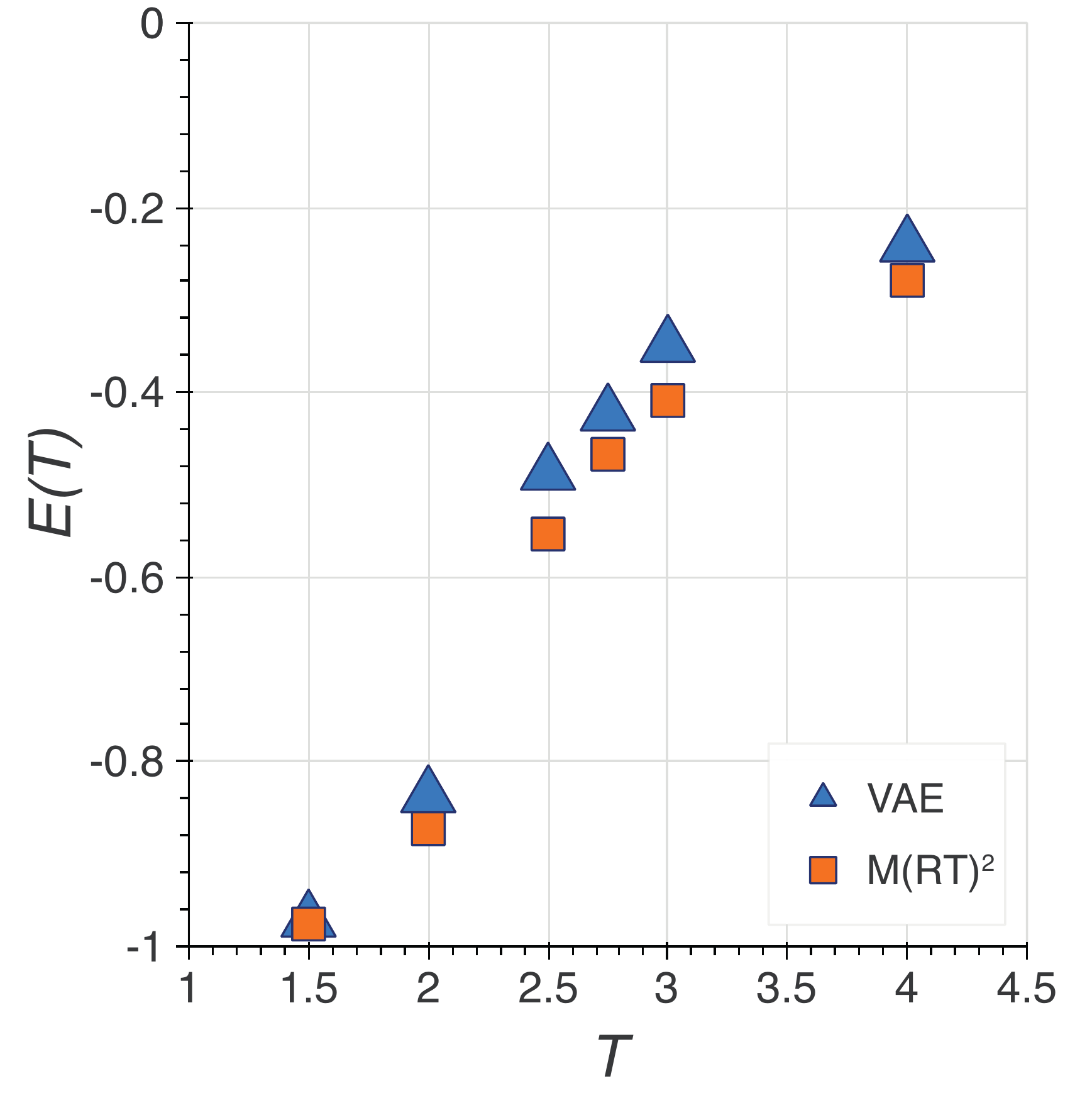}
\includegraphics[width=0.32\textwidth]{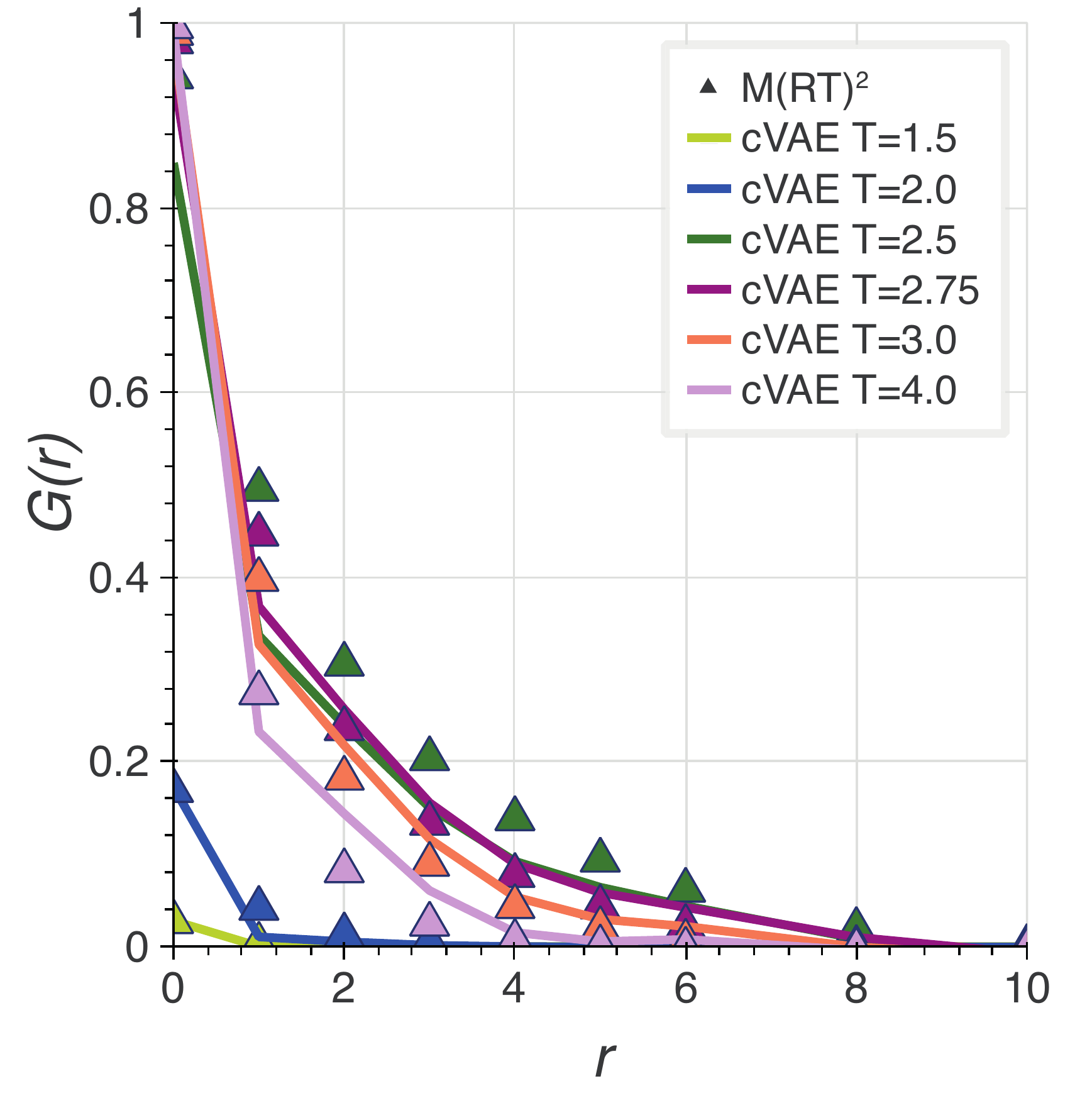}
\caption{\textbf{Physical features of non-convolutional VAE Ising samples.} We use non-convolutional cVAEs to generate $20\times 10^4$ samples of Ising configurations with $32\times 32$ spins for temperatures $T\in\{1.5,2,2.5,2.75,3,4\}$. We show the magnetization $M(T)$, energy $E(T)$, and correlation function $G(r,T)$ for neural-network- and corresponding M(RT)$^2$ samples. The non-convolutional cVAE was trained for all temperatures. Error bars are smaller than the markers.}
\label{fig:non_conVAE}
\end{figure*}
In addition to the convolutional architecture of cVAEs that we described in Sec.~\ref{sec:VAE}, we also considered their non-convolutational counterparts. We used an encoder that is composed of an input layer with 1024 units, one fully connected hidden layers with 700 units, and a sigmoid activation function. As for the convolutional VAEs, we also used dropout to avoid overfitting. The output layer has 400 units and represents a 200-dimensional Gaussian latent variable $\mb{z}$. We use a decoder with a symmetrical architecture composed of an input layer that consists of 200 units to represent the latent variable $\mb{z}$. There are six additional units that we use to encode the temperature of Ising samples. The remaining layers in the decoder are one fully connected hidden layer with 700 units and a sigmoid activation function. The output layer is sigmoidal and consists of 1024 units. 

We trained the non-convolutational cVAE in the same way as the convolutational cVAE (see Sec.~\ref{sec:VAE}). We then generated samples and determined their temperatures with the classifier that we describe in Sec.~\ref{sec:classifier}. In Fig.~\ref{fig:non_conVAE}, we show the resulting temperature dependence of magnetization, energy, and spin correlations. We observe that energy and correlations are not captured as well as for convolutational architectures and RBMs (see Fig.~\ref{fig:physical_features}).

\newpage
\setcounter{figure}{0}    
\section{Energy and magnetization scatter plots}
\label{app:scatter_plots}
\begin{figure*}[h]
\includegraphics[width=0.32\textwidth]{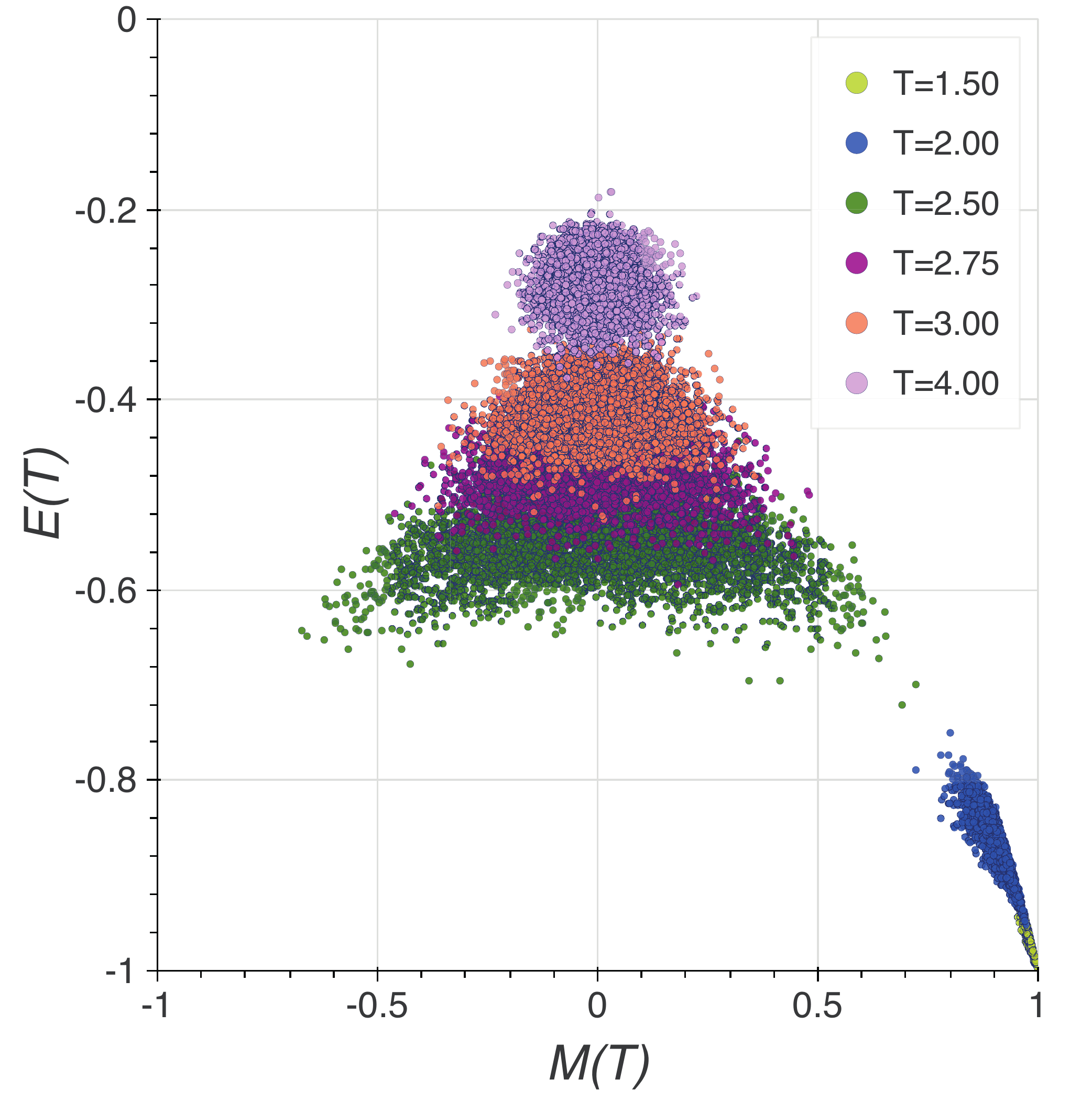}
\includegraphics[width=0.32\textwidth]{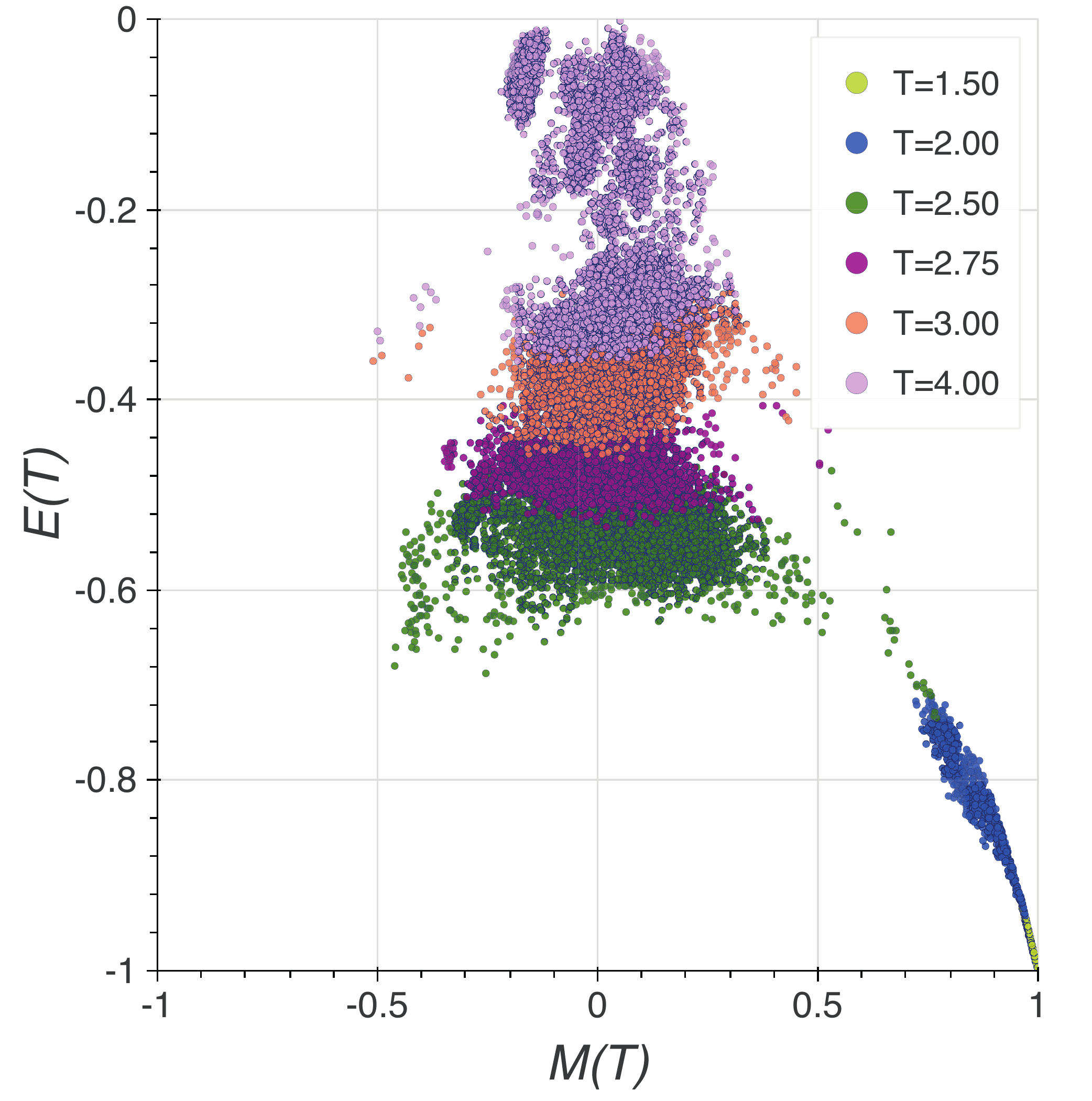}
\includegraphics[width=0.32\textwidth]{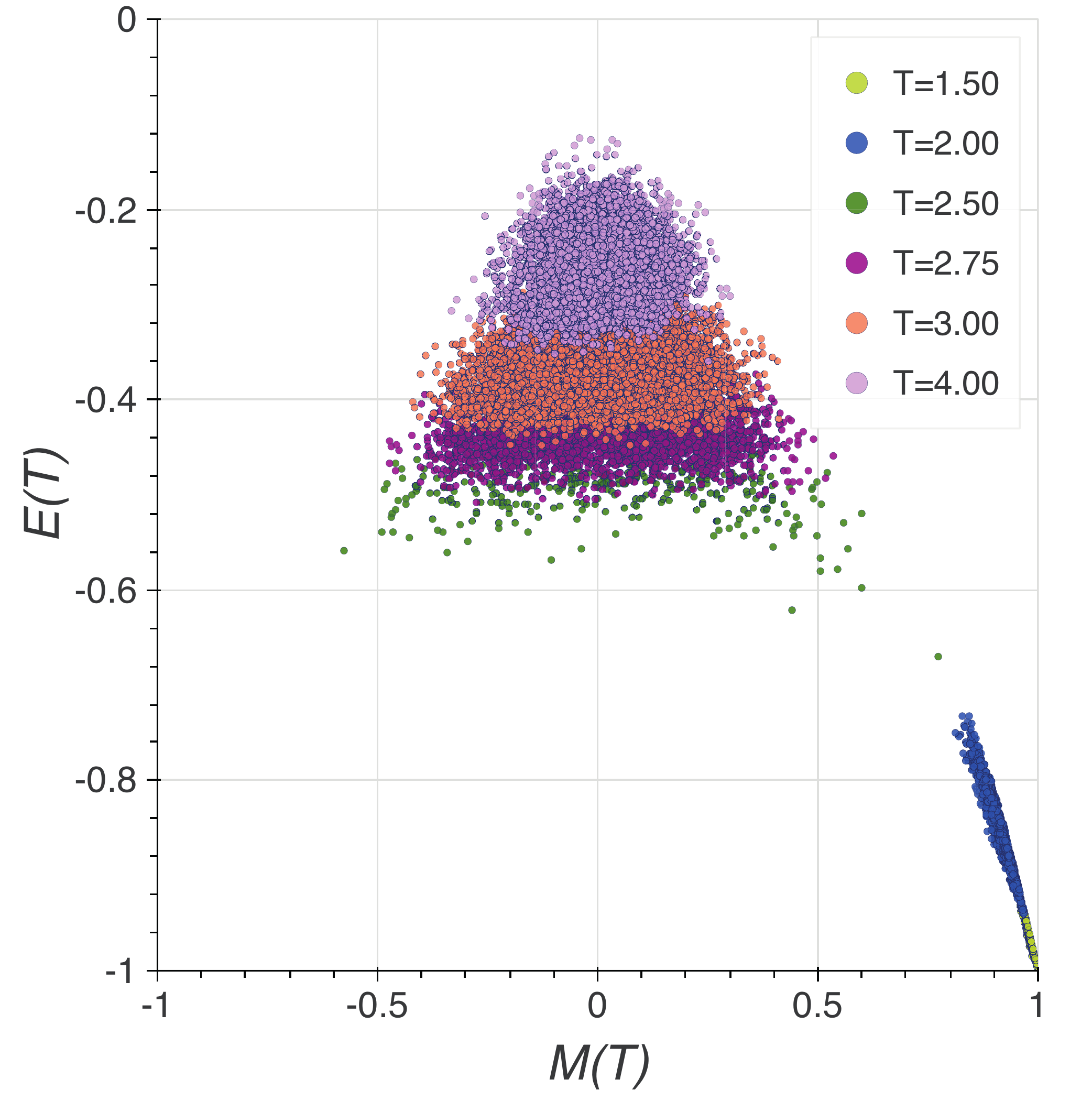}
\caption{\textbf{Two-dimensional visualization of Ising samples.} We show the distribution of $20\times 10^4$ Ising samples for temperatures $T\in\{1.5,2,2.5,2.75,3,4\}$. The data in the left, middle, and right panels are based on M(RT)$^2$, RBM, and convolutional cVAE samples, respectively.}
\label{fig:two_dim_manifold}
\end{figure*}
\begin{figure*}
\includegraphics[width=0.32\textwidth]{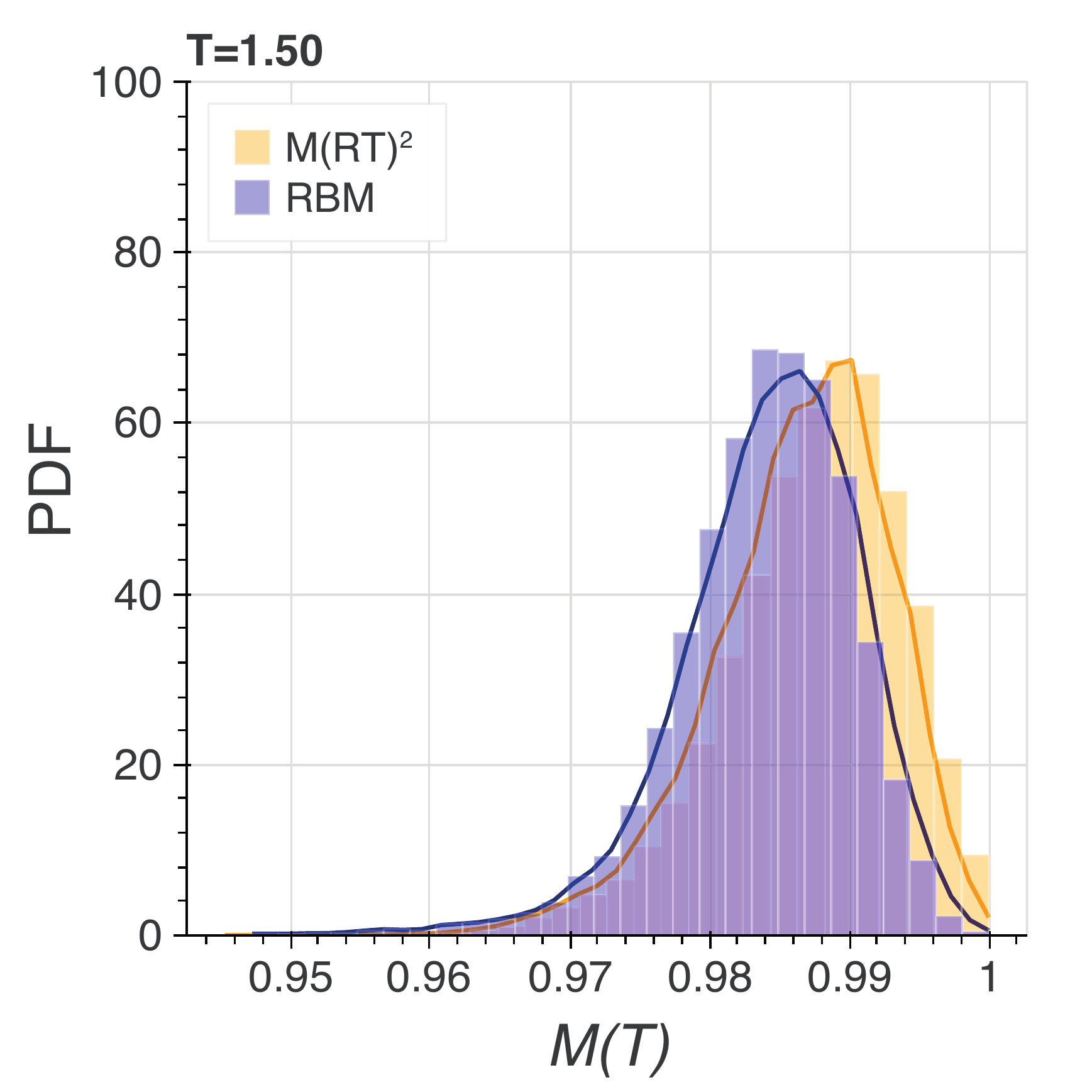}
\includegraphics[width=0.32\textwidth]{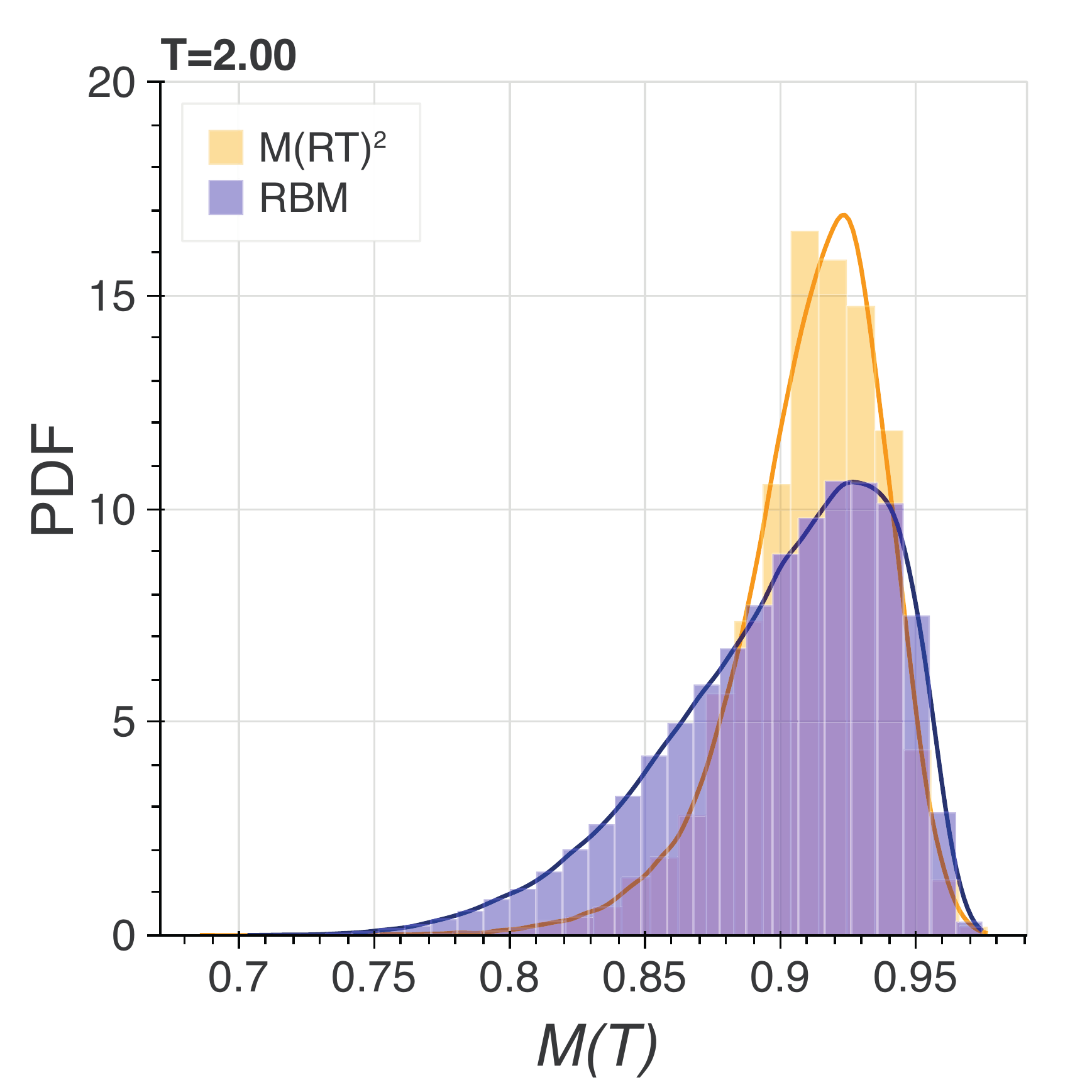}
\includegraphics[width=0.32\textwidth]{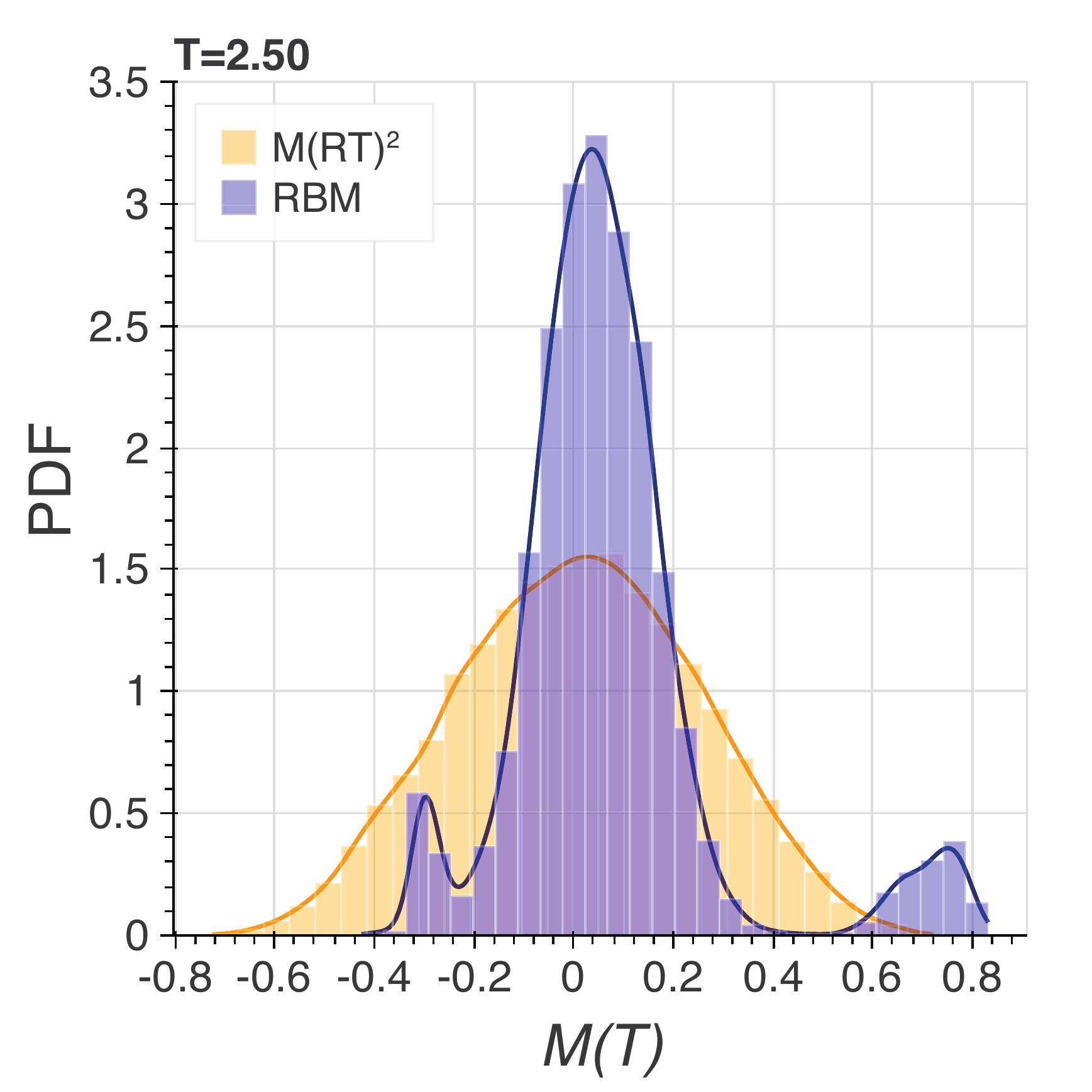}
\includegraphics[width=0.32\textwidth]{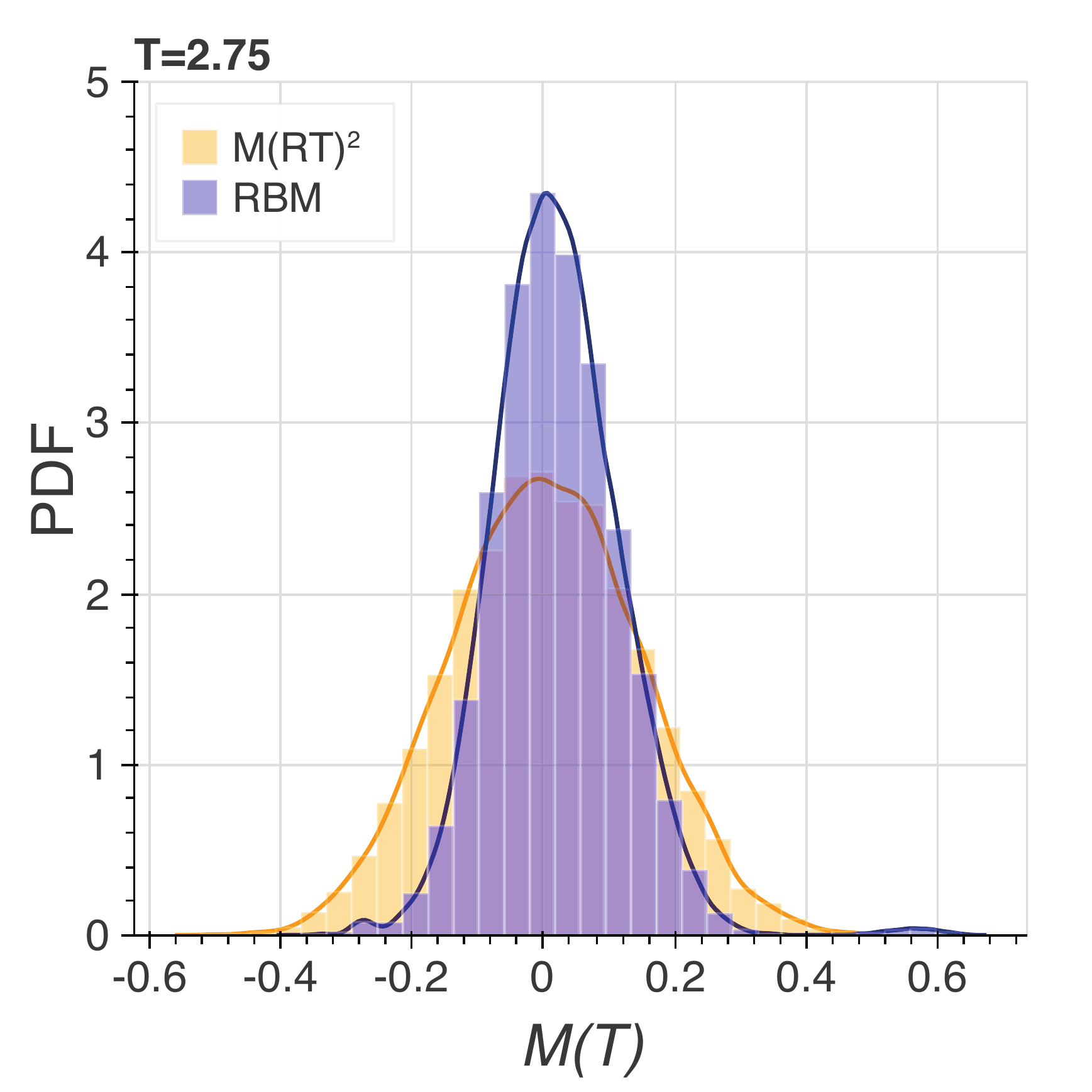}
\includegraphics[width=0.32\textwidth]{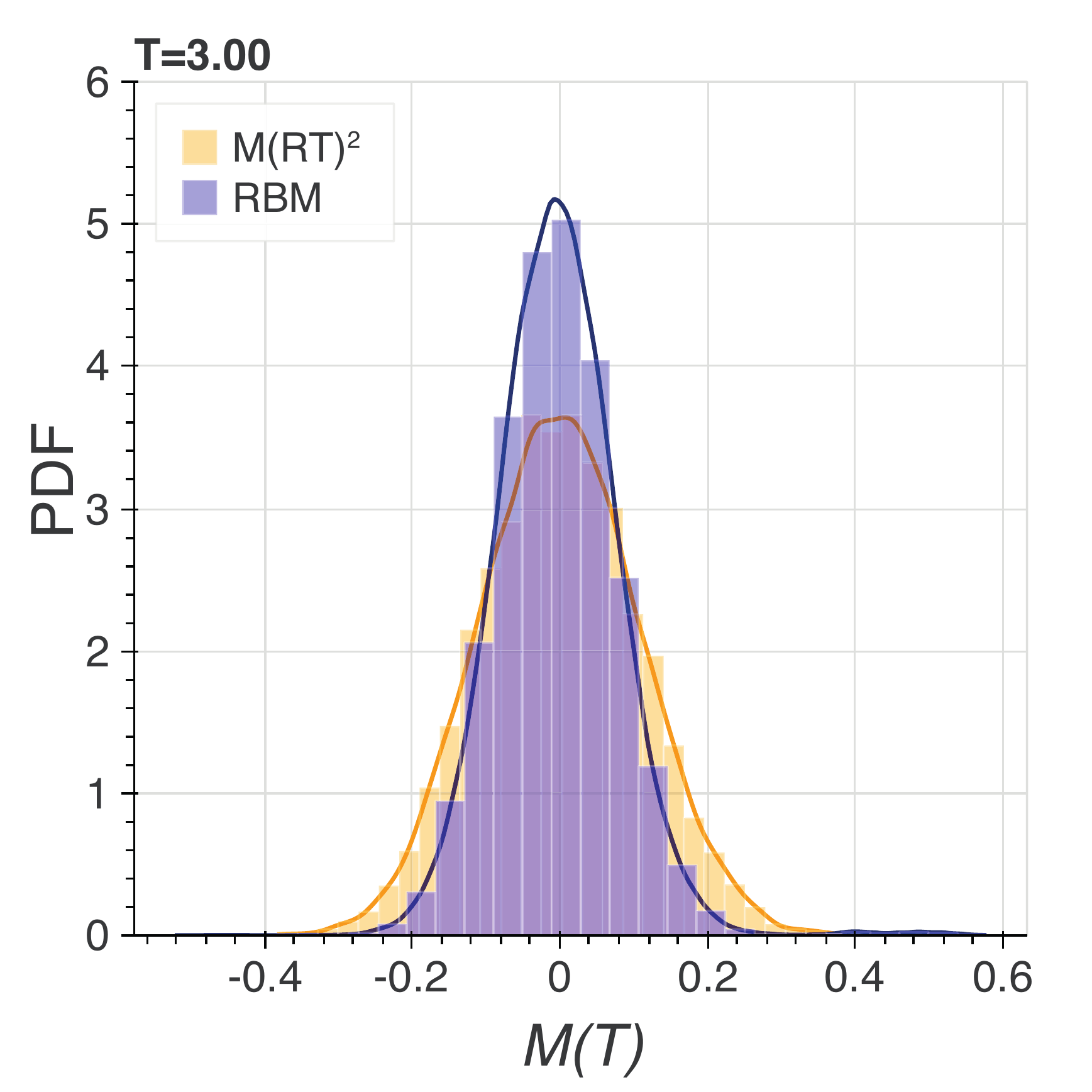}
\includegraphics[width=0.32\textwidth]{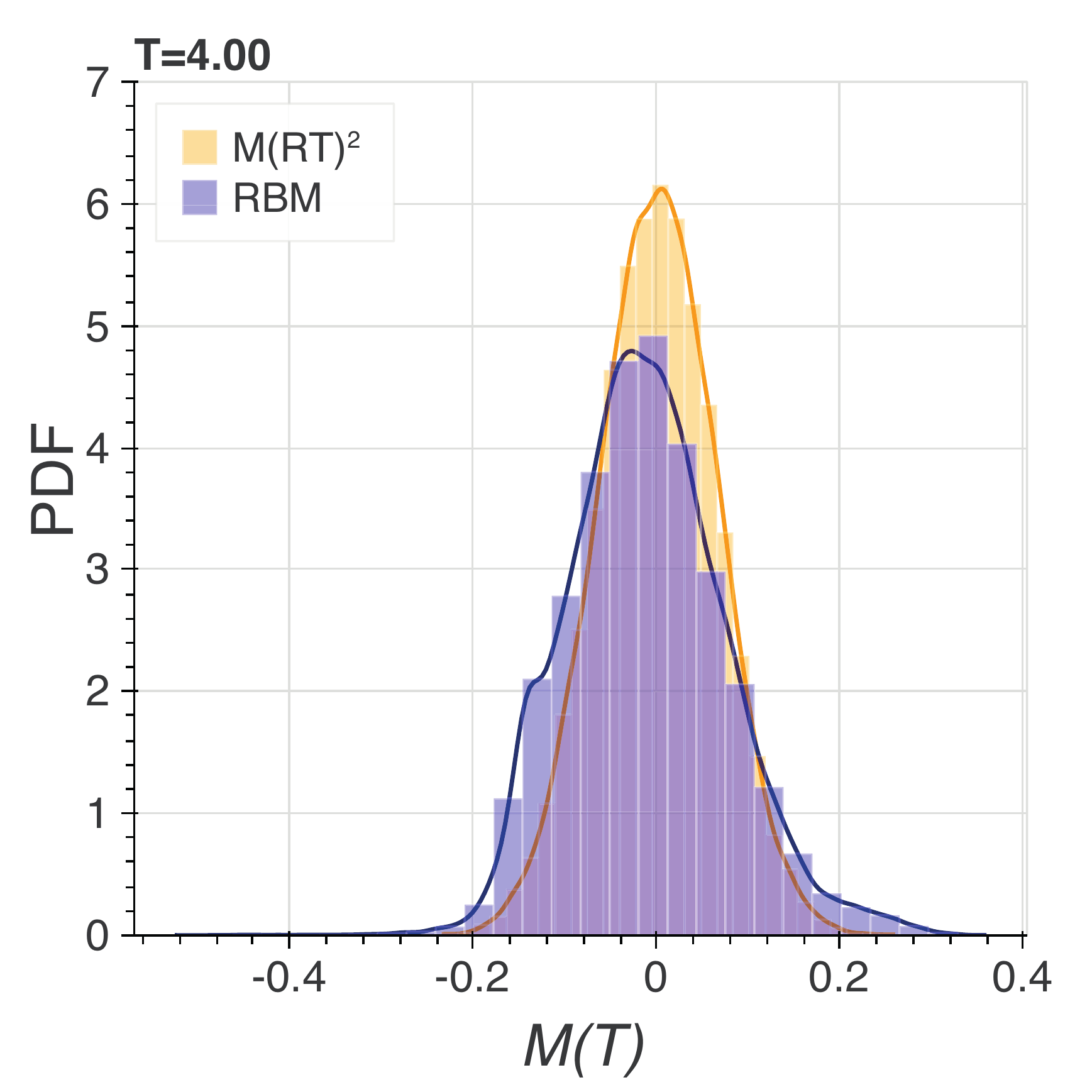}
\includegraphics[width=0.32\textwidth]{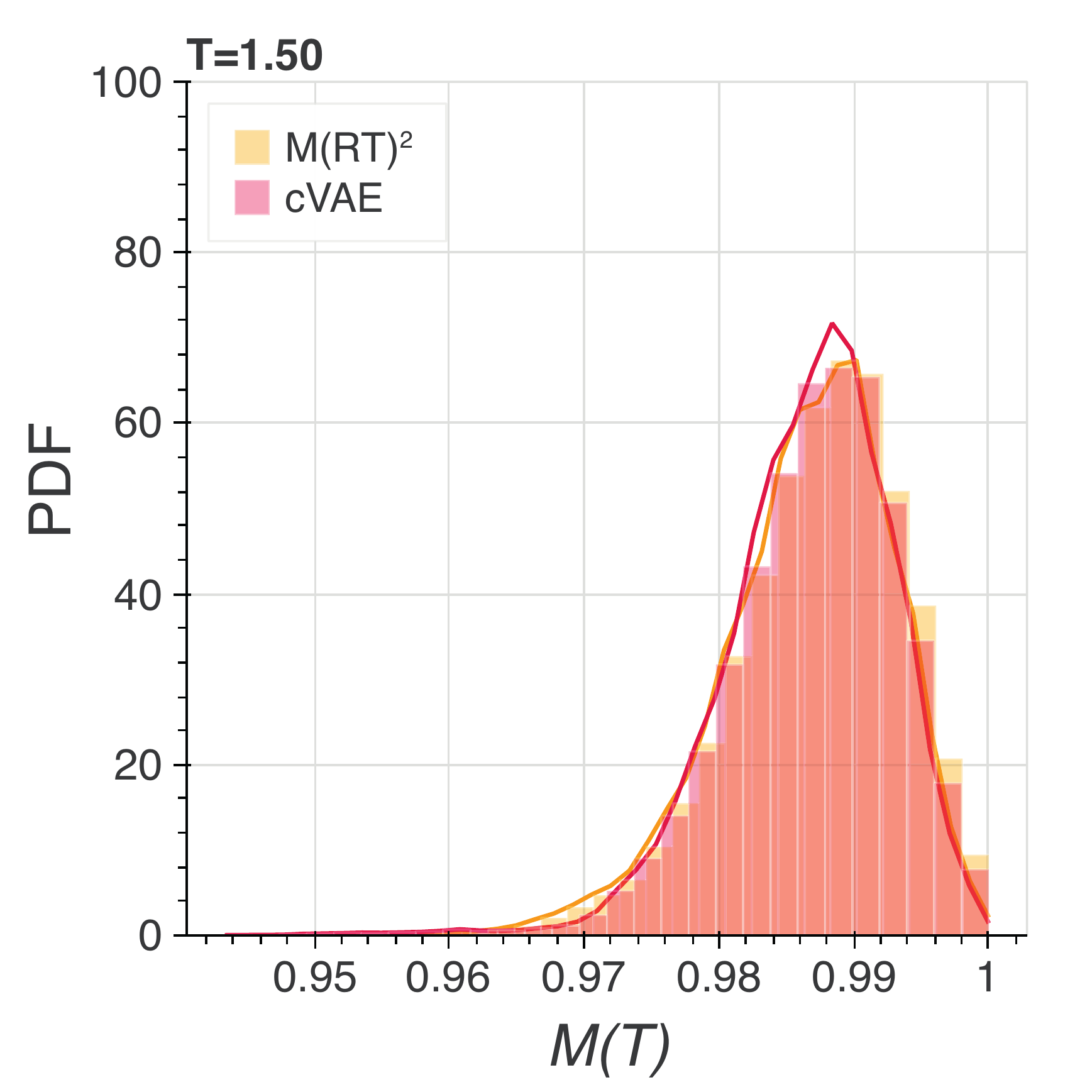}
\includegraphics[width=0.32\textwidth]{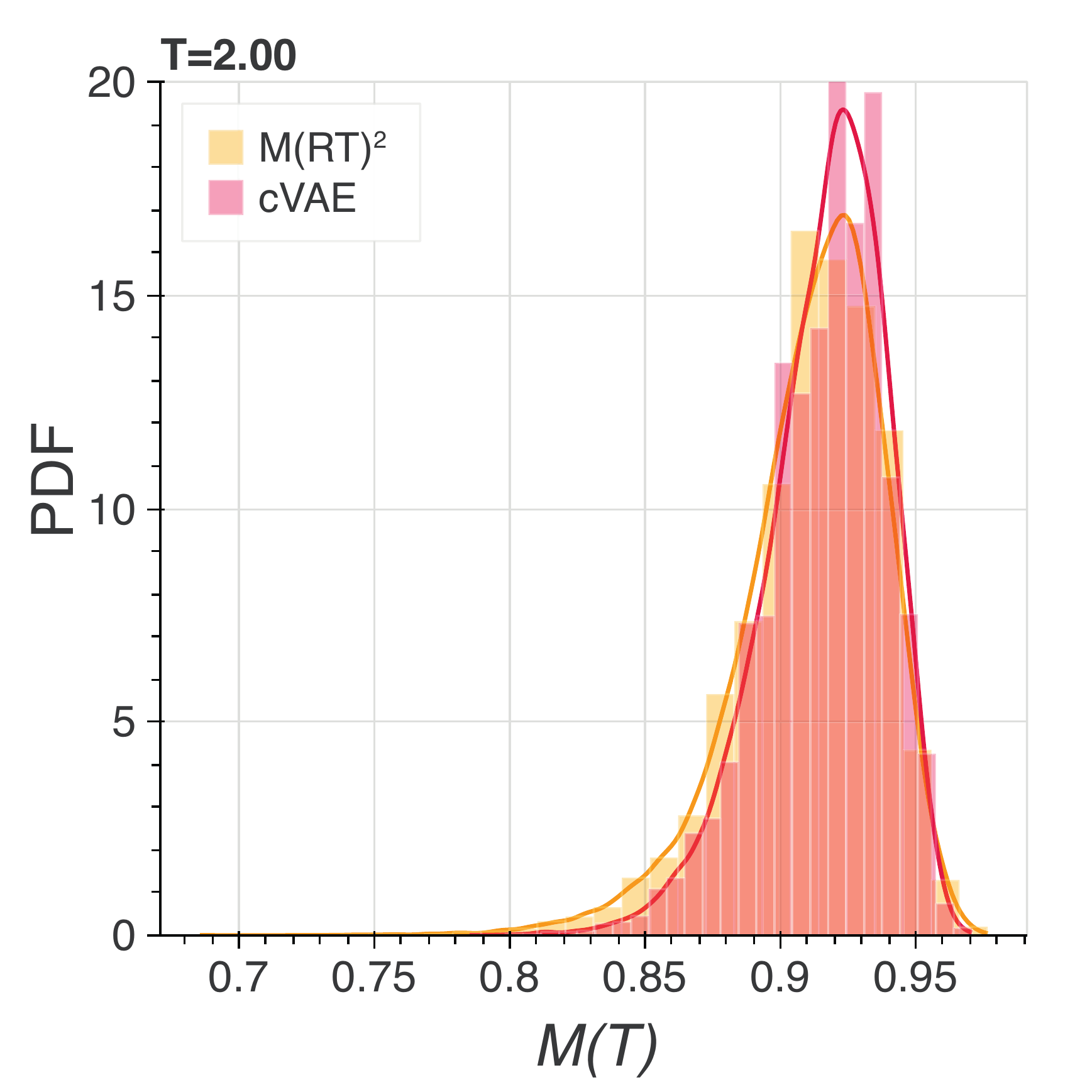}
\includegraphics[width=0.32\textwidth]{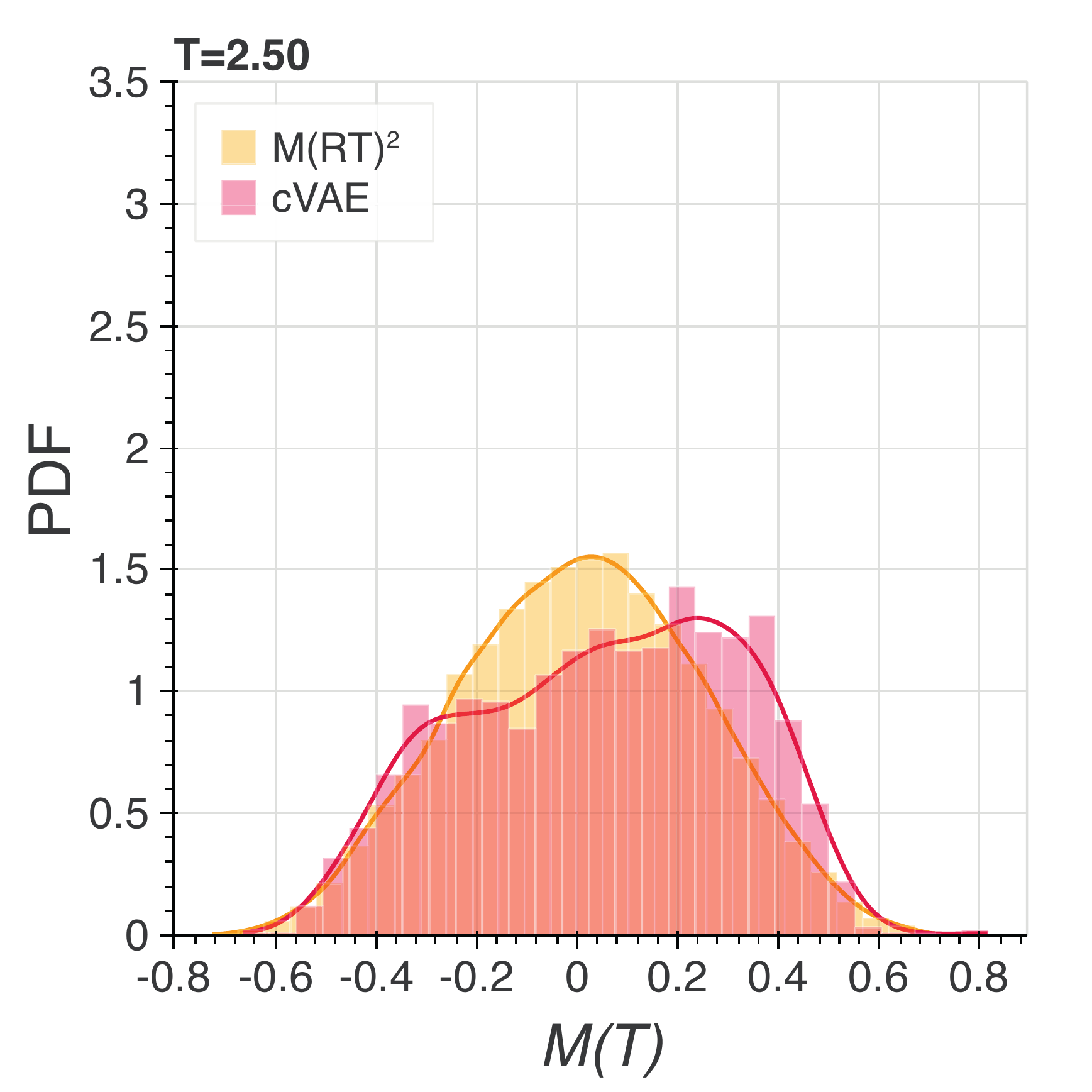}
\includegraphics[width=0.32\textwidth]{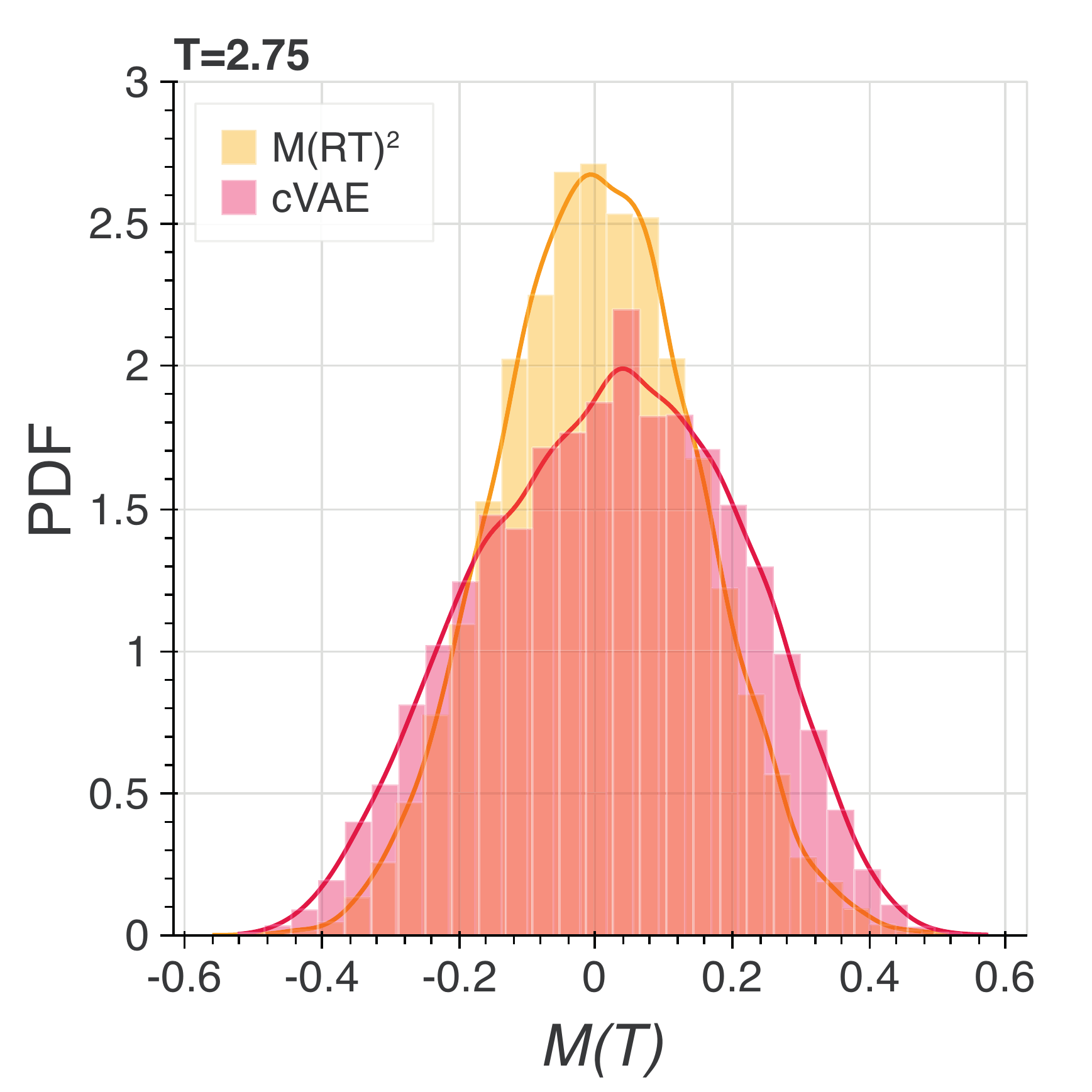}
\includegraphics[width=0.32\textwidth]{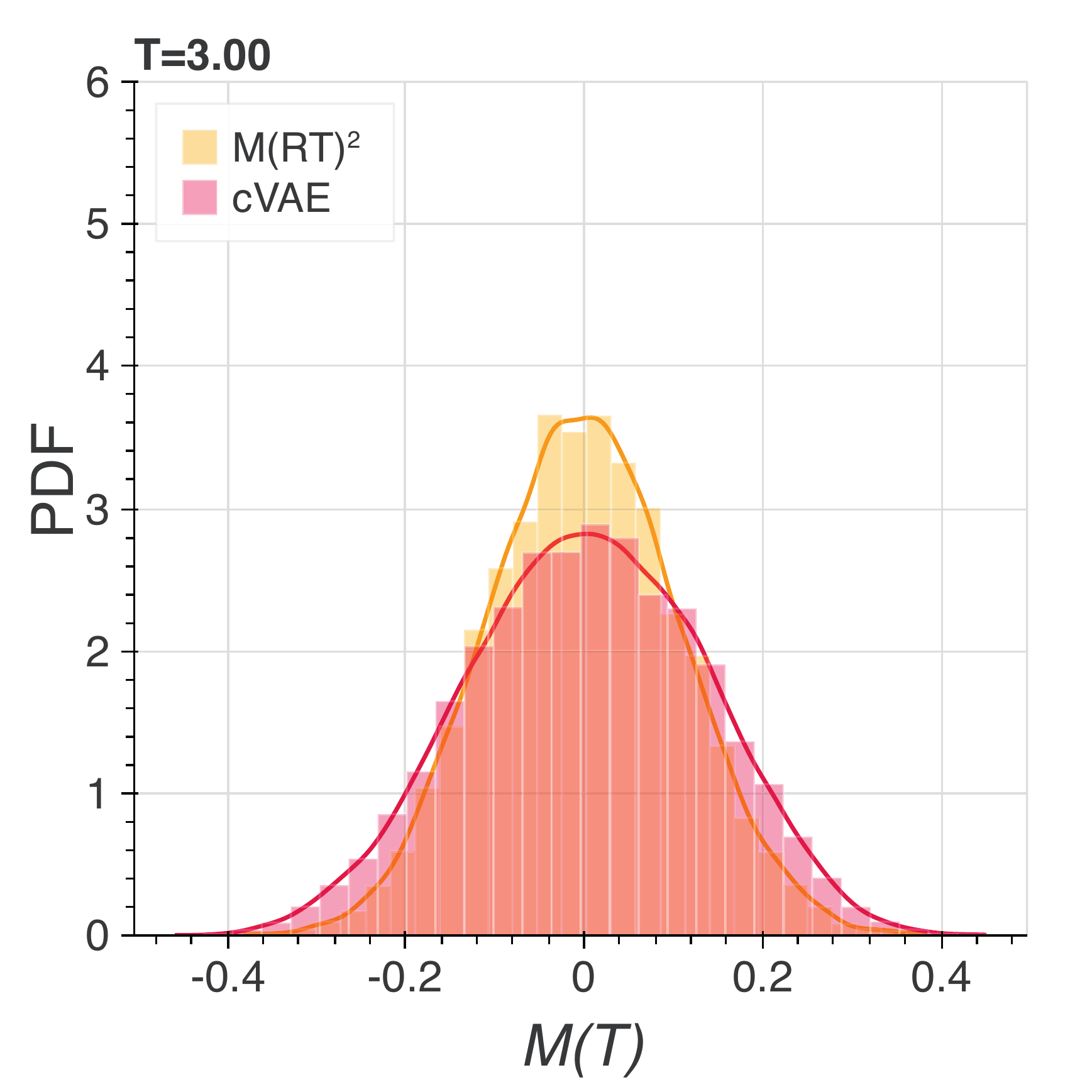}
\includegraphics[width=0.32\textwidth]{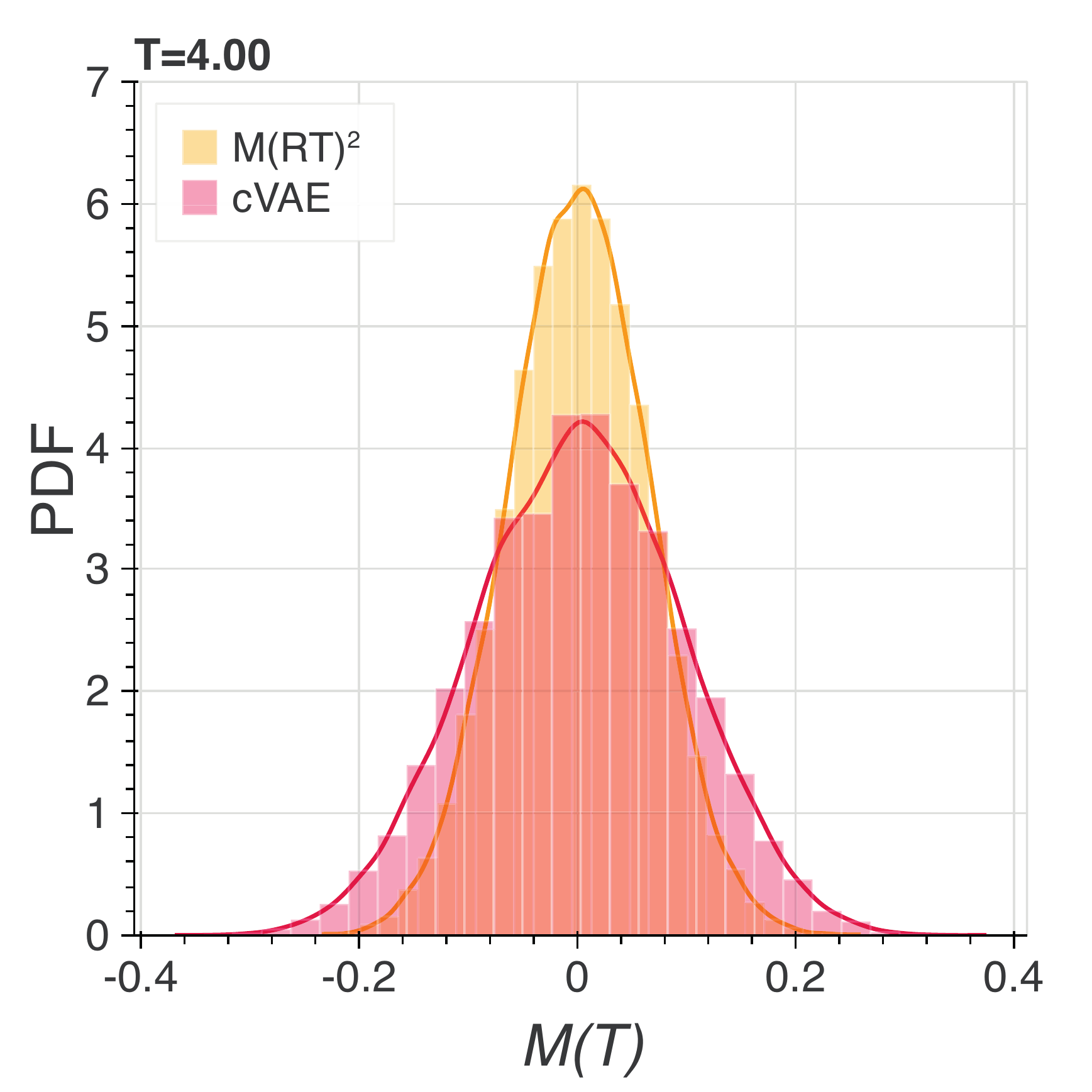}
\caption{\textbf{Distribution of magnetization at different temperatures.} We show the distribution of $20\times 10^4$ Ising samples for temperatures $T\in\{1.5,2.0,2.5,2.75,3,4\}$. The data are based on M(RT)$^2$ (yellow), RBM (blue), and convolutional cVAE (red) samples, respectively.}
\label{fig:distributions}
\end{figure*}

We generated $20\times 10^4$ samples with the RBMs and convolutional cVAEs that we trained for all temperatures $T\in\{1.5,2,2.5,2.75,3,4\}$. After determining the temperatures of the generated samples (see Sec.~\ref{sec:classifier}), we compute different physical features including energy $E(T)$ and magnetization $M(T)$. We show the two-dimensional scatter plots of $E(T)$ and $M(T)$ in Fig.~\ref{fig:two_dim_manifold}. The left panel of Fig.~\ref{fig:two_dim_manifold} shows the distribution of $E(T)$ and $M(T)$ in the original data. We find that the RBMs we consider have problems to reproduce the behavior at high temperatures (e.g., for $T=4$, see middle panel of Fig.~\ref{fig:two_dim_manifold}). As also apparent in Fig.~\ref{fig:counts}, the convolutational cVAE has difficulties to generate samples at temperatures close to $T_c$ (e.g., for $T=2.5$). These observations are also reflected in the magnetization distributions that we show in Fig.~\ref{fig:distributions}. As outlined in the main text and Tab.~\ref{tab:tableW}, the considered cVAEs are able to capture the magnetization distributions better than the considered RBMs in terms of smaller Wasserstein distances $\mathcal{W}$~\cite{wasserstein}. Still, the magnetization mean values of the considered RBMs are better aligned with the corresponding training data mean values for certain temperatures than is the case for the considered cVAEs (see Fig.~\ref{fig:physical_features}).

\newpage

\clearpage

\newpage

\setcounter{figure}{0}    
\section{Heterogeneous couplings}
\label{app:heterogeneous}
\begin{figure}
\centering
\includegraphics[width=0.32\textwidth]{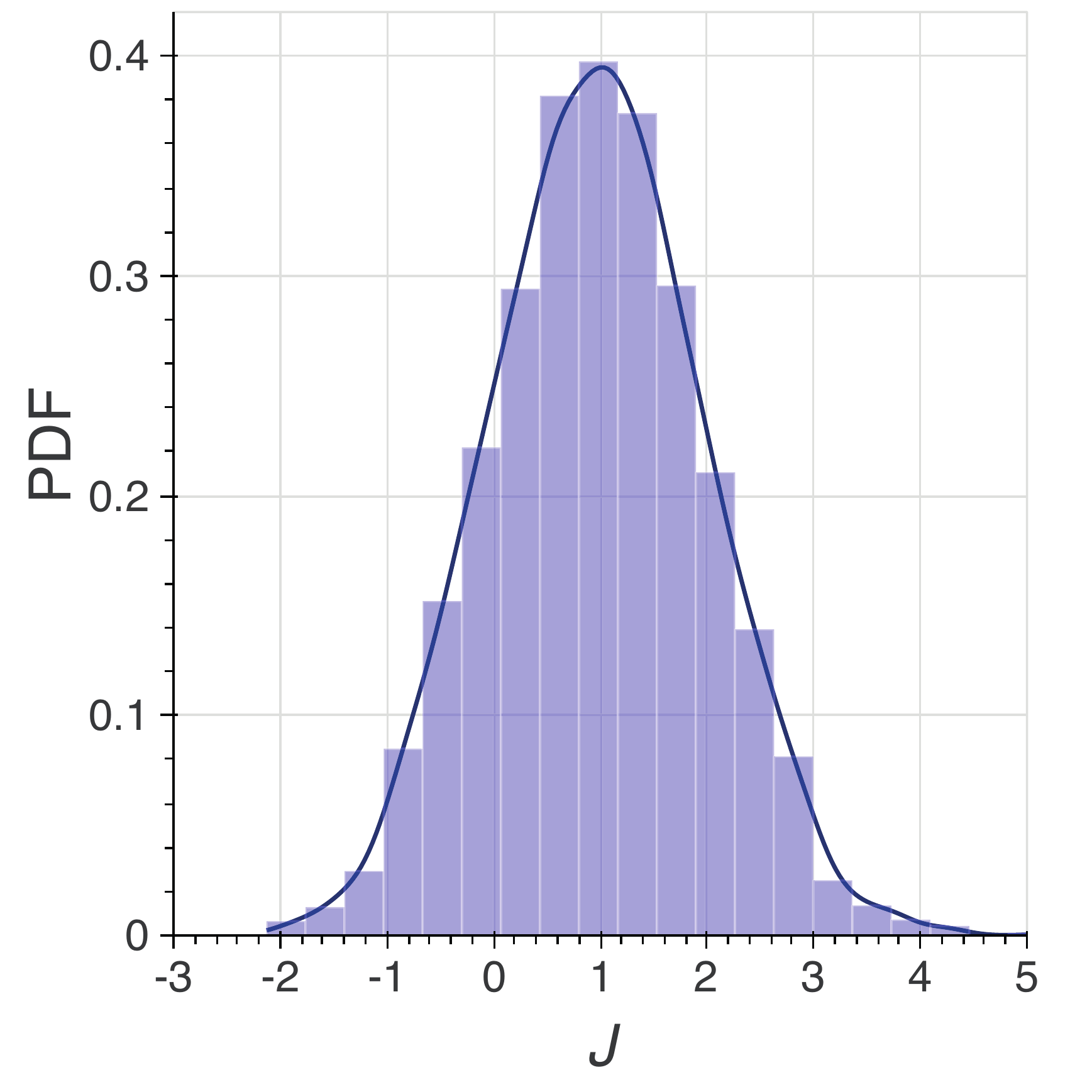}
\caption{\textbf{Gaussian coupling distribution}. Gaussian distribution of Ising spin couplings (see Eq.~\eqref{eq:ising_hamiltonian}). The mean and standard deviation of the Gaussian distribution are both equal to one.}
\label{fig:coupling_gauss}
\end{figure}
\begin{figure*}
\includegraphics[width=0.32\textwidth]{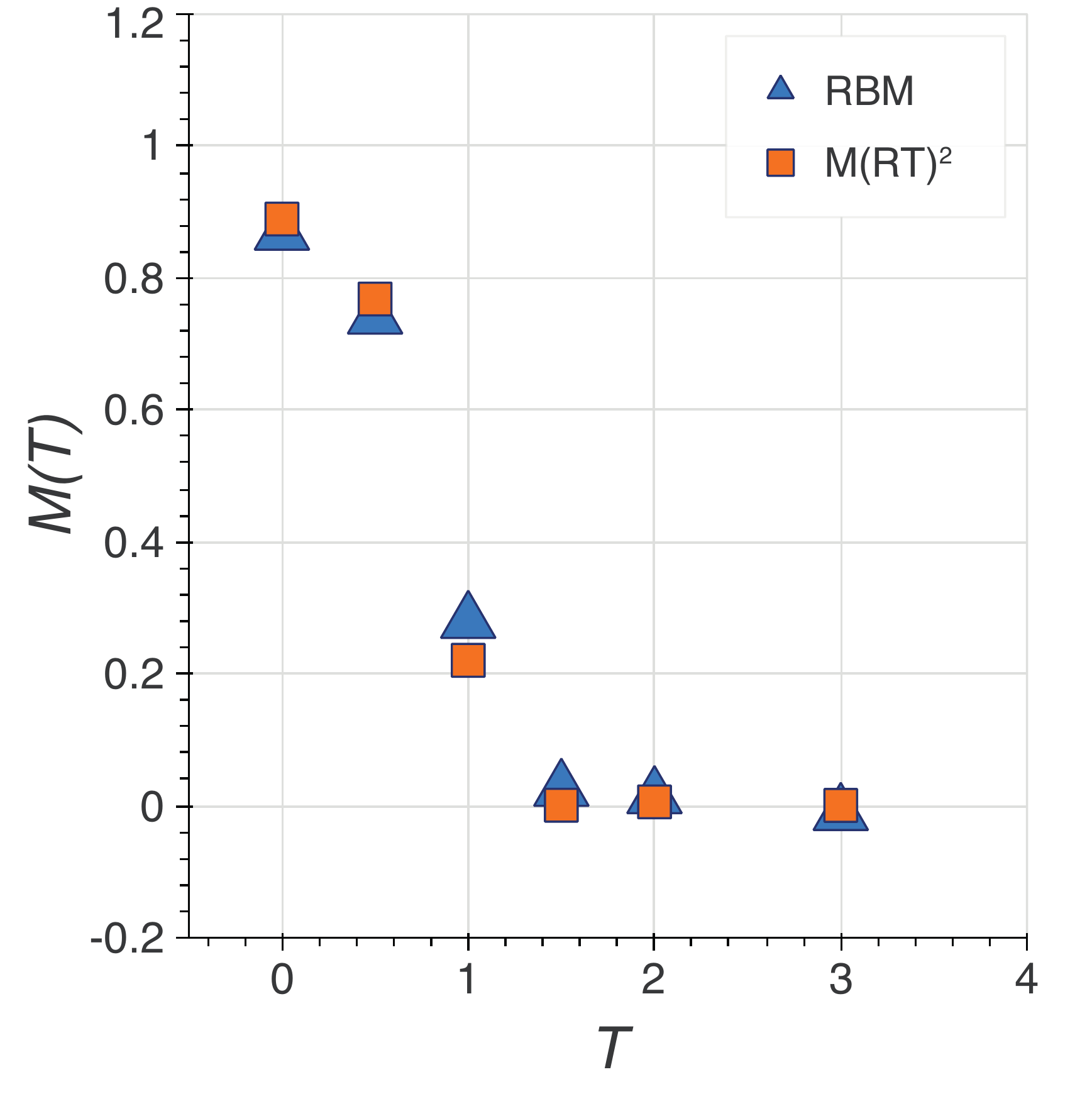}
\includegraphics[width=0.32\textwidth]{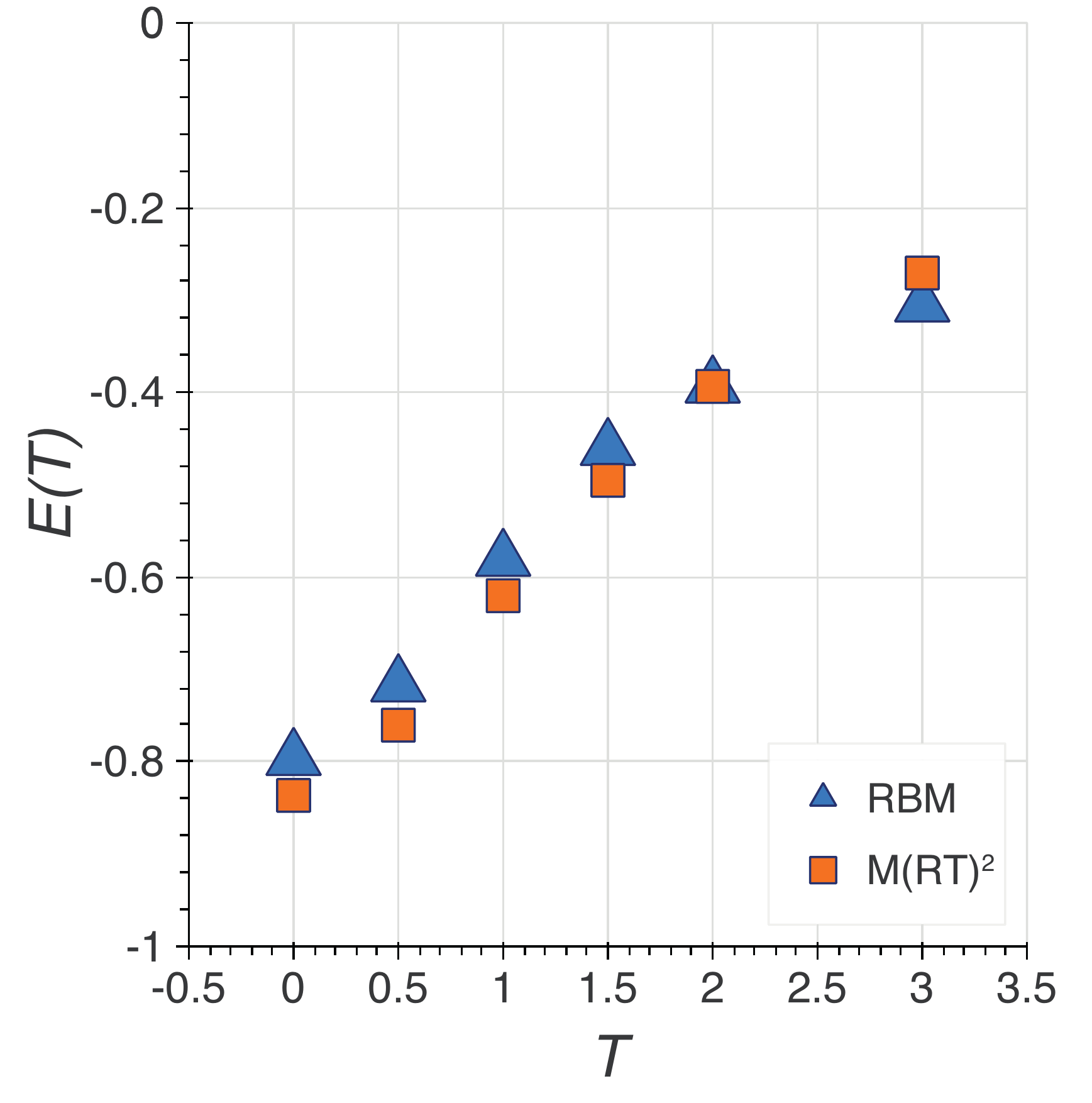}
\includegraphics[width=0.32\textwidth]{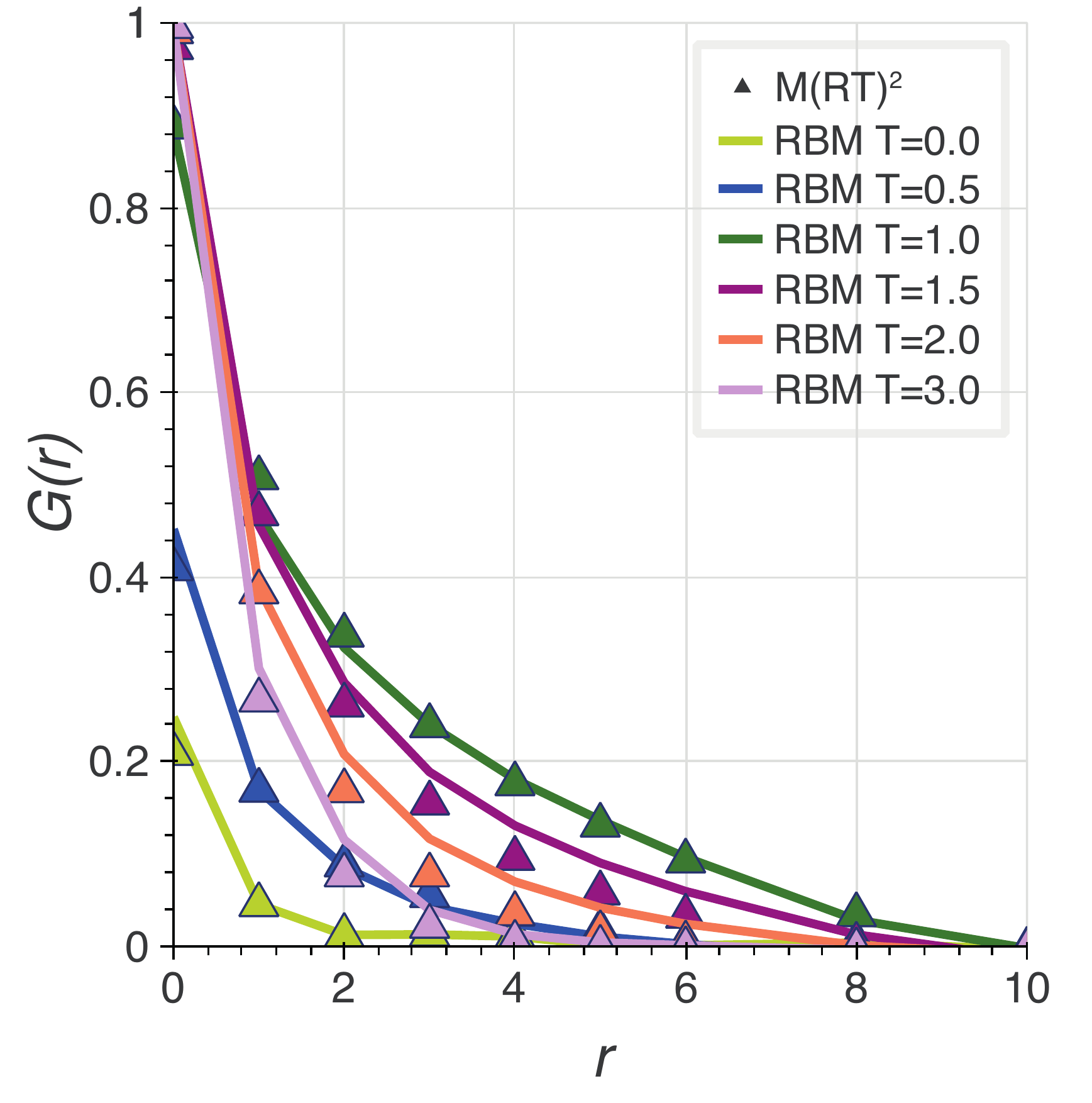}
\includegraphics[width=0.32\textwidth]{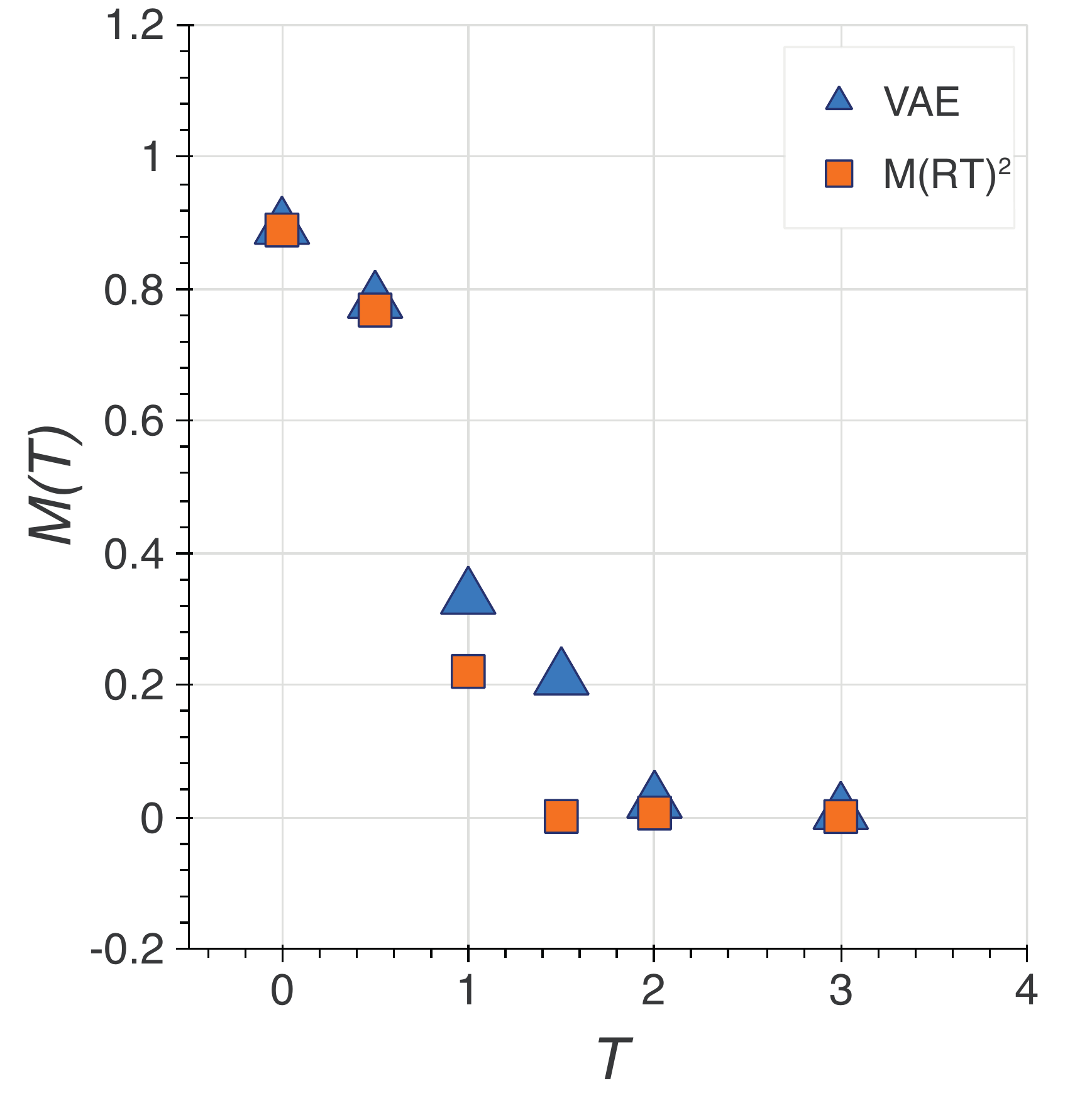}
\includegraphics[width=0.32\textwidth]{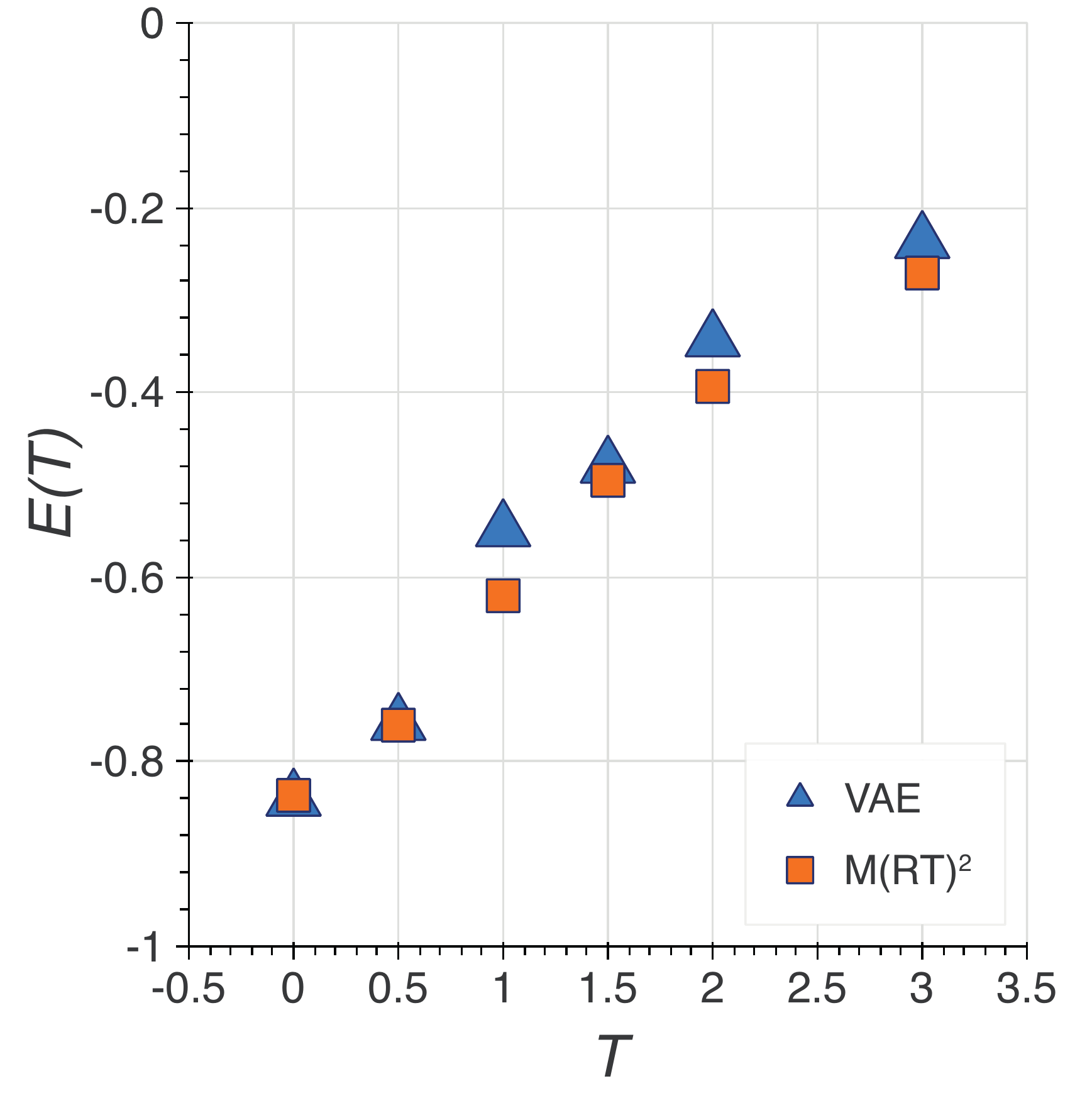}
\includegraphics[width=0.32\textwidth]{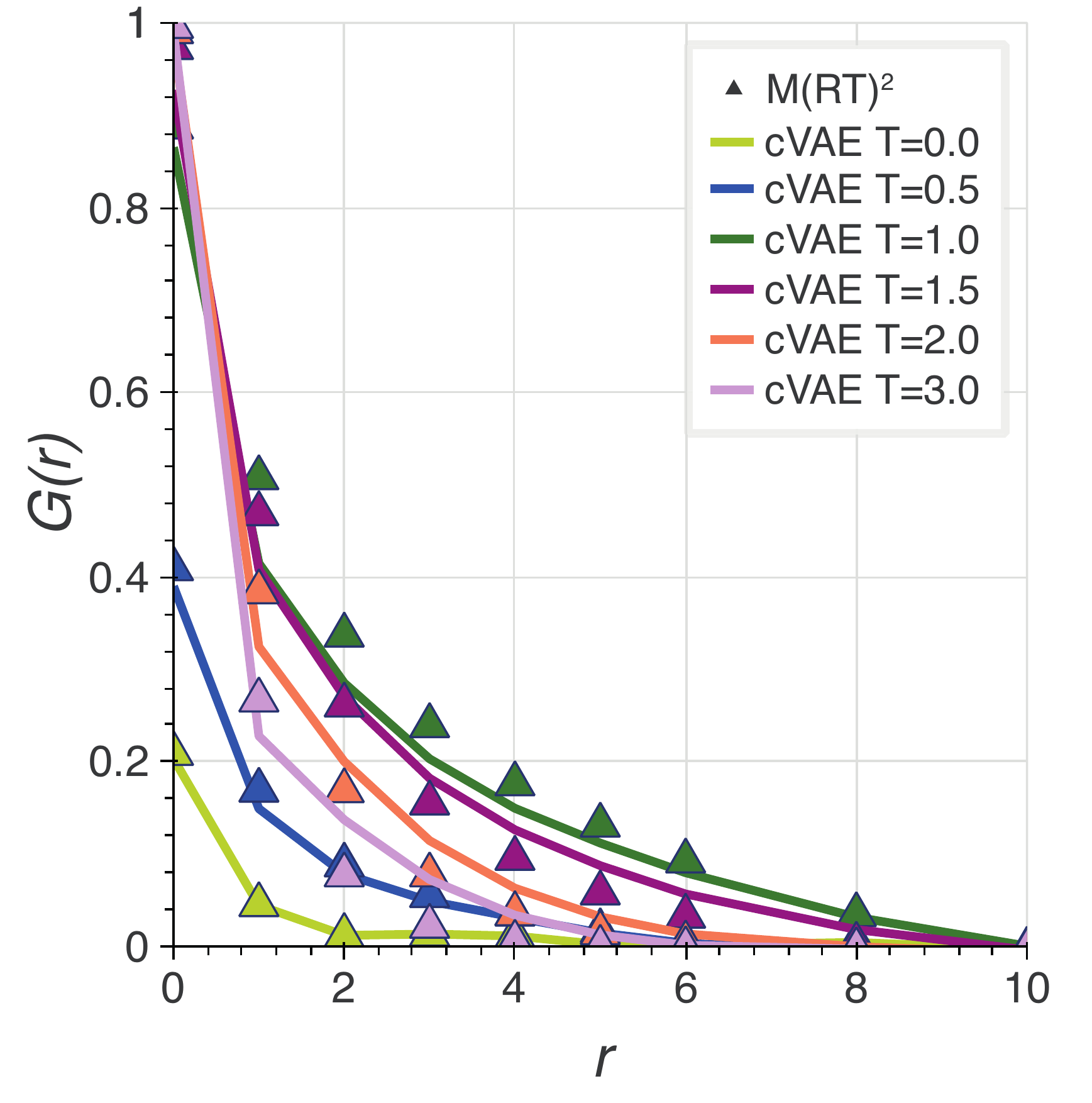}
\caption{\textbf{Physical features of RBM and convolutional VAE Ising samples for heterogeneous couplings}. We use RBMs and convolutional cVAEs to generate $20\times 10^4$ samples of Ising configurations with $32\times 32$ spins for temperatures $T\in\{10^{-6},0.5,1,1.5,2,3\}$ and a Gaussian coupling distribution whose mean and standard deviation are equal to one. We show the magnetization $M(T)$, energy $E(T)$, and correlation function $G(r,T)$ for neural network and corresponding M(RT)$^2$ samples. In the top panel, we used a single RBM for all temperatures; in the bottom panel, we show the behavior of $M(T)$, $E(T)$, and $G(r,T)$ for samples that are generated with a convolutional cVAE that was trained for all temperatures. Error bars are smaller than the markers.}
\label{fig:physical_features_heterogeneous}
\end{figure*}

We also apply RBMs and cVAEs to an Ising model with spin couplings that are distributed according to a Gaussian distribution whose mean and standard deviation are both equal to one (see Fig.~\ref{fig:coupling_gauss}). We use the RBM and cVAE architectures of Sec.~\ref{sec:learning} and generated $20\times 10^4$  M(RT)$^2$ samples for temperatures $T\in \{10^{-6},0.5,1,1.5,2,3\}$ to train the two neural networks. Training protocols are also as outlined in Sec.~\ref{sec:learning}. Only in the case of the RBM, we train over 90 epochs instead of 200 epochs, as in Sec.~\ref{sec:learning}, and repeat this procedure 10 times. After training both neural networks, we generated $20\times 10^4$ Ising configurations. We show the corresponding results in Fig.~\ref{fig:physical_features_heterogeneous} and observe that both neural networks are able to qualitatively capture the temperature dependence of magnetization, energy, and spin correlations.
\end{document}